%% file: Pamporovo.tex
\documentclass[openany]{book}
\usepackage{times}
\usepackage{fancyhdr}
\usepackage{epsf}

\newcommand{\dd}{\mbox{d}}
\newcommand{\DD}{\mbox{D}}
\newcommand{\bbox}[1]{\mbox{\boldmath $#1$}} 
\newcommand{\tfrac}[2]{{\textstyle\frac{#1}{#2}}}

\def\overlay#1#2{\ifmmode%
\setbox0=\hbox{$#1$}%
\setbox1=\hbox to\wd0{\hss$#2$\hss}\else%
\setbox0=\hbox{#1}%
\setbox1=\hbox to\wd0{\hss#2\hss}\fi%
#1\hskip-\wd0\box1 }
\begin{document}
\pagenumbering{roman}
\title{\Huge \bf Quantum Phase Transitions \\ in \\ 2d Quantum
Liquids\footnote{\normalsize Lectures presented at the first Pamporovo
International Winter Workshop on {\it Cooperative Phenomena in Condensed
Matter}, Pamporovo, Bulgarian, March 7-15, 1998.  } }  \author{\LARGE
\bf Adriaan M.\ J.\ Schakel\\ \\ \bf Institut f\"ur Theoretische Physik \\
\bf Freie Universit\"at Berlin \\ \bf Arnimallee 14, \bf 14195 Berlin }
\maketitle
\pagestyle{fancyplain}
\renewcommand{\chaptermark}[1]%
     {\markboth{Chapter \thechapter \hspace{.1cm} \ #1}{}}
\renewcommand{\sectionmark}[1]%
     {\markright{\thesection \hspace{.1cm} \ #1}}
\lhead[\thepage]{\bf\let\uppercase\relax\rightmark}
\rhead[\bf\let\uppercase\relax\leftmark]{\thepage}
\cfoot{}
\newcommand{\TheAuthor}{}
\newcommand{\Author}[1]{\renewcommand{\TheAuthor}{#1}}
\rfoot[\TheAuthor]{}
\lfoot[]{\TheAuthor}
\Author{\copyright \bf Amstex}
%
%
%
%
%
\tableofcontents
\include{Notation}

\setcounter{chapter}{0}
\pagenumbering{arabic}
\include{Prelude}

\include{Functional}

\include{Superfluid}
\include{Superconductor}
\include{FQHE}

\include{QPT}

\include{Bib}
\end{document}

%% file: Notation.tex
\chapter*{Notation\markboth{Notation}{Notation}}
\label{chap:not}
We adopt Feynman's notation and denote a spacetime point by $x=x_\mu =(t,{\bf
x})$, $\mu = 0,1, \cdots,d$, with $d$ the number of space dimensions, while
the energy $k_0$ and momentum ${\bf k}$ of a particle will be denoted by
$k=k_\mu = (k_0,{\bf k})$.  The time derivative $\partial_0 =
\partial/\partial t$ and the gradient $\nabla$ are sometimes combined in a
single vector $\tilde{\partial}_\mu = (\partial_0, -\nabla)$.  The tilde on
$\partial_\mu$ is to alert the reader for the minus sign appearing in the
spatial components of this vector.  We define the scalar product $k \cdot x
= k_\mu x_\mu = k_0 t - {\bf k} \cdot {\bf x}$ and use Einstein's summation
convention.  Because of the minus sign in the definition of the vector
$\tilde{\partial}_\mu$ it follows that $\tilde{\partial}_\mu a_\mu =
\partial_0 a_0 + \nabla \cdot {\bf a}$, with $a_\mu$ an arbitrary vector.

Integrals over spacetime are denoted by
$$
\int_{x} = \int_{t,{\bf x}} = \int \dd t \, \dd^d x,
$$
while those over energy and momentum by
$$
\int_k = \int_{k_0,{\bf k}} = \int \frac{\dd k_0}{2 \pi}
\frac{\dd^d k}{(2 \pi)^d}.
$$
When no integration limits are indicated, the integrals are assumed to
run over all possible values of the integration variables.

Natural units $\hbar = c = k_{\rm B} = 1$ are adopted throughout.

%% file: Prelude.tex
\chapter{Prelude \label{chap:intro}}
Continuous quantum phase transitions have attracted considerable
attention in this decade both from experimentalists as well as from
theorists.  (For reviews see Refs.\ \cite{UzunovB,LG,Sachdev,SGCS}.)
These transitions, taking place at the absolute zero of temperature, are
dominated by quantum and not by thermal fluctuations as is the case in
classical finite-temperature phase transitions.  Whereas time plays no
role in a classical phase transition, being an equilibrium phenomenon,
it becomes important in quantum phase transitions.  The dynamics is
characterized by an additional critical exponent, the so-called dynamic
exponent, which measures the asymmetry between the time and space
dimensions.  The natural language to describe these transitions is
quantum field theory.  In particular, the functional-integral approach,
which can also be employed to describe classical phase transitions,
turns out to be highly convenient.

The subject is at the border of condensed matter and statistical physics.
Typical systems being studied are superfluid and superconducting films,
quantum-Hall and related two-dimensional electron systems, as well as
quantum spin systems.  Despite the diversity in physical content, the
quantum critical behavior of these systems shows surprising similarities.  It
is fair to say that the present theoretical understanding of most of the
experimental results is scant.  

The purpose of these Lectures is to provide the reader with a framework
for studying quantum phase transitions.  A central role is played by a
repulsively interacting Bose gas at the absolute zero of temperature.
The universality class defined by this paradigm is believed to be of
relevance to most of the systems studied.  Without impurities and a
Coulomb interaction, the quantum critical behavior of this system turns
out to be surprisingly simple.  However, these two ingredients are
essential and have to be included.  Very general hyperscaling arguments
are powerful enough to determine the exact value of the dynamic exponent
in the presence of impurities and a Coulomb interaction, but the other
critical exponents become highly intractable.

The emphasis in these Lectures will be on effective theories, giving a
description of the system under study valid at low energy and small
momentum.  The rationale for this is the observation that the (quantum)
critical behavior of continuous phase transitions is determined by such
general features as the dimensionality of space, the symmetries
involved, and the dimensionality of the order parameter.  It does not
depend on the details of the underlying microscopic theory.  In the
process of deriving an effective theory starting from some microscopic
model, irrelevant degrees of freedom are integrated out and only those
relevant for the description of the phase transition are retained.
Similarities in critical behavior in different systems can, accordingly,
be more easily understood from the perspective of effective field
theories.

The ones discussed in these Lectures are so-called {\it phase-only}
theories.  The are the dynamical analogs of the familiar O(2) nonlinear
sigma model of classical statistical physics.  As in that model, the focus
will be on phase fluctuations of the order parameter.  The inclusion of
fluctuations in the modulus of the order parameter is generally believed not
to change the critical behavior.  Indeed, there are convincing arguments
that both the Landau-Ginzburg model with varying modulus and the nonlinear
O($n$) sigma model with fixed modulus belong to the same universality class.
For technical reasons a direct comparison is not possible, the
Landau-Ginzburg model usually being investigated in an expansion around four
dimensions, and the nonlinear sigma model in one around two.

In the case of a repulsively interacting Bose gas at the absolute zero
of temperature, the situation is particular simple as phase fluctuations
are the only type of field fluctuations present.

These Lectures cover exclusively lower-dimensional systems.  The reason
is that it will turn out that in three space dimensions and higher the
quantum critical behavior is in general Gaussian and therefore not very
interesting.

Since time and how it compares to the space dimensions is an important
aspect of quantum phase transitions, Galilei invariance will play an
important role in the discussion.

%% file: Functional.tex
\chapter{Functional Integrals \label{cap:funct}}
In these Lectures we shall adopt, unless stated otherwise, the
functional-integral approach to quantum field theory.  To illustrate the
use and power of functional integrals, let us consider one of the
simplest models of {\it classical} statistical mechanics: the Ising
model.  It is remarkable that functional integrals can not only be used
to describe quantum systems, governed by quantum fluctuations, but also
classical systems, governed by thermal fluctuations.  
\section{Ising Model}
The Ising model provides an idealized description of an uniaxial
ferromagnet.  To be specific, let us assume that the spins of some
lattice system can point only along one specific crystallographic axis.
The magnetic properties of this system can then be modeled by a lattice
with a spin variable $s({\bf x})$ attached to every site ${\bf x}$
taking the values $s({\bf x}) = \pm 1$.  For definiteness we will assume
a $d$-dimensional cubic lattice.  The Hamiltonian is given by
\begin{equation} \label{HIsing}
H = - \frac{1}{2} \sum_{{\bf x}, {\bf y}} J({\bf x}, {\bf y}) \, s({\bf
x}) \, s({\bf y}).
\end{equation}
Here, ${\bf x} = a \, x_i \, {\bf e}_i$, with $a$ the lattice
constant, $x_i$ integers labeling the sites, and ${\bf e}_i$ ($i = 1,
\cdots ,d$) unit vectors spanning the lattice.  The
sums over ${\bf x}$ and ${\bf y}$ extend over the entire lattice, and
$J({\bf x}, {\bf y})$ is a symmetric matrix representing the
interactions between the spins.  If the matrix element $J({\bf x}, {\bf
y})$ is positive, the energy is minimized when the two spins at site
${\bf x}$ and ${\bf y}$ are parallel---they are said to have a
ferromagnetic coupling.  If, on the other hand, the matrix element is
negative, anti-parallel spins are favored---the spins are said to have
an anti-ferromagnetic coupling.

The classical partition function $Z$ of the Ising model reads
\begin{equation}
Z =  \sum_{\{s({\bf x})\}} {\rm e}^{-\beta H}, 
\end{equation}
with $\beta = 1/T$ the inverse temperature.  The sum is over all spin
configurations $\{s({\bf x})\}$, of which there are $2^N$, with $N$
denoting the number of lattice sites.  To evaluate the partition
function we linearize the exponent by introducing an auxiliary
$\phi({\bf x})$ at each site via a so-called Hubbard-Stratonovich
transformation.  Such a transformation generalizes the Gaussian integral
\begin{equation}
\exp \left(-\tfrac{1}{2} \beta J s^2 \right) = \sqrt{\frac{\beta}{2 \pi
J}} \int_\phi
\exp \left( - \tfrac{1}{2} \beta J^{-1} \phi^2 + \beta \phi s \right),
\end{equation}  
where the integration variable $\phi$ runs from $-\infty$ to
$\infty$.  The generalization reads
\begin{eqnarray} 
\lefteqn{
\exp \left[\tfrac{1}{2} \beta  \sum_{{\bf x}, {\bf y}} J({\bf x}, {\bf y})
\, s({\bf x}) \, s({\bf y}) \right] = } \\ \nonumber &&  \prod_{\bf x} \int
\dd \phi({\bf x}) \exp \left[ -\tfrac{1}{2} \beta \sum_{{\bf x}, {\bf
y}} J^{-1}({\bf x}, {\bf y}) \, \phi({\bf x}) \, \phi({\bf y}) + \beta
\sum_{{\bf x} }\phi({\bf x}) s({\bf x}) \right]. 
\end{eqnarray} 
Here, $J^{-1}({\bf x}, {\bf y})$ is the inverse of the matrix $J({\bf
x}, {\bf y})$ and we ignored---as will be done throughout these
notes---an irrelevant normalization factor in front of the product at
the right-hand side.  The equation should not be taken too literally.
It is an identity only if $J({\bf x}, {\bf y})$ is a symmetric
positive definite matrix.  This is not true for the Ising model since
the diagonal matrix elements $J({\bf x},{\bf x})$ are all zero,
implying that the sum of the eigenvalues is zero.  We will
nevertheless use this representation and regard it as a formal one.
The partition function now reads
\begin{equation} \label{ZIsing}
Z = \sum_{\{s({\bf x})\}} \prod_{\bf x} \int
\dd \phi({\bf x}) \exp \left[ -\tfrac{1}{2} \beta \sum_{{\bf x}, {\bf
y}} J^{-1}({\bf x}, {\bf y}) \, \phi({\bf x}) \,\phi({\bf y}) + \beta
\sum_{\bf x} \phi({\bf x}) \, s({\bf x}) \right]. 
\end{equation}
The spins are decoupled in this representation, so that the sum over
the spin configurations is easily carried out with the result
\begin{equation} \label{Zphi}
Z = \prod_{\bf x} \int
\dd \phi({\bf x}) \exp \left( -\tfrac{1}{2} \beta \sum_{{\bf x}, {\bf
y}} J^{-1}({\bf x}, {\bf y}) \, \phi({\bf x}) \, \phi({\bf y}) +
\sum_{\bf x} \ln\{\cosh\, [\beta \phi({\bf x})]\}\right),
\end{equation} 
ignoring again an irrelevant constant.  

The auxiliary field $\phi({\bf x})$ is not devoid of physical
relevance.  To see this let us first consider its field equation:
\begin{equation} \label{fieldeq} 
\phi({\bf x}) =  \sum_{\bf y} J({\bf x}, {\bf y}) \,  s({\bf y}) ,
\end{equation}
which follows from (\ref{ZIsing}).  This shows that the auxiliary
field $\phi({\bf x})$ represents the effect of the other spins at
site ${\bf x}$.  To make this more intuitive let us study the expectation
value of the field.  For simplicity, we take only
nearest-neighbor interactions into account by setting
\begin{eqnarray} \label{nene} 
J({\bf x}, {\bf y}) = \left\{ \begin{array}{ll} J & \mbox{if site
${\bf x}$ and ${\bf y}$ are nearest neighbors} \nonumber \\  0  &
\mbox{otherwise,} \end{array} \right. 
\end{eqnarray} 
with $J$ positive, so that we have a ferromagnetic coupling between
the spins.  The model is now translational invariant and the
expectation value $\langle s({\bf x}) \rangle$ is independent of ${\bf x}$:
\begin{equation}
\langle s({\bf x}) \rangle = M.
\end{equation}
We will refer to $M$ as the magnetization.  Upon taking the expectation
value of the field equation (\ref{fieldeq}),
\begin{equation} \label{expvalue}
\langle \phi({\bf x}) \rangle = 2d J M,
\end{equation} 
where $2d$ is the number of nearest neighbors, we see that the
expectation value of the auxiliary field represents the magnetization.

A useful approximation often studied is the so-called mean-field
approximation.  It corresponds to approximating the integral over
$\phi({\bf x})$ in (\ref{Zphi}) by the saddle point---the value of the
integrand for which the exponent is stationary.  This is the case for
$\phi({\bf x})$ satisfying the field equation
\begin{equation} \label{phieq}
- \sum_{\bf y} J^{-1}({\bf x}, {\bf y}) \, \phi({\bf y}) + \tanh \,
[\beta \phi({\bf x})] = 0.
\end{equation}
We will denote the solution by $\phi_{\rm mf}$.  In this approximation,
the auxiliary field is no longer a fluctuating field taking all possible
real values, but a classical one having the value determined by the
field equation (\ref{phieq}).  Being a nonfluctuating field, the
expectation value $\langle \phi_{\rm mf}({\bf x})
\rangle = \phi_{\rm mf}({\bf x})$, and (\ref{phieq}) yields a
self-consistent equation for the magnetization
\begin{equation} \label{self-con}
M = \tanh (2d \beta J M),
\end{equation} 
where we assumed a uniform field solution and invoked Eq.\
(\ref{expvalue}).  It is easily seen graphically that the equation has a
nontrivial solution when $2d \beta J > 1$.  If, on the other hand, $2d
\beta J < 1$ it has only a trivial solution.  It follows that
\begin{equation} 
\beta^{-1}_0 =  2d J
\end{equation}
is the critical temperature separating the ordered low-temperature state
with a nonzero magnetization from the high-temperature disordered state
where the magnetization is zero.

Let us continue by expanding the Hamiltonian in powers of $\phi$.  To this
end we note that the term $\ln[\cosh (\beta \phi)]$ in (\ref{Zphi}) has the
Taylor expansion
\begin{equation}
\ln[\cosh (\beta \phi)] = \tfrac{1}{2} \beta^2 \phi^2 -
\tfrac{1}{12} \beta^4 \phi^4 + \cdots . 
\end{equation}
Before considering the other term in (\ref{Zphi}), $\sum_{{\bf x}, {\bf
y}} J^{-1}({\bf x}, {\bf y}) \, \phi({\bf x}) \, \phi({\bf y})$, let us
first study the related object $\sum_{{\bf x}, {\bf y}} J({\bf x}, {\bf
y}) \, s({\bf x}) \, s({\bf y})$ which shows up in the original Ising
Hamiltonian (\ref{HIsing}).  With our choice (\ref{nene}) of the
interaction, the Taylor expansion of this object becomes
\begin{equation} 
\sum_{{\bf x}, {\bf y}} J({\bf x}, {\bf y}) \, s({\bf x}) \,
s({\bf y}) = J \sum_{\bf x} s({\bf x}) \, (2d + a^2 \nabla^2 + \cdots)
\, s({\bf x}),
\end{equation}
neglecting higher orders in derivatives.  From this it follows that 
\begin{equation}
\sum_{{\bf x}, {\bf y}} J^{-1}({\bf x}, {\bf y}) \, \phi({\bf x}) \,
\phi({\bf y}) = J^{-1} \sum_{\bf x} \phi({\bf x}) \left(\frac{1}{2d} -
\frac{1}{4d^2} a^2 \nabla^2 + \cdots \right) \phi({\bf x}),
\end{equation} 
and the partition function (\ref{Zphi}) becomes in the small-$\phi$
approximation 
\begin{equation} \label{smallZ}
Z = \prod_{\bf x} \int \dd \phi({\bf x}) \, {\rm e}^{- \beta H},
\end{equation}
with $H$ the so-called Landau-Ginzburg Hamiltonian
\begin{equation} \label{LandauO}
H = \sum_{\bf x} \left[ \frac{a^2}{8 d^2  J}  (\nabla
\phi)^2 + \frac{1}{2} \left( \frac{1}{2d J} - \beta \right) \phi^2 +
\frac{\beta^3}{12} \phi^4 \right].
\end{equation}
The model has a classical phase transition when the coefficient of the
$\phi^2$-term changes sign.  This happens when $\beta = 1/2 dJ$ in accord
with the conclusion obtained by inspecting the self-consistent equation
for the magnetization (\ref{self-con}).  

In the mean-field approximation, the thermal fluctuations around the
mean-field configuration are ignored, so that $\phi$ becomes a
nonfluctuating field.  The functional integral $\prod_{\bf x} \int \dd \phi
({\bf x})$ is approximated by the saddle point.  

\begin{figure}
\vspace{-.5cm}
\begin{center}
\epsfxsize=6.cm
\mbox{\epsfbox{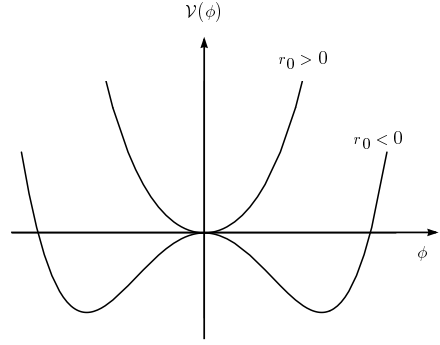}}
\end{center}
\vspace{-1.cm}
\caption{The potential ${\cal V}(\phi)$ of the Ising model in the
high-temperature ($r_0 >0$) and low-temperature ($r_0<0$)
phase. \label{fig:Isingpot}}
\end{figure}

For future reference we go over to the continuum by letting $a
\rightarrow 0$.  To this end we replace the discrete sum $\sum_i$ by
the integral $a^{-d} \int_{\bf x}$, and rescale the field
$\phi({\rm x})$,
\begin{equation} 
\phi ({\bf x}) \rightarrow  \phi'({\bf x})= \sqrt{\frac{\beta a
^{2-d}}{4 d^2 J}} \phi ({\bf x}),
\end{equation}
such that the coefficient of the gradient term in the Hamiltonian takes
the canonical form of $\tfrac{1}{2}$.  In this way the Hamiltonian
becomes
\begin{equation} \label{LandauC}
\beta H = \int_{\bf x} \left[ \frac{1}{2}  (\nabla \phi)^2 + \frac{1}{2}
r_0 \phi^2 + \frac{1}{4!} \lambda_0 \phi^4 \right],
\end{equation}
where we dropped the prime on the field; the parameter $r_0$ and the 
coupling constant $\lambda_0$ are given by
\begin{equation}
r_0 = \frac{\beta_0^{-2}}{J a^2} (\beta_0-\beta), \;\;\;
\lambda_0 =
\frac{4 \beta^2}{J^2 \beta_0^4 } a^{d-4}.
\end{equation}
The partition function now reads
\begin{equation} \label{Zfunct}
Z = \int \DD \phi \, {\rm e}^{- \beta H},
\end{equation} 
where the functional integral $\int \DD \phi$ denotes the continuum
limit of the product of integrals $\prod_{\bf x} \int \dd \phi({\bf
x})$.  The last two terms in the integrand of (\ref{LandauC}) constitute
the potential ${\cal V}(\phi)$,
\begin{equation}  
{\cal V}(\phi) = \frac{1}{2} r_0 \phi^2 + \frac{1}{4!} \lambda_0 \phi^4.
\end{equation} 
In Fig.\ \ref{fig:Isingpot}, the potential ${\cal V}(\phi)$ is depicted
in the high-temperature phase where $r_0 > 0$, and also in the
low-temperature phase where $r_0<0$.  The minimum of the potential in
the low-temperature phase is obtained for a value $\phi \neq 0$, whereas
in the high-temperature phase the minimum is always at $\phi=0$.
\section{Derivative Expansion} 
\label{sec:der}
We are interested in taking into account field fluctuations around the
mean field $\phi_{\rm mf}$, which is the solution of the field
equation obtained from (\ref{LandauC}).  To this end we set $\phi =
\phi_{\rm mf} + \tilde{\phi}$, and expand the Hamiltonian around the
mean field up to second order in $\tilde{\phi}$:
\begin{equation}  \label{Hexp0}
\beta H = \beta H_{\rm mf} + \frac{1}{2} \int_{\bf x} \left[
(\nabla \tilde{\phi})^2 + (r_0 + \tfrac{1}{2} \lambda_0
\phi_{\rm mf}^2 ) \tilde{\phi}^2  \right],
\end{equation}
where $H_{\rm mf}$ denotes the value of the Hamiltonian
(\ref{LandauC}) for $\phi= \phi_{\rm mf}$.  Because of the change of
variables, the functional integral $\int \DD \phi$ changes to $\int
\DD \tilde{\phi}$.  Since we neglected higher-order terms, the
functional integral is Gaussian and  easily carried out.  The
partition function (\ref{Zfunct}) becomes in this approximation
\begin{eqnarray} \label{Zapp}
Z &=& {\rm e}^{- \beta H_{\rm mf}} \int  \DD \tilde{\phi} \, \exp \left\{-
\frac{1}{2} \int_{\bf x} \left[ (\nabla \tilde{\phi})^2 +
(r_0 + \tfrac{1}{2} \lambda_0 \phi_{\rm mf}^2 ) \tilde{\phi}^2
\right] \right\} \nonumber \\ 
&=&  {\rm e}^{- \beta H_{\rm mf}} \, {\rm Det}^{-1/2} ( {\bf p}^2 +
r_0 + \tfrac{1}{2} \lambda_0 \phi_{\rm mf}^2 ),
\end{eqnarray}
with the derivative ${\bf p} = - i \nabla$.  The determinant represents the
first corrections to the mean-field expression $\exp (-\beta H_{\rm mf})$ of
the partition function due to fluctuations.  Using the identity
$\mbox{Det(A)} = \exp \left[ \mbox{Tr} \, \ln (A)\right]$, we can collect
them in the effective Hamiltonian
\begin{equation} \label{Heff}
\beta H_{\rm eff} = \tfrac{1}{2} {\rm Tr} \ln [ {\bf p}^2 + r_0 +
\tfrac{1}{2} \lambda_0 \phi_{\rm mf}^2 ({\bf x}) ],
\end{equation}
so that to this order
\begin{equation}
Z = {\rm e}^{-\beta(H_{\rm mf} + H_{\rm eff})}.
\end{equation} 
As indicated, the mean field $\phi_{\rm mf} ({\bf x})$ may be space
dependent.

We next specify the meaning of the trace Tr appearing in (\ref{Heff}).
Explicitly,
\begin{equation} \label{Hexplicit}
\beta H_{\rm eff} = \frac{1}{2} \int_{\rm x} \ln\left\{ \left[ {\bf
p}^2 + r_0 + \tfrac{1}{2} \lambda_0 \phi_{\rm mf}^2 ({\bf x}) \right]
\delta ({\bf x} - {\bf y})\bigr|_{{\bf y} = {\bf x}} \right\}.
\end{equation}
The delta function arises because the expression in parenthesis at the
right-hand side of (\ref{Zapp}) is obtained as a functional derivative of
the Hamiltonian (\ref{Hexp0}),
\begin{equation}  
\frac{\delta^{2} \beta H}{\delta {\tilde \phi}^2 ({\bf x})} =
\left[  {\bf p}^2 + r_0 +
\tfrac{1}{2} \lambda_0 \phi_{\rm mf}^2 ({\bf x}) \right] \,  
\delta ({\bf x} - {\bf y}) \bigr|_{{\bf y} = {\bf x}},
\end{equation} 
which gives a delta function.  Since it is the unit operator in function
space, the delta function may be taken out of the logarithm and we can write
for (\ref{Hexplicit})
\begin{eqnarray}  \label{Trexplicit}
\beta H_{\rm eff} &=& \frac{1}{2} \int_{\bf x} 
\ln \left[ {\bf p}^2 + r_0 +
\tfrac{1}{2} \lambda_0 \phi_{\rm mf}^2 ({\bf x}) \right] 
\delta ({\bf x} - {\bf y}) \bigr|_{{\bf y} = {\bf x}}  \nonumber  \\ &=&
\frac{1}{2} \int_{\bf x} \int_{\bf k}
\mbox{e}^{-i{\bf k} \cdot {\bf x}} \, \ln \left[ {\bf p}^2 + r_0 +
\tfrac{1}{2} \lambda_0 \phi_{\rm mf}^2 ({\bf x}) 
\right]  \mbox{e}^{i {\bf k} \cdot {\bf x}}.
\end{eqnarray}
In the last step, we used the integral representation of the delta
function:
\begin{equation}
\delta ({\bf x}) = \int_{\bf k} {\rm e}^{i {\bf k} \cdot {\bf x}},
\end{equation}
shifted the exponential function $\exp (-i {\bf k} \cdot {\bf y})$ to the
left, which is justified because the derivative ${\bf p}$ does not operate
on it, and, finally, set ${\bf y}$ equal to ${\bf x}$.  We thus see that the
trace Tr in (\ref{Trexplicit}) stands for the trace over discrete indices as
well as the integration over space and over momentum.  The integral
$\int_{\bf k}$ arises because the effective Hamiltonian calculated here is a
one-loop result with ${\bf k}$ the loop momentum.

The integrals in (\ref{Trexplicit}) cannot in general be evaluated in
closed form because the logarithm contains momentum operators and
space-dependent functions in a mixed order.  To disentangle the integrals
resort has to be taken to a derivative expansion \cite{FAF} in which the
logarithm is expanded in a Taylor series.  Each term contains powers of the
momentum operator ${\bf p}$ which acts on every space-dependent function to
its right.  All these operators are shifted to the left by repeatedly
applying the identity
\begin{equation} \label{commu}
f({\bf x}) {\bf p} g({\bf x}) = ({\bf p} + i \nabla) f({\bf x}) g({\bf x}),
\end{equation} 
where $f({\bf x})$ and $g({\bf x})$ are arbitrary functions and the
derivative $\nabla$ acts {\it only} on the next object to the right.  One
then integrates by parts, so that all the ${\bf p}$'s act to the left where
only a factor $\exp(-i {\bf k} \cdot {\bf x})$ stands.  Ignoring total
derivatives and taking into account the minus signs that arise when
integrating by parts, one sees that all occurrences of ${\bf p}$ (an
operator) are replaced with ${\bf k}$ (an integration variable).  The
exponential function $\exp(i {\bf k} \cdot {\bf x})$ can at this stage be
moved to the left where it is annihilated by the function $\exp(-i {\bf k}
\cdot {\bf x})$.  The momentum integration can now in principle be carried
out and the effective Hamiltonian be cast in the form of an integral over a
local density ${\cal H}_{\rm eff}$:
\begin{equation}   
H_{\rm eff} = \int_{\bf x} {\cal H}_{\rm eff}.
\end{equation} 
This is in a nutshell how the derivative expansion works. 

Let us illustrate the method by applying it to (\ref{Heff}).  When we assume
$\phi_{\rm mf}$ to be a constant field $\bar{\phi}$, the effective
Hamiltonian (\ref{Heff}) may be evaluated in closed form:
\begin{equation}    \label{Veff0}
\beta {\cal V}_{\rm eff} = \frac{1}{2} \int_{\bf k} \ln ( {\bf k}^2 +
M^2) = \frac{\Gamma(1-d/2)}{d (4 \pi)^{d/2} M^d},
\;\;\;\;\;\;\;\; M = \sqrt{r_0 + \tfrac{1}{2}\lambda_0 \bar{\phi}^2},
\end{equation} 
where instead of an Hamiltonian we introduced a potential ${\cal V}_{\rm
eff}$ to indicate that we are working with a space-independent field
$\bar{\phi}$.  To obtain the last equation, we first differentiated $\ln(k^2
+ M^2)$ with respect to $M^2$ and used the dimensional-regularized integral
\begin{equation} 
\int_{\bf k} \frac{1}{({\bf k}^2 + M^2)^\alpha} = \frac{\Gamma(\alpha
-d/2)}{(4 \pi)^{d/2} \Gamma(\alpha)} \frac{1}{\left(M^2\right)^{\alpha-d/2}}
\end{equation} 
to suppress irrelevant ultraviolet divergences, and finally integrated again
with respect to $M^2$.  To illustrate the power of dimensional
regularization, let us consider the case $d=3$ in detail.  Introducing a
momentum cutoff, we find in the large-$\Lambda$ limit
\begin{equation} 
\beta {\cal V}_{\rm eff} = \frac{1}{8 \pi^2} \lambda_0 \bar{\phi}^2 \Lambda -
\frac{1}{12 \pi} M^3 + {\cal O} \left(\frac{1}{\Lambda}\right),
\end{equation} 
where we ignored irrelevant, $\bar{\phi}$-in\-depen\-dent constants
proportional to powers of $\Lambda$.  We see that in (\ref{Veff0}) only the
finite part emerges.  That is, all terms that diverge with a strictly
positive power of the momentum cutoff are suppressed in dimensional
regularization.  These contributions, which come from the ultraviolet
region, cannot physically be very relevant because the simple
Landau-Ginzburg model (\ref{LandauC}) stops being valid here and new
theories are required.  It is a virtue of dimensional regularization that
these irrelevant divergences are suppressed.

Expanded up to fourth order in $\bar{\phi}$, (\ref{Veff0}) becomes
\begin{equation} \label{Veffexp}
\beta {\cal V}_{\rm eff} = - \frac{1}{12 \pi} r_0^{3/2}  - \frac{1}{16 \pi}
\lambda_0 r_0^{1/2} \bar{\phi}^2 - \frac{1}{128 \pi} \frac{\lambda^2_0}{r_0^{1/2}}
\bar{\phi}^4 + \cdots, 
\end{equation}
where the first term is an irrelevant $\bar{\phi}$-independent constant.
These one-loop contributions, when added to the mean-field potential
\begin{equation}
\beta {\cal V}_0 = \frac{1}{2} r_0 \bar{\phi}^2+ \frac{1}{4!} \lambda_0
\bar{\phi}^4, 
\end{equation} 
lead to a renormalization of the bare parameters
\begin{equation} 
\lambda = \lambda_0 - \frac{3}{16 \pi} \frac{\lambda_0^2}{r_0^{1/2}}, \;\;\;
r = r_0 - \frac{1}{8 \pi} \lambda_0 r_0^{1/2}.
\end{equation} 

In the case $\phi_{\rm mf}$ is not a constant field, we write the mean field
$\phi_{\rm mf}({\bf x})$, solving the field equation, as $\phi_{\rm mf}({\bf
x}) = \bar{\phi} + {\hat \phi}({\bf x})$, where $\bar{\phi}$ is the constant
field introduced above (\ref{Veff0}), and expand the logarithm at the
right-hand side of (\ref{Heff}) to second order in ${\hat \phi}$:
\begin{equation} \label{Heffexpand}
\beta {\hat H}_{\rm eff} = \frac{1}{4} \lambda_0 {\rm Tr} \frac{1}{{\bf p}^2
+ M^2} (2 \bar{\phi} {\hat \phi} + {\hat \phi}^2 ) - \frac{1}{8}
\lambda^2_0 \bar{\phi}^2 {\rm Tr} \frac{1}{{\bf p}^2 + M^2} {\hat \phi}
\frac{1}{{\bf p}^2 + M^2} {\hat \phi},
\end{equation} 
with
\begin{eqnarray} 
{\hat H}_{\rm eff} &:=& H_{\rm eff} (\bar{\phi} + {\hat \phi}) - H_{\rm eff}
(\bar{\phi}) \nonumber \\ &=& \int_{\bf x} \left[ \frac{\partial {\cal
V}_{\rm eff}}{\partial \bar{\phi}} \hat{\phi} + \frac{1}{2}
\frac{\partial^2 {\cal V}_{\rm eff}}{\partial \bar{\phi}^2} \hat{\phi}^2 +
\frac{1}{2} {\cal Z}(\bar{\phi}) (\nabla \hat{\phi})^2 + \cdots \right] .
\end{eqnarray}  
Moving the momentum operator ${\bf p}$ to the left by using (\ref{commu}),
we obtain
\begin{equation} \label{H2nd}
\beta {\hat H}_{\rm eff} = \frac{1}{4} \lambda_0 {\rm Tr} \frac{1}{{\bf p}^2
+ M^2} (2 \bar{\phi} {\hat \phi} + {\hat \phi}^2 ) - \frac{1}{8}
\lambda^2_0 \bar{\phi}^2 {\rm Tr} \frac{1}{{\bf p}^2 + M^2} 
\frac{1}{({\bf p}- i \nabla)^2 + M^2} {\hat \phi}{\hat \phi},
\end{equation} 
where we recall the definition of the derivative $\nabla$ as operating only
on the first object to its right.  Using the integral
\begin{equation}
\int_{\bf k} \frac{1}{{\bf k}^2 + M^2} \frac{1}{({\bf k} + {\bf q})^2 + M^2}
= \frac{1}{4 \pi |{\bf q}| } \arctan\left(\frac{|{\bf q}|}{2 M} \right),
\end{equation}
with ${\bf q} = -i \nabla$, we obtain for (\ref{H2nd})
\begin{equation}
\beta {\hat H}_{{\rm eff}} = - \frac{1}{16 \pi} \lambda_0 M (2 \bar{\phi}
{\hat \phi} + {\hat \phi}^2 ) - \frac{1}{32 \pi} \lambda^2_0 \bar{\phi}^2 \,
{\hat \phi} \left[\frac{1}{|{\bf q}|}
\arctan\left(\frac{|{\bf q}|}{2 M} \right) \right]{\hat \phi}.
\end{equation}
We note that only terms with an even number of derivatives appear in the
expansion of this expression.  The coefficient of the linear term is
$\partial \beta {\cal V}_{\rm eff}/\partial \bar{\phi}$, while that of the
two quadratic terms independent of ${\bf q}$ is $\tfrac{1}{2} \partial^2
\beta {\cal V}_{\rm eff}/\partial \bar{\phi}^2$, as it should be.  For
${\cal Z}$ we obtain
\begin{equation}  
{\cal Z}(\bar{\phi}) = \frac{1}{192 \pi} \frac{\lambda_0^2 \bar{\phi}^2}{M^3}.
\end{equation}  
Other terms involving higher powers of ${\hat\phi}$, obtained from
expanding the logarithm in (\ref{Heff}) to higher orders, can be treated
in a similar fashion.

%% file: Superfluid.tex
\chapter{Superfluidity \label{chap:super}}
A central role in these Lectures is played by an interacting Bose gas.
In this chapter we wish to study some of its salient features, notably
its ability to become superfluid below a critical temperature.  We shall
derive the zero-temperature effective theory of the superfluid state,
and discuss the effect of the inclusion of impurities and of a $1/|{\bf
x}|$-Coulomb potential.  Finally, vortices both at the absolute zero of
temperature and at finite temperature are studied.
\section{Bogoliubov Theory}
The system of an interacting Bose gas is defined by the the Lagrangian
\cite{GrPi}
\begin{equation} \label{eff:Lagr}
{\cal L} = \phi^* \bigl[i \partial_0 - \epsilon(-i \nabla) + \mu_0\bigr]
\phi - \lambda_0 |\phi|^4,
\end{equation} 
where the complex scalar field $\phi$ describes the atoms of mass $m$,
$\epsilon(-i \nabla) = - \nabla^2/2m$ is the kinetic energy operator, and
$\mu_0$ is the chemical potential.  The last term with positive coupling
constant, $\lambda_0 > 0$, represents a repulsive contact interaction.  The
(zero-temperature) grand-canonical partition function $Z$ is obtained by
integrating over all field configurations weighted with an exponential
factor determined by the action $S = \int_x {\cal L}$:
\begin{equation} 
Z = \int \DD \phi^* \DD \phi \, {\rm e}^{i S}.
\end{equation} 
This is the quantum analog of Eq.\ (\ref{Zfunct})---the
functional-integral representation of a classical partition function.

\begin{figure}
\begin{center}
\epsfxsize=8.cm
\mbox{\epsfbox{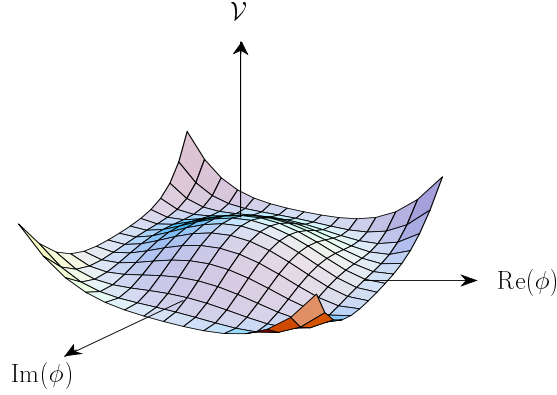}}
\end{center}
\caption{Graphical representation of the potential 
(\protect\ref{eff:V}). \label{fig:potential}}
\end{figure}

The theory (\ref{eff:Lagr}) possesses a global U(1) symmetry under which
\begin{equation} 
\phi(x) \rightarrow {\rm e}^{i \alpha} \phi(x),
\end{equation} 
with $\alpha$ a constant transformation parameter.  At zero temperature,
this symmetry is spontaneously broken by a nontrivial ground state, and
the system is in its superfluid phase.  Most of the startling phenomena
of a superfluid follow from this symmetry breakdown.  The nontrivial
groundstate can be easily seen by considering the shape of the potential
\begin{equation} \label{eff:V}
{\cal V} = - \mu_0 |\phi|^2 + \lambda_0 |\phi|^4,
\end{equation} 
depicted in Fig.\ \ref{fig:potential}. It is seen to have a minimum away
from the origin $\phi = 0$.
To account for this, we shift $\phi$ by a (complex) constant $\bar{\phi}$ and
write
\begin{equation}  \label{eff:newfields}
\phi(x) = {\rm e}^{i \varphi(x)} \, [\bar{\phi} + \tilde{\phi}(x)].
\end{equation}
The phase field $\varphi(x)$ represents the Goldstone mode accompanying the
spontaneous breakdown of the global U(1) symmetry.  At zero temperature, the
constant value
\begin{equation}  \label{eff:min}
|\bar{\phi}|^2 = \frac{1}{2} \frac{\mu_0}{\lambda_0 }                     
\end{equation}
minimizes the potential energy.  It physically represents the number
density of particles contained in the condensate for the total
particle number density is given by
\begin{equation} 
n(x) = |\phi(x)|^2 .
\end{equation} 
Because $\bar{\phi}$ is a constant, the condensate is a uniform,
zero-momentum state.  That is, the particles residing in the ground state
are in the ${\bf k}=0$ mode.  We will be working in the Bogoliubov
approximation which amounts to including only the quadratic terms in
$\tilde{\phi}$ and ignoring the higher-order ones.  These terms may be cast
in the matrix form
\begin{equation}  \label{eff:L0}
{\cal L}^{(2)} = \tfrac{1}{2} \tilde{\Phi}^{\dagger} M_0(p,x)
\tilde{\Phi}, \;\;\;\;\;\;  \tilde{\Phi} = \left(\begin{array}{l}
\tilde{\phi} \\ 
\tilde{\phi}^* \end{array} \right),
\end{equation}
with
\begin{eqnarray}  \label{eff:M} 
\lefteqn{M_0(p,x) =} \\ \nonumber && \!\!\!\!\!\!\!
\left( \begin{array}{cc}
p_0 - \epsilon({\bf p}) + \mu_0 - U(x) - 4 \lambda_0 |\bar{\phi}|^2 & 
\!\!\!\! - 2 \lambda_0 \bar{\phi}^2  \\
- 2 \lambda_0 \bar{\phi}^*\mbox{}^2    & \!\!\!\! -p_0 - \epsilon ({\bf p}) 
+ \mu_0 - U(x) - 4 \lambda_0 |\bar{\phi}|^2
\end{array} \right),
\end{eqnarray} 
where $U$ stands for the combination
\begin{equation}  \label{eff:U}
U(x) = \partial_0 \varphi(x) + \frac{1}{2m} [\nabla \varphi(x)]^2.
\end{equation} 
In writing (\ref{eff:M}) we have omitted a term $\nabla^2 \varphi$
containing two derivatives which is irrelevant in the regime of low
momentum in which we shall be interested.  We also omitted a term of the
form $\nabla \varphi \cdot {\bf j}$, where ${\bf j}$ is the Noether
current associated with the global U(1) symmetry,
\begin{equation} 
{\bf j} =  \frac{1}{2 i m}  \phi^*
\stackrel{\leftrightarrow}{\nabla} \phi. 
\end{equation}  
This term, which after a partial integration becomes $- \varphi \nabla
\cdot {\bf j}$, is irrelevant too at low energy and small momentum
because in a first approximation the particle number density is
constant, so that the classical current satisfies the condition
\begin{equation} 
\nabla \cdot {\bf j} =0.
\end{equation}
The spectrum $E({\bf k})$ obtained from the matrix $M_0$ with the 
field $U$ set to zero is the famous single-particle Bogoliubov
spectrum \cite{Bogoliubov},
\begin{eqnarray}  \label{eff:bogo}
E({\bf k}) &=& \sqrt{ \epsilon ^2({\bf k}) + 2 \mu_0 \epsilon({\bf k}) }
\nonumber \\ &=& \sqrt{ \epsilon ^2({\bf k}) + 4 \lambda_0 |\bar{\phi}|^2
\epsilon({\bf k}) }. 
\end{eqnarray} 
The most notable feature of this spectrum is that it is gapless,
behaving for small momentum as
\begin{equation} \label{eff:micror}
E({\bf k}) \sim u_0 \, |{\bf k}|, 				
\end{equation} 
with $u_0 = \sqrt{\mu_0/m}$ a velocity which is sometimes referred to as the
microscopic sound velocity.  It was first shown by Beliaev \cite{Beliaev} that
the gaplessness of the single-particle spectrum persists at the one-loop
order.  This was subsequently proven to hold to all orders in
perturbation theory by Hugenholtz and Pines \cite{HP}.  For large
momentum, the Bogoliubov spectrum takes a form
\begin{equation} \label{eff:med}
E({\bf k}) \sim \epsilon({\bf k}) + 2 \lambda_0 |\bar{\phi}|^2
\end{equation} 
typical for a nonrelativistic particle with mass $m$ moving in a medium.  To
highlight the condensate we have chosen here the second form in
(\ref{eff:bogo}) where $\mu_0$ is replaced with $2 \lambda_0 |\bar{\phi}|^2$.
\section{Effective Theory} \label{sec:ET}
Since gapless modes in general require a justification for their existence,
we expect the gaplessness of the single-particle spectrum to be a result of
Goldstone's theorem.  This is corroborated by the relativistic version of
the theory.  There, one finds two spectra, one corresponding to a massive
Higgs particle which in the nonrelativistic limit becomes too heavy and
decouples from the theory, and one corresponding to the Goldstone mode of
the spontaneously broken global U(1) symmetry \cite{BBD}.  The latter
reduces in the nonrelativistic limit to the Bogoliubov spectrum.  Also, when
the theory is coupled to an electromagnetic field, one finds that the
single-particle spectrum acquires an energy gap.  This is what one expects
to happen with the spectrum of a Goldstone mode when the Higgs mechanism is
operating.  The equivalence of the single-particle excitation and the
collective density fluctuation has been proven to all orders in perturbation
by Gavoret and Nozi\`eres \cite{GN}.

\begin{figure}
\begin{center}
\epsfxsize=8.cm
\mbox{\epsfbox{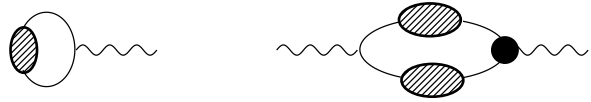}}
\end{center}
\caption{Graphical representation of the effective theory
(\protect\ref{eff:Leff}). The symbols are explained in the
text. \label{fig:effective}}
\end{figure}

Let us derive the effective theory governing the Goldstone mode at low
energy and small momentum by integrating out the fluctuating field
$\tilde{\Phi}$ \cite{effbos}.  The effective theory is graphically
represented by Fig.\ \ref{fig:effective}.
A line with a shaded bubble inserted stands for $i$ times the {\it full}
Green function $G$ and the black bubble denotes $i$ times the {\it full}
interaction $\Gamma$ of the $\tilde{\Phi}$-field with the field $U$
which is denoted by a wiggly line.  Both $G$ and $\Gamma$ are $2 \times 2$
matrices.  The full interaction is obtained from the inverse Green function
by differentiation with respect to the chemical potential,
\begin{equation}  \label{bcs:defga}
\Gamma = - \frac{\partial G^{-1}}{\partial \mu}.		
\end{equation}
This follows because $U$, as defined in (\ref{eff:U}), appears in the theory
only in the combination $\mu_0 - U$.  To lowest order, the inverse
propagator is given by the matrix $M_0$ in (\ref{eff:M}) with $U(x)$ set to
zero.  It follows that the vertex of the interaction between the
$\tilde{\Phi}$ and $U$-fields is minus the unit matrix.  Because in terms of
the full Green function $G$, the particle number density reads
\begin{equation}
\bar{n} = \frac{i}{2} \, {\rm tr} \int_k G (k),
\end{equation}
we conclude that the first diagram in Fig.\ \ref{fig:effective} stands
for $-\bar{n} U$.  The bar over $n$ is to indicate that the particle
number density obtained in this way is a constant, representing the
density of the uniform system with $U(x)$ set to zero.  The second
diagram without the wiggly lines denotes $i$ times the (0 0)-component
of the {\it full} polarization tensor, $\Pi_{0 0}$, at zero energy
transfer and low momentum ${\bf q}$,
\begin{equation} \label{eff:pi}
i \lim_{{\bf q} \rightarrow 0} \Pi_{0 0}(0,{\bf q}) = -\frac{1}{2}
\lim_{{\bf q} \rightarrow 0} {\rm tr} \int_k G \, \Gamma \, G \, (k_0,{\bf
k}+ {\bf q}).
\end{equation} 
The factor $\tfrac{1}{2}$ is a symmetry factor which arises because the two
Bose lines are identical.  We proceed by invoking an argument due to Gavoret
and Nozi\`eres \cite{GN} to relate the left-hand side of (\ref{eff:pi}) to
the sound velocity.  By virtue of relation (\ref{bcs:defga}) between the
full Green function $G$ and the full interaction $\Gamma$, the (0
0)-component of the polarization tensor can be cast in the form
\begin{eqnarray}  \label{bcs:cruc}
\lim_{{\bf q} \rightarrow 0} \Pi_{0 0} (0,{\bf q}) &=& - \frac{i}{2}
\lim_{{\bf q} \rightarrow 0} {\rm tr} \int_k G \, \frac{\partial
G^{-1}}{\partial \mu} \, G (k_0,{\bf k}+ {\bf q}) \nonumber \\ &=& 
\frac{i}{2} \frac{\partial }{\partial \mu} \lim_{{\bf q} \rightarrow 0} {\rm
tr} \int_k G (k_0,{\bf k}+ {\bf q}) \nonumber \\ &=& \frac{\partial
\bar{n}}{\partial \mu} = - \frac{1}{V} \frac{\partial \Omega}{\partial \mu^2},
\end{eqnarray} 
where $\Omega$ is the thermodynamic potential and $V$ the volume of the
system.  The right-hand side of (\ref{bcs:cruc}) is $\bar{n}^2 \kappa$, with
$\kappa$ the compressibility.  Because it is related to the macroscopic
sound velocity $c$ via
\begin{equation}
\kappa = \frac{1}{m \bar{n} c^2},
\end{equation}
we conclude that the (0 0)-component of the full polarization tensor
satisfies the so-called compressibility sum rule of statistical
physics \cite{GN} 
\begin{equation}          \label{bec:rel}           
\lim_{{\bf q} \rightarrow 0} \Pi_{0 0} (0,{\bf q}) = \bar{n}^2 \kappa =
\frac{\bar{n}}{m c^2}. 
\end{equation}
Putting the pieces together, we infer that the diagrams in Fig.\
\ref{fig:effective} stand for the effective theory
\begin{equation} \label{eff:Leff}  
{\cal L}_{{\rm eff}} = -\bar{n}\left[\partial_{0}\varphi +
\frac{1}{2m}( {\bf \nabla} \varphi)^{2} \right] + \frac{\bar{n}}{2m
c^{2}}\left[\partial_{0}\varphi + \frac{1}{2m}( {\bf
\nabla}\varphi)^{2}\right]^{2},
\end{equation} 
where we recall that ${\bar n}$ is the particle number density of the
fluid at rest.  The theory describes a nonrelativistic sound wave, with
the dimensionless phase field $\varphi$ representing the Goldstone
mode of the spontaneously broken global U(1) symmetry.  It has the
gapless dispersion relation $E^2({\bf k}) = c^2 {\bf k}^2$.  The
effective theory gives a complete description of the superfluid valid at
low energies and small momenta.  The same effective theory appears in
the context of (neutral) superconductors \cite{effBCS} (see next
chapter) and also in that of classical hydrodynamics \cite{hydro}.

The chemical potential $\mu$ is represented in the effective theory
(\ref{eff:Leff}) by \cite{PWA}
\begin{equation} \label{jo-pwa}
\mu(x) = - \partial_0 \varphi(x),
\end{equation} 
so that 
\begin{equation} 
\frac{\partial {\cal L}_{\rm eff}}{\partial \mu} = - \frac{\partial {\cal
L}_{\rm eff}}{\partial \partial_0 \phi} = n(x),
\end{equation} 
as required.  It also follows from this equation that the particle
number density $n(x)$ is canonical conjugate to $-\phi(x)$.
 
The most remarkable aspect of the effective theory (\ref{eff:Leff}) is that
it is nonlinear.  The nonlinearity is necessary to provide a
Galilei-invariant description of a gapless mode, as required in a
nonrelativistic context.  Under a Galilei boosts,
\begin{equation} \label{boost} 
t \rightarrow t' = t, \;\; {\bf x} \rightarrow {\bf x}' = {\bf x} -
{\bf u} t; \;\;\;\;
\partial_0 \rightarrow \partial_0' = \partial_0 + {\bf u} \cdot
\nabla, \;\; \nabla \rightarrow \nabla' = \nabla,
\end{equation}    
with ${\bf u}$ a constant velocity, the Goldstone field $\varphi(x)$
transforms as
\begin{equation} 
\frac{1}{m} \varphi(x) \rightarrow \frac{1}{m} \varphi'(x') =
\frac{1}{m} \varphi(x) - {\bf u} \cdot {\bf x} + \tfrac{1}{2} {\bf u}^2
t.
\end{equation}  
As a result, the superfluid velocity ${\bf v}_{\rm s} = \nabla
\varphi/m$ and the chemical potential (per unit mass) $\mu/m = -
\partial_0 \varphi/m$ transform under a Galilei boost in the correct
way,
\begin{equation} 
{\bf v}_{\rm s}(x) \rightarrow {\bf v}'_{\rm s}(x') = {\bf v}_{\rm s}(x)
- {\bf u}, \;\;\; \mu(x)/m \rightarrow \mu'(x')/m = \mu (x)/m - {\bf u}
\cdot {\bf v}_{\rm s}(x) + \tfrac{1}{2} {\bf u}^2.
\end{equation} 
It is readily checked that the field $U(x)$ defined in (\ref{eff:U}) and
therefore the effective theory (\ref{eff:Leff}) is invariant under Galilei
boosts.

Since the Goldstone field in (\ref{eff:Lagr}) is always accompanied by a
derivative, we see that that the nonlinear terms carry additional factors of
$|{\bf k}|/mc$, with $|{\bf k}|$ the wave number.  They can therefore be
ignored provided the wave number is smaller than the inverse coherence
length $\xi^{-1} = mc$,
\begin{equation}  \label{zwd}
|{\bf k}| < 1/\xi.
\end{equation} 
For example, in the case of $^4$He the coherence length, or Compton
wavelength, is about 10 nm.  In this system, the bound (\ref{zwd}), below
which the nonlinear terms can be neglected, coincide with the region where
the spectrum is linear and the description in terms of solely a sound mode is
applicable.

The alert reader might be worrying about an apparent mismatch in the number
of degrees of freedom in the normal and the superfluid phase.  Whereas the
normal phase is described by a complex $\phi$-field, the superfluid phase is
described by a real scalar field $\varphi$.  The resolution of this paradox
lies in the spectrum of the modes \cite{Leutwyler}.  In the normal phase,
the spectrum $E({\bf k}) = {\bf k}^2/2m$ is linear in $E$, so that only
positive energies appear in the Fourier decomposition, and one needs---as is
well known from standard quantum mechanics---a complex field to describe a
single particle.  In the superfluid phase, where the spectrum, $E^2({\bf k})
= c^2 {\bf k}^2$, is quadratic in $E$, the counting goes differently.  The
Fourier decomposition now contains positive as well as negative energies and
a single real field suffice to describe this mode.  In other words, although
the number of fields is different, the number of degrees of freedom is the
same in both phases.

The particle number density and current that follows from (\ref{eff:Leff})
read
\begin{eqnarray}  
n(x) &=& \bar{n} -\frac{\bar{n}}{m c^{2}} \left\{ \partial_{0}
\varphi(x) + \frac{1}{2 m} [ {\bf \nabla} \varphi(x)]^{2}\right\}
\label{roh1} \\ {\bf j}(x) &=& n(x) {\bf v}_{\rm s}(x). \label{roh2}
\end{eqnarray} 
Physically, (\ref{roh1}) reflects Bernoulli's principle which states that in
regions of rapid flow, the density and therefore the pressure is low.

The diagrams of Fig.~\ref{fig:effective} can be evaluated in a loop
expansion to obtain explicit expressions for the particle number density
$\bar{n}$ and the sound velocity $c$ to any given order \cite{effbos}.  In
doing so, one encounters---apart from ultraviolet divergences which will be
dealt with shortly---also infrared divergences because the Bogoliubov
spectrum is gapless.  When however all one-loop contributions are added
together, these divergences are seen to cancel \cite{effbos}.  One finds for
$d=2$ to the one-loop order
\begin{equation} \label{eff:nc}
\bar{n} =  \frac{1}{2} \frac{\mu}{\lambda}, \;\;\; 
c^2 = 2 \frac{\lambda \bar{n}}{m},
\end{equation} 
where $\mu$ and $\lambda$ are the renormalized parameters.  Following
Ref.\
\cite{NP}, we adopted a dimensional regularization scheme, in which after
the integrals over the loop energies have been carried out, the remaining
integrals over the loop momenta are analytically continued to arbitrary
space dimensions $d$.  As renormalization prescription we employed the
modified minimal subtraction scheme.  This leads to the following relation
between the bare ($\lambda_0$) and renormalized coupling constant [see
Eq. (\ref{Vd=2}) below]
\begin{equation} \label{eff:lambdar}
\frac{1}{\lambda_0} = \frac{1}{\kappa^\epsilon} \left(\frac{1}{\hat{\lambda}} -
\frac{m}{\pi \epsilon}\right),
\end{equation}  
where $\epsilon = 2-d$, and $\kappa$ is an arbitrary renormalization
group scale parameter introduced to give the renormalized coupling
constant $\hat{\lambda}$ the same engineering dimension as in $d=2$.
The chemical potential is not renormalized to this order.

Incidentally, from the vantage point of renormalization, the mass $m$ is an
irrelevant parameter in nonrelativistic theories and can be scaled away
(see, e.g., Ref.\ \cite{NP}).

The form of the effective theory (\ref{eff:Leff}) can also be derived from
general symmetry arguments \cite{GWW}.  More specifically, it follows from
making the presence of a gapless Goldstone mode compatible with Galilei
invariance which demands that the mass current and the momentum density are
equal.  The latter observation leads to the conclusion that the U(1)
Goldstone field $\varphi$ can only appear in the combination (\ref{eff:U}).
To obtain the required linear spectrum for the Goldstone mode it is
necessary then to have the form (\ref{eff:Leff}).  Given the form of the
effective theory, the particle number density and sound velocity can then
more easily be obtained directly from the thermodynamic potential $\Omega$
via
\begin{equation}  \label{bec:thermo}
\bar{n} = - \frac{1}{V} \frac{\partial \Omega }{\partial \mu}; \;\;\;\;\;\;
\frac{1}{c^2} = - \frac{1}{V} \frac{m}{\bar{n}} \frac{\partial^2 \Omega
}{\partial \mu^2},
\end{equation} 
where $V$ is the volume of the system.  In this approach, one only has to
calculate the thermodynamic potential which at zero temperature and in the
Bogoliubov approximation in which we are working is given by the sum ${\cal
V}$ of the classical potential ${\cal V}_0$ and the effective potential
${\cal V}_{\rm eff}$ corresponding to the theory (\ref{eff:Lagr}):
\begin{equation} \label{eff:Omega}
\Omega = \int_{\bf x} ({\cal V}_0 + {\cal V}_{\rm eff}),
\end{equation} 
where ${\cal V}_0$ is given by (\ref{eff:V}) with $\phi$ replaced by
$\bar{\phi}$.  The effective potential for the uniform system is obtained as
follows.  In the Bogoliubov approximation of ignoring higher than second
order in the fields, the integration over $\tilde{\Phi}$ is Gaussian.
Carrying out this integral, we obtain for the zero-temperature partition
function
\begin{eqnarray} \label{bec:Z}
Z &=& {\rm e}^{-i \int_x {\cal V}_0} \int \DD \phi^* \DD \phi \exp
\left(i \int_x {\cal L}^{(2)} \right) 
\nonumber \\ &=& {\rm e}^{- i \int_x {\cal V}_0} \, {\rm Det}^{-1/2} 
(M_0),
\end{eqnarray}  
where $M_0$ stands for the matrix introduced in (\ref{eff:M}).  Setting
\begin{equation} \label{Zeff}
Z = \exp\left[i  \left(-\int_x{\cal V}_0 + S_{\rm eff}\right)\right], 
\end{equation} 
we conclude from (\ref{bec:Z}) that the effective action in the Bogoliubov
approximation is given to the one-loop order by
\begin{equation} \label{bec:Seff}
S_{\rm eff} = \tfrac{1}{2} i {\rm Tr} \ln[M_0(p,x)],
\end{equation} 
where we again used the identity Det($A$) = exp[Tr ln($A$)].
The trace Tr appearing here stands besides for the trace over discrete
indices now also for the integral $\int_x$ over spacetime as well as the one
$\int_k$ over energy and momentum.  The latter integral reflects the fact
that the effective action calculated here is a one-loop result with $k_\mu$
the loop energy and momentum.  To disentangle the integrals one has to carry
out similar steps as the ones outlined in Sec.\ \ref{sec:der} and repeatedly
apply the identity
\begin{equation} 
f(x) p_\mu g(x) = (p_\mu - i \tilde{\partial}_\mu) f(x) g(x),
\end{equation} 
where $f(x)$ and $g(x)$ are arbitrary functions of spacetime and the
derivative $\tilde{\partial}_\mu = (\partial_0,-\nabla)$ acts only on the
next object to the right.  The method outlined there can easily be
transcribed to the present case where the time dimension is included.

If the field $U(x)$ in $M_0$ is set to zero, things simplify because $M_0$
now depends on $p_\mu$ only.  The effective action then becomes $S_{\rm eff}
= - \int_x {\cal V}_{\rm eff}$ with
\begin{equation} \label{eff:Veff}
{\cal V}_{\rm eff}  = -\frac{i}{2} {\rm tr} \int_k \ln[M_0(k)]
\end{equation}
the effective potential.  The easiest way to evaluate the integral over
the loop variable $k$ is to first differentiate the expression with
respect to the chemical potential $\mu_0$
\begin{equation}
\frac{\partial}{\partial \mu_0}  {\rm tr} \, \int_k \ln[M_0(k)] = -2 \,
\int_k \frac{\epsilon({\bf k})}{k_0^2 - E^2({\bf k}) + i \eta },
\end{equation}
with $E({\bf k})$ the Bogoliubov spectrum (\ref{eff:bogo}).  The
integral over $k_0$ can be carried out with the help of a contour
integration, yielding
\begin{equation}
\int_k \frac{\epsilon({\bf k})}{k_0^2 - E^2({\bf k}) + i \eta } =
- \frac{i}{2} \, \int_{\bf k} \frac{\epsilon({\bf k})}{E({\bf k})}.
\end{equation} 
This in turn is easily integrated with respect to $\mu_0$.  Putting the
pieces together, we obtain 
\begin{equation} \label{Veff}
{\cal V} = - \frac{\mu_0^2}{4 \lambda_0} + \frac{1}{2} \int_{\bf
k} E({\bf k}).
\end{equation} 
The integral over the loop momentum in arbitrary space dimension $d$
yields
\begin{equation} \label{regularized}
{\cal V} = - \frac{\mu_0^2}{4 \lambda_0} - L_d m^{d/2} \mu_0^{d/2 + 1},
\;\;\; L_d = \frac{\Gamma(1-d/2) \Gamma(d/2 + 1/2)}{2 \pi^{d/2 + 1/2}
\Gamma(d/2+2)} 
\end{equation}  
where we employed the integral representation of the Gamma function
\begin{equation}  \label{gamma}
\frac{1}{a^z} = \frac{1}{\Gamma(z)} \int_0^\infty \frac{\dd \tau}{\tau}
\tau^z {\rm e}^{-a \tau}
\end{equation} 
together with dimensional regularization to suppress irrelevant ultraviolet
divergences.  

For comparison, let us also evaluate the integral in (\ref{Veff}) over the
loop momentum in three dimensions by introducing a momentum cutoff
$\Lambda$
\begin{eqnarray} \label{bec:Vnon}
\lefteqn{{\cal V}_{\rm eff} = \frac{1}{2} \int_{\bf k} E({\bf k}) =
\frac{1}{4 \pi^2} \int_0^\Lambda k^2 
E(|{\bf k}|) = } \nonumber \\ && \frac{1}{4 \pi^2} \left(\frac{1}{10}
\frac{\Lambda^5}{m} +\frac{1}{3} \mu_0 \Lambda^3 - m \mu_0^2 \Lambda +
\frac{32}{15} m^{3/2} \mu_0^{5/2} \right) + {\cal O}
\left(\frac{1}{\Lambda}\right). 
\end{eqnarray} 
From (\ref{regularized}), we obtain by setting $d=3$ only the finite part,
so that all terms diverging with a strictly positive power of the momentum
cutoff are suppressed.  As we remarked in Sec.\ \ref{sec:der}, these
contributions, which come from the ultraviolet region, cannot be physically
very relevant because the simple model (\ref{eff:Lagr}) breaks down here.
On account of the uncertainty principle, stating that large momenta
correspond to small distances, these terms are always local and can be
absorbed by redefining the parameters appearing in the Lagrangian
\cite{Donoghue}.  Since $\mu_0 = 2 \lambda_0 |\bar{\phi}|^2$, we see that
the first diverging term in (\ref{bec:Vnon}) is an irrelevant constant,
while the two remaining diverging terms can be absorbed by introducing the
renormalized parameters
\begin{eqnarray} 
\mu &=& \mu_0 - \frac{1}{6\pi^2} \lambda_0 \Lambda^3 \label{bec:renmu} \\
\lambda &=& \lambda_0 - \frac{1}{\pi^2} m \lambda_0^2
\Lambda. \label{bec:renla} 
\end{eqnarray} 
Because the diverging terms are---at least to this order---of a form already
present in the original Lagrangian, the theory is called ``renormalizable''.
The renormalized parameters are the physical ones that are to be identified
with those measured in experiment.  In this way, we see that the
contributions to the loop integral stemming from the ultraviolet region are
of no importance.  What remains is the finite part
\begin{equation} \label{bec:finite}
{\cal V}_{\rm eff} = \frac{8}{15 \pi^2} m^{3/2} \mu_0^{5/2},
\end{equation}
which, as we have seen, is obtained directly without
renormalization when using dimensional regularization.  In this scheme,
divergences proportional to powers of the cutoff never show up.  Only
logarithmic divergences appear as $1/\epsilon$ poles, where $\epsilon$
is the deviation from the upper critical dimension ($d=2$ in the present
case).  These logarithmic divergences $\ln(\Lambda/E)$, with $E$ an
energy scale, are relevant also in the infrared because for fixed cutoff
$\ln(\Lambda/E) \rightarrow -\infty$ when $E$ is taken to zero.

In so-called ``nonrenormalizable'' theories, the ultraviolet-diverging terms
are still local but not of a form present in the original Lagrangian.
Whereas in former days such theories were rejected because there supposed
lack of predictive power, the modern view is that there are no fundamental
theories and that there is no basic difference between renormalizable and
nonrenormalizable theories \cite{CaoSc}.  Even a renormalizable theory like
(\ref{eff:Lagr}) should be extended to include all higher-order terms such
as a $|\phi|^6$-term which are allowed by symmetry.  These additional terms
render the theory ``nonrenormalizable''.  This does not however change the
predictive power of the theory.  The point is that when describing the
physics at an energy scale $E$ far below the cutoff, the higher-order terms
are suppressed by powers of $E/\Lambda$, as follows from dimensional
analysis.  Therefore, far below the cutoff, the nonrenormalizable terms are
negligible.

That $d=2$ is the upper critical dimension of the problem at hand can be
seen by noting that $L_d$ in (\ref{regularized}) diverges when $d$ tends to
2. Special care has to be taken for this case.  For $d \neq 2$, we obtain
with the help of (\ref{bec:thermo})
\cite{Weichman}
\begin{equation}  \label{bec:n}
\bar{n} = \frac{\mu_0}{2 \lambda_0} \left[1 + (d + 2) L_d  m^{d/2} \lambda_0
\mu_0^{d/2-1}\right] 
\end{equation} 
and 
\begin{equation} 
c^2 = \frac{\mu_0}{m} \left[1 -  (d-2) (d/2+1) L_d
m^{d/2}\lambda_0 \mu_0^{d/2-1} \right],
\end{equation} 
where to arrive at the last equation an expansion in the coupling constant
$\lambda_0$ is made.  Up to this point, we have considered the chemical
potential to be the independent parameter, thereby assuming the presence of
a reservoir that can freely exchange particles with the system under study.
The system can thus have any number of particles, only the average number is
fixed by external conditions.  From the experimental point of view it is,
however, often more realistic to consider the particle number fixed.  If
this is the case, the particle number density $\bar{n}$ should be considered
as independent variable and the chemical potential should be expressed in
terms of it. This can be achieved by inverting relation (\ref{bec:n}):
\begin{equation} 
\mu_0 = 2 \lambda_0 \bar{n} \left[1 - 2  (d-2) (d/2+1) L_d
m^{d/2} \lambda_0 (2 \lambda_0 \bar{n})^{d/2-1} \right].
\end{equation} 
The sound velocity expressed in terms of the particle number density reads
\begin{equation} \label{bec:c} 
c^2 = \frac{2 \lambda_0 \bar{n}}{m} \left[1 -  d (d/2+1)
L_d m^{d/2} \lambda_0 (2 \lambda_0 \bar{n})^{d/2-1} \right].
\end{equation} 
These formulas reproduce the known results in $d=3$ \cite{FW} and $d=1$
\cite{Lieb}.

To investigate the case $d=2$, we expand the potential (\ref{regularized})
around $d=2$:
\begin{equation}  \label{Vd=2}
{\cal V} = - \frac{\mu_0^2}{4 \lambda_0} - \frac{1}{4
\pi \epsilon} \frac{m \mu_0^2}{\kappa^\epsilon} + {\cal O}(\epsilon^0),
\end{equation} 
with $\epsilon = 2-d$.  This expression is seen to diverge in the limit $d
\rightarrow 2$.  The theory can be rendered finite by
introducing a renormalized coupling constant via (\ref{eff:lambdar}).
We also see that the chemical potential is not renormalized to this order.
The beta function $\beta(\hat{\lambda})$ follows as
\cite{Uzunov}
\begin{equation} 
\beta(\hat{\lambda}) = \kappa \left. \frac{\partial \hat{\lambda}}{\partial
\kappa} \right|_{\lambda_0} = -\epsilon \hat{\lambda} + \frac{m}{\pi}
\hat{\lambda}^2.
\end{equation}  
In the upper critical dimension, this yields only one fixed point, viz.\
the infrared-stable (IR) fixed point $\hat{\lambda}^* = 0$.  Below
$d=2$, this point is shifted to $\hat{\lambda}^* = \epsilon \pi/m$.  It
is now easily checked that Eqs.\ (\ref{bec:n}) and (\ref{bec:c})
also reproduce the two-dimensional results (\ref{eff:nc}).

In the one-loop approximation there is no field renormalization; this is the
reason why in (\ref{eff:Lagr}) we gave only the bare parameters $\mu_0$ and
$\lambda_0$ an index 0, and not $\phi$. 

We proceed by calculating the fraction of particles residing in the
condensate.  In deriving the Bogoliubov spectrum (\ref{eff:bogo}), we
set $|\bar{\phi}|^2 = \mu_0/2 \lambda_0$ thereby fixing the number
density of particles contained in the condensate,
\begin{equation}   \label{bec:n0}
\bar{n}_0 = |\bar{\phi}|^2,
\end{equation}
in terms of the chemical potential.  For our present consideration we
have to keep $\bar{\phi}$ as independent variable.  The spectrum of the
elementary excitation expressed in terms of $\bar{\phi}$ is
\begin{equation} \label{bec:bogog}
E({\bf k}) = \sqrt{\bigl[ \epsilon({\bf k})  - \mu_0 + 4 \lambda_0
|\bar{\phi}|^2 \bigr]^2 - 4 \lambda_0^2 |\bar{\phi}|^4 } \, .  	
\end{equation}
It reduces to the Bogoliubov spectrum when the mean-field value
(\ref{eff:min}) for $\bar{\phi}$ is inserted.  Equation (\ref{eff:Veff})
for the effective potential is still valid, and so is
(\ref{eff:Omega}).  We thus obtain for the particle number density
\begin{equation} 
\bar{n} = \left.  |\bar{\phi}|^2 - \frac{1}{2} \frac{\partial}{\partial \mu_0} 
\int_{\bf k} E ({\bf k})  \right|_{|\bar{\phi}|^2 = \mu_0/2 \lambda_0},
\end{equation} 
where the mean-field value for $\bar{\phi}$ is to be substituted after the
differentiation with respect to the chemical potential has been carried out.
We find
\begin{equation} 
\bar{n} = |\bar{\phi}|^2 - 2^{d/2-2} \frac{d^2-4}{d-1} L_d m^{d/2}
\lambda_0^{d/2} |\bar{\phi}|^d 
\end{equation} 
or for the so-called depletion of the condensate \cite{TN}
\begin{equation}  \label{depl}
\frac{\bar{n}}{\bar{n}_0} -1 \approx - 2^{d/2-2} \frac{d^2-4}{d-1} L_d m^{d/2}
\lambda^{d/2} n^{d/2-1}, 
\end{equation} 
where in the last term we replaced the bare coupling constant with the
(one-loop) renormalized one.  This is consistent to this
order since this term is already a one-loop result.  Equation
(\ref{depl}) shows that even at zero temperature not all the particles
reside in the condensate.  Due to the interparticle repulsion, particles
are removed from the zero-momentum ground state and put in states of
finite momentum.  It has been estimated that in bulk superfluid
$^4$He---a strongly interacting system---only about 8\% of the particles
condense in the zero-momentum state \cite{PeOn}.  For $d=2$, the
right-hand side of Eq.\ (\ref{depl}) reduces to
\begin{equation} 
\frac{\bar{n}}{\bar{n}_0} -1 \approx \frac{m \lambda}{2 \pi},
\end{equation}  
which is seen to be independent of the particle number density.

Despite the fact that not all the particles reside in the condensate,
they all participate in the superfluid motion at zero temperature
\cite{NoPi}.  Apparently, the condensate drags the normal fluid along
with it.  To show this, let us assume that the entire system moves with
a velocity ${\bf u}$ relative to the laboratory system.  As is known
from standard hydrodynamics the time derivate in the frame following the
motion of the fluid is $\partial_0 + {\bf u} \cdot \nabla$ [see Eq.\
(\ref{boost})].  If we insert this in the Lagrangian
(\ref{eff:Lagr}) of the interacting Bose gas, it becomes
\begin{equation}   \label{bec:Lagu}
{\cal L} = \phi^* [i \partial_0 - \epsilon(-i \nabla) + \mu_0 - {\bf u}
\cdot (-i \nabla)] \phi - \lambda_0 |\phi|^4,      
\end{equation} 
where the extra term features the total momentum $\int_{\bf x}
\phi^* (-i \nabla) \phi$ of the system.  The velocity $-{\bf u}$
multiplying this is on the same footing as the chemical potential $\mu_0$
multiplying the particle number $\int_{\bf x} |\phi|^2$.  Whereas
$\mu_0$ is associated with particle number conservation, ${\bf u}$ is
related to the conservation of momentum.

In the two-fluid picture, the condensate can move with a different velocity
${\bf v}_{\rm s}$ as the rest of the system.  To bring this out we introduce
new fields, cf.\ (\ref{eff:newfields}) 
\begin{equation} 
\phi (x) \rightarrow \phi'(x) = {\rm e}^{im {\bf v}_{\rm s} \cdot {\bf x}}
\phi (x) 
\end{equation} 
in terms of which the Lagrangian becomes \cite{Brown}
\begin{equation}   \label{bec:Lagus}
{\cal L} = \phi^* \bigl[i\partial_0 - \epsilon(-i \nabla) + \mu_0 -
\tfrac{1}{2} m {\bf v}_{\rm s} \cdot ({\bf v}_{\rm s} - 2 {\bf u}) -
({\bf u} - {\bf v}_{\rm s}) \cdot (-i\nabla) \bigr] \phi - \lambda_0
|\phi|^4,
\end{equation} 
where we dropped the primes on $\phi$ again.  Both velocities appear in
this expression.  Apart from the change ${\bf u} \rightarrow {\bf u} -
{\bf v}_{\rm s}$ in the second last term, the field transformation
resulted in a change of the chemical potential
\begin{equation} \label{bec:mureplacement}
\mu_0 \rightarrow \mu_{\rm eff} :=
\mu_0 - \tfrac{1}{2} m {\bf v}_{\rm s} \cdot ({\bf v}_{\rm s} - 2 {\bf u}) 
\end{equation} 
where $\mu_{\rm eff}$ may be considered as an effective chemical potential.

The equations for the Bogoliubov spectrum and the thermodynamic
potential are readily written down for the present case with these two
changes are kept in mind.  In particular, the effective potential is
given by (\ref{Veff}) with the replacement Eq.\
(\ref{bec:mureplacement}).  The momentum density, or equivalently, the
mass current ${\bf g}$ of the system is obtained in this approximation
by differentiating the effective potential with respect to $-{\bf u}$.
We find, using the equation
\begin{equation}
\frac{\partial \mu_{\rm eff}}{\partial {\bf u}} = m {\bf v}_{\rm s}
\end{equation} 
that it is given by
\begin{equation} \label{bec:j}
{\bf g} = \rho_{\rm s} {\bf v}_{\rm s} ,
\end{equation}
with $\rho_{\rm s} = m \bar{n}$ the superfluid mass density.  This
equation, comprising the total particle number density $\bar{n}$, shows
that at zero temperature indeed all the particles are involved in the
superflow, despite the fact that only a fraction of them resides in the
condensate \cite{NoPi}.  The superfluid mass density $\rho_{\rm s}$,
obtained by evaluating the response of the system to an externally
imposed velocity field ${\bf u}$, should not be confused with the number
density $\bar{n}_0$ of particles contained in the condensate introduced in
Eq.\ (\ref{bec:n0}).

Let us close this section by pointing out a quick trail to arrive at the
effective theory (\ref{eff:Leff}) starting from the microscopic model
(\ref{eff:Lagr}).  To this end we set
\begin{equation}  
\phi(x) = {\rm e}^{i \varphi(x)} \, [\sqrt{\bar{n}} + \tilde{\phi}(x)],
\end{equation}
and expand the Lagrangian (\ref{eff:Lagr}) up to quadratic terms in
$\tilde{\phi}$.  This leads to
\begin{equation} 
{\cal L}^{(2)} = - {\cal V}_0 - \bar{n} U - \sqrt{\bar{n}} U
(\tilde{\phi} +
\tilde{\phi}^*) - \lambda_0 \bar{n} (\tilde{\phi} + \tilde{\phi}^*)^2,
\end{equation} 
where we used the mean-field equation $\mu_0 = 2 \lambda_0 \bar{n}$.  We next
integrate out the tilde fields---which is tantamount to substituting the
field equation for these fields back into the Lagrangian---to obtain
\begin{equation} \label{eff:quick}
{\cal L}_{\rm eff} = -  \bar{n} U(x) + \frac{1}{4} U(x) \frac{1}{\lambda_0}
U(x),
\end{equation} 
apart from the irrelevant constant term ${\cal V}_0$.  This form of the
effective theory is equivalent to the one found before in (\ref{eff:Lagr}).
We have cast the last term in a form that can be easily generalized to
systems with long-ranged interactions.  A case of particular interest to us
is the Coulomb potential
\begin{equation} 
V({\bf x}) = \frac{e_0^2}{|{\bf x}|},
\end{equation} 
whose Fourier transform in $d$ space dimensions reads
\begin{equation} 
V({\bf k}) = 2^{d-1} \pi^{(d-1)/2} \Gamma\left[\tfrac{1}{2}(d-1)\right]
\frac{e_0^2}{|{\bf k}|^{d-1}}.
\end{equation} 
The simple contact interaction $L_{\rm i} = - \lambda_0 \int_{\bf x} 
|\phi(x)|^4$ in (\ref{eff:Lagr}) gets now replaced by
\begin{equation}  
L_{\rm i} = - \frac{1}{2} \int_{{\bf x}, {\bf y}} |\phi(t,{\bf x})|^2
V({\bf x} - {\bf y}) |\phi(t,{\bf y})|^2.
\end{equation} 
The rationale for using the three-dimensional Coulomb potential even when
considering charges confined to move in a lower dimensional space is that
the electromagnetic interaction remains three-dimensional.  The effective
theory (\ref{eff:quick}) now becomes in the  Fourier representation
\begin{equation}  \label{effCoul}
{\cal L}_{\rm eff} = - \bar{n} U(k)  + \frac{1}{2} U(k_0,{\bf k})
\frac{1}{V({\bf k})} U(k_0,-{\bf k})
\end{equation}
and leads to the dispersion relation
\begin{equation}
E^2({\bf k}) =  2^d \pi^{(d-1)/2} \Gamma\left[\tfrac{1}{2}(d-1)\right]
\frac{\bar{n} e_0^2}{m} |{\bf k}|^{3-d}.
\end{equation}
For $d=3$, this yields the famous plasma mode with an energy gap given by
the plasma frequency $\omega_{\rm p}^2 = 4 \pi \bar{n} e_0^2/m$.

To appreciate under which circumstances the Coulomb interaction becomes
important, we note that for electronic systems $1/|{\bf x}| \sim k_{\rm
F}$ for dimensional reasons and the fermion number density $\bar{n} \sim
k_{\rm F}^d$, where $k_{\rm F}$ is the Fermi momentum.  The ratio of the
Coulomb interaction energy $\epsilon_{\rm C}$ to the Fermi energy
$\epsilon_{\rm F} = k_{\rm F}^2/2m$ is therefore proportional to
$\bar{n}^{-1/d}$.  This means that the lower the electron number
density, the more important the Coulomb interaction becomes.
\section{Quenched Impurities}
In most of the quantum systems we will be considering, impurities plays
an important role.  The main effect of impurities is typically to
localize states.  Localization counteracts the tendency of the system to
become superfluid.  We shall therefore now include impurities in the
interacting Bose gas to see whether this leads to localization and
whether the system still has a superfluid phase.  It is expected that on
increasing the strength of the disorder for a given repulsive
interparticle interaction, the superfluid undergoes a zero-temperature
phase transition to an insulating phase of localized states.  The
location and nature of this transition will be the subject of
Ch. \ref{chap:qpt}.

We shall assume that the impurities are fixed and that their
distribution is not affected by the host system.  This type of
impurities is called quenched impurities and is to be distinguished from
so-called annealed impurities which change with and depend on the host
system.  To account for impurities, we add to the theory
(\ref{eff:Lagr}) the term
\begin{equation} \label{Dirt:dis}
{\cal L}_{\Delta} = \psi({\bf x}) \, |\phi(x)|^2,
\end{equation} 
with $\psi({\bf x})$ a random field whose distribution is assumed to be
Gaussian \cite{Ma}
\begin{equation}  \label{random}
P(\psi) = \exp \left[-\frac{1}{\Delta_0} \int_{\bf x} \, \psi^2({\bf x})
\right],
\end{equation}
and characterized by the disorder strength $\Delta_0$.  The engineering
dimension of the random field is the same as that of the chemical
potential which is one, $[\psi]=1$, while that of the parameter
$\Delta_0$ is $[\Delta_0] = 2-d$ so that the exponent in (\ref{random})
is dimensionless.  Since $\psi({\bf x})$ depends only on the $d$ spatial
dimensions, the impurities it describes should be considered as grains
randomly distributed in space.  The quantity
\begin{equation} \label{Dirt:Z}
Z[\psi] = \int \DD \phi^* \DD \phi \, \exp\left(i \int_x \, {\cal L}
\right),
\end{equation} 
where now ${\cal L}$ stands for the Lagrangian (\ref{eff:Lagr}) with the
term (\ref{Dirt:dis}) added, is the zero-temperature partition function
for a given impurity configuration $\psi$.  In the case of quenched
impurities, the average of an observable $O(\phi^*,\phi)$ is obtained as
follows
\begin{equation} 
\langle O(\phi^*,\phi) \rangle = \int \DD \psi P(\psi) \langle
O(\phi^*,\phi) \rangle_\psi,
\end{equation} 
where $\langle O(\phi^*,\phi) \rangle_\psi$ indicates the
grand-canonical average for a given impurity configuration.  In other
words, first the ensemble average is taken, and only after that the
averaging over the random field is carried out.

In terms of the shifted field, the added term reads
\begin{equation}  
{\cal L}_{\Delta} = \psi({\bf x}) (|\bar{\phi}|^2 + |\tilde{\phi}|^2 +
\bar{\phi} \tilde{\phi}^* + \bar{\phi}^* \tilde{\phi}  ).
\end{equation} 
The first two terms lead to an irrelevant change in the chemical
potential, so that we only have to consider the last two terms, which we can
cast in the form
\begin{equation}
{\cal L}_{\Delta} = \psi({\bf x}) \, \bar{\Phi}^\dagger \tilde{\Phi},
\;\;\;\;\;\;\;
\bar{\Phi} = \left(\begin{array}{l} \bar{\phi} \\ \bar{\phi}^*
\end{array} \right). 
\end{equation} 
The integral over $\tilde{\Phi}$ is Gaussian in the Bogoliubov
approximation and is easily performed to yield an additional term to the
effective action
\begin{equation} 
S_{\Delta} = -\frac{1}{2} \int_{x,y} \psi({\bf x}) \bar{\Phi}^\dagger \, G_0(x-y)
\bar{\Phi} \psi({\bf y}),
\end{equation} 
where the propagator $G_0$ is the inverse of the matrix $M_0$ introduced
in (\ref{eff:M}) with the field $U(x)$ set to zero.  Let us
first Fourier transform the fields,
\begin{eqnarray} 
G_0(x-y) &=& \int_k {\rm e}^{-i k \cdot (x-y)} \, G_0(k)    \\
\psi({\bf x}) &=& \int_{\bf k} {\rm e}^{i {\bf k} \cdot {\bf x}} \psi({\bf k}).
\end{eqnarray} 
The contribution to the effective action then appears in the form
\begin{equation} \label{S_d}
S_{\Delta} = -\frac{1}{2} \int_{\bf k} |\psi({\bf k})|^2
\bar{\Phi}^\dagger G(0,{\bf k}) \bar{\Phi}.
\end{equation} 
Since the random field is Gaussian distributed [see (\ref{random})], the
average over this field representing quenched impurities yields,
\begin{equation} 
\langle |\psi({\bf k})|^2 \rangle = \tfrac{1}{2} V \Delta_0.
\end{equation} 
The remaining integral over the loop momentum in (\ref{S_d}) is readily
carried out to yield 
\begin{equation} \label{L_D}
\langle {\cal L}_\Delta \rangle = \frac{1}{2} \Gamma(1-d/2)
\left(\frac{m}{2 \pi} \right)^{d/2} |\bar{\phi}|^2 (6 \lambda_0
|\bar{\phi}|^2 - \mu_0)^{d/2-1} \Delta_0. 
\end{equation} 
This contribution is seen to diverge in the limit $d \rightarrow 2$:
\begin{equation} \label{L2_D}
\langle {\cal L}_\Delta \rangle = \frac{1}{4 \pi} \frac{m
\mu_0}{\lambda_0 \kappa^\epsilon} \frac{\Delta_0}{\epsilon},
\end{equation} 
where we substituted the mean-field value $\mu_0 = 2 \lambda_0
|\bar{\phi}|^2$.  Recall that $\kappa$ is an arbitrary scale parameter
introduced for dimensional reasons; the engineering dimension of the
right-hand side in (\ref{L2_D}) has the correct value $3 - \epsilon$ this
way.  The result (\ref{L2_D}) is a first indication of the importance of
impurities in $d=2$, showing that in order to render the random theory finite
a modified renormalized coupling constant $\hat{\lambda}$ has to be
introduced via, cf.\ (\ref{eff:lambdar}),
\begin{equation} 
\frac{1}{\lambda_0} = \frac{1}{\kappa^\epsilon} 
\left[ \frac{1}{\hat{\lambda}} -
\frac{m}{\pi\epsilon} \left(1 - \frac{\hat{\Delta}}{\mu \hat{\lambda}}
\right)  \right],
\end{equation} 
which depends on the disorder strength.  The renormalized parameter
$\hat{\Delta}$ is defined in the same way as $\hat{\lambda}$.

In the previous section we saw that due to the interparticle repulsion,
not all the particles reside in the condensate.  We expect that the
random field causes an additional depletion of the condensate.  To
obtain this, we differentiate (\ref{L_D}) with respect to the chemical
potential.  This gives \cite{pla}
\begin{equation}  \label{depDelta}
\bar{n}_\Delta = \frac{\partial {\langle \cal L
}_\Delta \rangle}{\partial \mu} =
\frac{2^{d/2-5}\Gamma(2-d/2)}{\pi^{d/2}} m^{d/2} \lambda^{d/2-2}
\bar{n}_0^{d/2-1} \Delta,
\end{equation}   
where $\bar{n}_0$ denotes the density of particles residing in the
condensate.  We have here again replaced the bare parameters with the
(one-loop) renormalized ones.  This is consistent to this order since
(\ref{depDelta}) is already a one-loop result.

The divergence in the limit $\lambda \rightarrow 0$ for
$d <4$ signals the collapse of the system when the interparticle
repulsion is removed.  Note that in $d=2$, the depletion is independent
of the condensate density $\bar{n}_0$ \cite{GPS}:
\begin{equation} 
\bar{n}_\Delta = \frac{1}{16 \pi} \frac{m}{\lambda} \Delta .
\end{equation}    
The total particle number density $\bar{n}$ is given by
\begin{equation} 
\bar{n} = \bar{n}_0 \left(1 + \frac{m \lambda}{2 \pi}  \right) +
\frac{1}{16 \pi} \frac{m}{\lambda} \Delta.
\end{equation} 

We next calculate the mass current ${\bf g}$ to determine the superfluid
mass density, i.e., the mass density flowing with the superfluid
velocity ${\bf v}_{\rm s}$.  As we have seen in the preceding section,
in the absence of impurities and at zero temperature all the particles
participate in the superflow and move on the average with the velocity
${\bf v}_{\rm s}$.  We expect this to no longer hold in the presence of
impurities.  To determine the change in the superfluid mass density due
to impurities, we replace $\mu_0$ with $\mu_{\rm eff}$ as defined in
(\ref{bec:mureplacement}) and $i\partial_0$ with $i\partial_0 - ({\bf u}
- {\bf v}_{\rm s}) \cdot (-i \nabla)$ in the contribution (\ref{S_d}) to
the effective action, and differentiate it with respect to $-{\bf
u}$---the externally imposed velocity.  We find to linear order in the
difference ${\bf u}- {\bf v}_{\rm s}$:
\begin{equation} 
{\bf g} = \rho_{\rm s} {\bf v}_{\rm s} + \rho_{\rm n} {\bf u},
\end{equation} 
with the superfluid and normal mass density \cite{pla}
\begin{equation} 
\rho_{\rm s} = m\left(\bar{n} -  \frac{4}{d} \bar{n}_\Delta \right), \;\;\;\;
\rho_{\rm n} = \frac{4}{d} m \bar{n}_\Delta.
\end{equation}  
We see that the normal density is a factor $4/d$ larger than the mass
density $m\bar{n}_\Delta$ knocked out of the condensate by the impurities.
(For $d=3$ this gives the factor $\tfrac{4}{3}$ first found in Ref.\
\cite{HM}.)  Apparently, part of the zero-momentum states belongs for $d < 4$
not to the condensate, but to the normal fluid.  Being trapped by the
impurities, this fraction of the zero-momentum states are localized.  This
shows that the phenomenon of localization can be accounted for in the
Bogoliubov theory of superfluidity by including a random field.
\section{Vortices}
We shall now include vortices in the system.  A vortex in two space
dimensions may be pictured as a point-like object at scales large
compared to their core seize.  It is characterized by the winding number
$w$ of the map
\begin{equation} 
\varphi({\bf x}) : {\rm S}^1_{\bf x} \rightarrow {\rm S}^1
\end{equation} 
of a circle S$^1_{\bf x}$ around the vortex into the internal circle
S$^1$ parameterized by the Goldstone field $\varphi$.  In the
microscopic theory (\ref{eff:Lagr}), the asymptotic solution of a static
vortex with winding number $w$ located at the origin is well known
\cite{Fetter}
\begin{equation}  \label{qm:sol}
\phi({\bf x}) = \sqrt{\frac{\mu_0}{2 \lambda_0}} \left(1 - \xi_0^2
\frac{w^2}{4 {\bf x}^2}\right) {\rm e}^{i w \theta} + {\cal
O}\left(\frac{1}{{\bf x}^4} \right),
\end{equation} 
where $\theta$ is the azimuthal angle and $\xi_0 =
1/\sqrt{m\mu_0}=1/mc_0$ is the coherence length.  The density profile
$n({\bf x})$ in the presence of this vortex follows from taking
$|\phi({\bf x})|^2$.

To incorporate vortices in the effective theory we employ the powerful
principle of defect gauge symmetry developed by Kleinert
\cite{GFCM,KleinertPl,KleinertCam}.  In this approach, one introduces a
so-called vortex gauge field $\varphi_\mu^{\rm P} = (\varphi_0^{\rm P},
\bbox{\varphi}^{\rm P})$ in the effective theory (\ref{eff:Leff}) via
minimally coupling to the Goldstone field:
\begin{equation} \label{hydro:minimal}
\tilde{\partial}_\mu \varphi \rightarrow \tilde{\partial}_\mu \varphi +
\varphi_\mu^{\rm P}, 
\end{equation}
with $\tilde{\partial}_\mu = (\partial_0,-\nabla)$.  If there are $N$
vortices with winding number $w_\alpha$ ($\alpha=1, \cdots, N$) 
centered at ${\bf X}^1(t), \cdots , {\bf X}^{N}(t)$, the plastic field
satisfies the relation
\begin{equation} \label{qm:pla}
\nabla \times \bbox{\varphi}^{\rm P}(x) = - 2 \pi \sum_\alpha w_\alpha
\delta[{\bf x} - {\bf X}^\alpha(t)],
\end{equation} 
so that we obtain for the superfluid velocity field
\begin{equation} \label{qm:vort}
\nabla \times {\bf v}_{\rm s} = \sum_\alpha \gamma_\alpha
\delta[{\bf x} - {\bf X}^\alpha(t)],
\end{equation}
as required.  Here, $\gamma_\alpha = (2 \pi/m) w_\alpha$ is the
circulation of the $\alpha$th vortex which is quantized in units of $2
\pi/m$.  A summation over the indices labeling the vortices will always be
made explicit.  The combination $\tilde{\partial}_\mu \varphi +
\varphi_\mu^{\rm P}$ is invariant under the local gauge transformation
\begin{equation} 
\varphi(x) \rightarrow \varphi(x) + \alpha(x); \;\;\;\;\;
\varphi^{\rm P}_\mu \rightarrow \varphi^{\rm P}_\mu - \tilde{\partial}_\mu
\alpha(x),
\end{equation} 
with $\varphi^{\rm P}_\mu$ playing the role of a gauge field.    

In the  gauge $\varphi^{\rm P}_0=0$, Eq.\ (\ref{qm:pla}) can be solved to yield
\begin{equation} \label{eff:pla}
\varphi^{\rm P}_i(x) = 2 \pi \epsilon_{ij} \sum_\alpha w_\alpha 
\delta_j[x,L_\alpha(t)] 
\end{equation}
where $\epsilon_{ij}$ is the antisymmetric Levi-Civita symbol in two
dimensions, with $\epsilon_{12}=1$, and $\bbox{\delta} [x,L_\alpha(t)]$ is a
delta function on the line $L_\alpha(t)$ starting at the center ${\bf
X}^\alpha(t)$ of the $\alpha$th vortex and running to spatial infinity along
an arbitrary path:
\begin{equation} 
\delta_i [x,L_\alpha(t)] = \int_{L_\alpha(t)} \dd y_i \, \delta({\bf x} -
{\bf y}).
\end{equation} 

Let us for the moment concentrate on static vortices.  The field equation
obtained from the effective theory (\ref{eff:Leff}) with $\nabla \varphi$
replaced by the covariant derivative $\nabla \varphi - \bbox{\varphi}^{\rm
P}$ and $\partial_0 \varphi$ set to zero simply reads
\begin{equation}
\nabla \cdot {\bf v}_{\rm s} = 0, \;\;\;\; {\rm or} \;\;\;\; \nabla \cdot (\nabla
\varphi - \bbox{\varphi}^{\rm P}) = 0,
\end{equation}
when the fourth-order term is neglected.  It can be easily solved to
yield
\begin{equation} \label{qm:solution}
\varphi ({\bf x}) = - \int_{\bf y}  G({\bf x} - {\bf y}) \nabla \cdot
\bbox{\varphi}^{\rm P}({\bf y}),
\end{equation}
where $G({\bf x})$ is the Green function of the Laplace operator
\begin{equation}
G({\bf x}) = \int_{\bf k} \frac{ {\rm e}^{i {\bf k}
\cdot {\bf x}}}{{\bf k}^2} = - \frac{1}{2 \pi} \ln( |{\bf x}|).
\end{equation}
For the velocity field we obtain in this way the well-known expression
\cite{Lamb}
\begin{equation} \label{qm:vortices}
v_i({\bf x}) = \frac{1}{2 \pi} \epsilon_{ij} \sum_{
\alpha=1}^{N} \gamma_\alpha \frac{x_j- X^\alpha_j}{|
{\bf x}-{\bf X}^\alpha | ^{2}}  ,
\end{equation} 
which is valid for ${\bf x}$ sufficiently far away from the vortex cores.
Let us now specialize to the case of a single static vortex at the origin.
On substituting the corresponding solution in (\ref{roh1}), we find for the
density profile in the presence of a static vortex asymptotically
\begin{equation} 
n({\bf x}) = \bar{n} \left(1 - \xi_0^2 \frac{w^2}{2{\bf x}^2} \right).
\end{equation} 
This is the same formula as the one obtained from the solution
(\ref{qm:sol}) of the microscopic theory.  This exemplifies that with
the aid of the defect gauge symmetry principle, vortices are correctly
accounted for in the effective theory.

Let us proceed to investigate the dynamics of vortices in this formalism and
derive the action which governs it.  We consider only the first part of
the effective theory (\ref{eff:Leff}).  In ignoring the higher-order terms,
we approximate the superfluid by an incompressible fluid for which the
particle number density is constant, $n(x) = \bar{n}$, see Eq.\
(\ref{roh1}).  We again work in the gauge $\varphi^{\rm P}_0=0$ and replace
$\nabla \varphi$ by the covariant derivative $\nabla \varphi -
\bbox{\varphi}^{\rm P}$, with the plastic field given by (\ref{qm:pla}).  The
solution of the resulting field equation for $\varphi$ is again of the form
(\ref{qm:solution}), but now it is time-dependent because the plastic field
is.  Substituting this in the action $S_{\rm eff} = \int_x {\cal
L}_{\rm eff}$, we find after some straightforward calculus
\begin{equation} \label{qm:action}
S_{\rm eff} = m \bar{n} \int_t \left[\frac{1}{2} \sum_\alpha \gamma_\alpha
{\bf X}^\alpha \times \dot{\bf X}^\alpha + \frac{1}{2\pi} \sum_{\alpha <
\beta} \gamma_\alpha \gamma_\beta \ln(|{\bf X}^\alpha - {\bf X}^\beta|/a)
\right].
\end{equation} 
The constant $a$ has the dimension of a length and is included in the
argument of the logarithm for dimensional reasons.  Physically, it
represents the core size of a vortex.  The first term in
(\ref{qm:action}) leads to a twisted canonical structure which is
reminiscent of that found in the so-called Landau problem of a charged
particle confined to move in a plane perpendicular to an applied
magnetic field $H$.

To display the canonical structure, let us rewrite the first term of the
Lagrangian corresponding to (\ref{qm:action}) as
\begin{equation} 
L_1 = m \bar{n} \sum_\alpha \gamma_\alpha X^\alpha_1 \dot{X}^\alpha_2,
\end{equation} 
where we ignored a total derivative.  It follows that the canonical
conjugate to the second component $X_2^\alpha$ of the center coordinate ${\bf
X}^\alpha$ is essentially its first component \cite{YM}
\begin{equation} 
\frac{\partial L_1}{\partial \dot{X}_2^\alpha} =  m \bar{n} \gamma_\alpha
X^\alpha_1.
\end{equation} 
It implies that phase space coincides with real space and gives rise to the
commutation relation
\begin{equation} 
[X_2^\alpha, X_1^\beta ] = \frac{i}{w_\alpha} \ell^2 \delta^{\alpha
\beta},
\end{equation} 
where
\begin{equation} \label{qm:ell}
\ell = 1/\sqrt{2 \pi \bar{n}}
\end{equation} 
is a characteristic length whose definition is such that $2 \pi \ell^2$ is
the average area occupied by a particle of the superfluid film.  The
commutation relation leads to an uncertainty in the location of the vortex
centers given by
\begin{equation} 
\Delta X_1^\alpha \Delta X_2^\alpha \geq \frac{\ell^2}{2 |w_\alpha|} ,
\end{equation}   
which is inverse proportional to the particle number density.

From elementary quantum mechanics we know that to each unit cell (of area
$h$) in phase space there corresponds one quantum state.  That is, the
number of states in an area $S$ of phase space is given by
\begin{equation} 
\mbox{\# states in} \; S = \frac{1}{h} \int_S \dd p \, \dd q,
\end{equation} 
where $p$ and $q$ are a pair of canonically conjugate variables.  For the
case at hand, this implies that the available number of states in an area
$S_\alpha$ of {\it real} space is
\begin{equation} 
\mbox{\# states in} \; S_\alpha = |w_\alpha| \, \bar{n}
S_\alpha ,
\end{equation} 
or, equivalently, that the number of states per unit area available to the
$\alpha$th vortex is $|w_\alpha| \, \bar{n}$.  

This phenomenon that phase space coincides with real space is known to
also arise in the Landau problem.  There, it leads to the well-known
degeneracy $|e_\alpha| H/h$ of each Landau level, where $e_\alpha =
v_\alpha e_0$ is the electric charge of the particle, with $e_0 (>0)$
the unit of charge.  In terms of the magnetic flux quantum $\Phi_0 =
h/e_0$, the Landau degeneracy can be rewritten as $|v_\alpha| H/\Phi_0 =
|v_\alpha| \bar{n}_\otimes$, with $\bar{n}_\otimes$ the flux number
density.  In other words, whereas the degeneracy in the case of vortices
in a superfluid film is given by the particle number density, here it is
given by the flux number density.  Using this analogy, we see that the
characteristic length (\ref{qm:ell}) translates into $\ell_H = 1/\sqrt{2
\pi \bar{n}_\otimes}$ which is precisely the magnetic length of the Landau
problem.

The first term in the action (\ref{qm:action}) is also responsible for
the so-called geometrical phase \cite{Berry} acquired by the
wavefunction of a vortex when it traverses a closed path.  Let us first
discuss the case of a charged particle moving adiabatically around a
close path $\Gamma_\alpha$.  Its wavefunction picks up an extra
Aharonov-Bohm phase factor given by the Wilson loop:
\begin{equation} \label{qm:Berry}
W(\Gamma_\alpha) = \exp[i \gamma(\Gamma_\alpha)] = \exp\left(\frac{i
e_\alpha}{\hbar}
\oint_{\Gamma_\alpha} \dd {\bf x} \cdot {\bf A}\right) = \exp \left[2
\pi i v_\alpha \frac{H S(\Gamma_\alpha)}{\Phi_0}\right]
\end{equation} 
where ${\bf A}$ is the vector potential describing the external magnetic
field and $H S(\Gamma_\alpha)$ is the magnetic flux through the area
$S(\Gamma_\alpha)$ spanned by the loop $\Gamma_\alpha$.  The geometrical
phase $\gamma(\Gamma_\alpha)$ in (\ref{qm:Berry}) is seen to be ($2 \pi
v_\alpha$ times) the number of flux quanta enclosed by the path
$\Gamma_\alpha$.

On account of the above analogy, it follows that the geometrical phase
picked up by the wavefunction of a vortex when it is moved adiabatically
around a closed path in the superfluid film is ($2 \pi w_\alpha$ times) the
number of superfluid particles enclosed by the path \cite{HW}.

The second term in the action (\ref{qm:action}) represents the long-ranged
interaction between two vortices mediated by the exchange of Goldstone
quanta.  The action yields the well-known equations of motion for point
vortices in an incompressible two-dimensional superfluid \cite{Lamb,Lund}:
\begin{equation} 
\dot{X}_i^\beta(t) = \frac{\epsilon_{ij}}{2 \pi}  \sum_{\alpha \neq \beta}
\gamma_\alpha \frac{X^\beta_j(t) - X^\alpha_j(t)}{| {\bf X}^\beta(t)-{\bf
X}^\alpha(t) | ^{2}} .
\end{equation} 
Note that $\dot{X}_i^\beta(t) = v_i\left[{\bf X}^\beta(t)\right]$, where
${\bf v}(x)$ is the superfluid velocity (\ref{qm:vortices}) with the
time-dependence of the centers of the vortices included.  This nicely
illustrates a result due to Helmholtz for ideal fluids, stating that
a vortex moves with the fluid, i.e., at the local velocity produced by the
other vortices in the system.  Experimental support for this conclusion has
been reported in Ref.\ \cite{YP}.
\section{Kosterlitz-Thouless Phase Transition \label{sec:KT}}
Although we are interested mainly in quantum phase transitions in these
Lectures, there is one classical phase transition special to two
dimensions which turns out to be relevant for our discussion later
on---the so-called Kosterlitz-Thouless phase transition.  It is well
known that a superfluid film undergoes such a phase transition at a
temperature well below the bulk transition temperature.  The superfluid
low-temperature state is characterized by tightly bound
vortex-antivortex pairs which at the Kosterlitz-Thouless temperature
unbind and thereby disorder the superfluid state.  The disordered state,
at temperatures still below the bulk transition temperature, consists of
a plasma of unbound vortices.

Since the phase transition is an equilibrium transition, we can ignore
any time dependence.  The important fluctuations here, at temperatures
below the bulk transition temperature, are phase fluctuations so that we
can consider the London limit, where the phase of the $\phi(x)$-field is
allowed to vary in spacetime while the modulus is kept fixed, and take as
Hamiltonian
\begin{equation} \label{kt:HHe}
{\cal H} = \tfrac{1}{2} \rho_{\rm s} {\bf v}^2_{\rm s},
\end{equation} 
where $\rho_{\rm s}$ is the superfluid mass density which we assume to be
constant and ${\bf v}_{\rm s}$ is the superfluid velocity
\begin{equation} \label{kt:vs}
{\bf v}_{\rm s} = \frac{1}{m} (\nabla \varphi - \bbox{\varphi}^{\rm P}),
\end{equation}
with the vortex gauge field $\bbox{\varphi}^{\rm P}$ included to account
for possible vortices in the system.  We shall restrict ourselves to
vortices of unit winding number, so that $w_\alpha = \pm 1$ for a vortex
and antivortex, respectively.

The canonical partition function describing the equilibrium configuration of
$N_+$ vortices and $N_-$ antivortices in a superfluid film is given by
\begin{equation}  \label{kt:Zorig}
Z_N = \frac{1}{N_+! N_-!} \prod_\alpha \int_{{\bf
X}^\alpha} \int \DD \varphi \, \exp\left(-\beta \int_{\bf x} {\cal
H}\right),  
\end{equation} 
with ${\cal H}$ the Hamiltonian (\ref{kt:HHe}) and $N = N_+ + N_-$ the
total number of vortices and antivortices.  The factors $N_+!$ and
$N_-!$ arise because the vortices and antivortices are
indistinguishable, and $\prod_\alpha \int_{{\bf X}^\alpha}$ denotes the
integration over the positions of the vortices.  The functional integral
over $\varphi$ is Gaussian and therefore easily carried out, with the
result
\begin{equation}   \label{kt:Zx}
Z_N = \frac{1}{N_+! N_-!} \prod_\alpha \int_{{\bf X}^\alpha}
\exp\left[\pi  \frac{\beta \rho_{\rm s}}{m^2} \sum_{\alpha, \beta}
w_\alpha w_\beta \ln\left(|{\bf X}^\alpha - {\bf
X}^\beta|/a\right) \right].
\end{equation} 
Apart from an irrelevant normalization factor, Eq.~(\ref{kt:Zx}) is the
canonical partition function of a two-dimensional Coulomb gas with
charges $q_\alpha = q w_\alpha = \pm q$, where
\begin{equation} 
q = \sqrt{2 \pi \rho_{\rm s} }/m.
\end{equation}
Let us rewrite the sum in the exponent appearing in (\ref{kt:Zx}) as
\begin{eqnarray} 
\lefteqn{\sum_{\alpha, \beta} q_\alpha
q_\beta \ln\left(|{\bf X}^\alpha - {\bf X}^\beta|/a\right) =}
\nonumber \\ && \sum_{\alpha, \beta} q_\alpha q_\beta \left[
\ln\left(|{\bf X}^\alpha - {\bf X}^\beta|/a\right) - \ln(0) \right] +
\ln(0) \left(\sum_\alpha q_\alpha \right)^2,
\end{eqnarray} 
where we isolated the self-interaction in the last term at the
right-hand side.  Since $\ln(0) = -\infty$, the charges must add up to
zero so as to obtain a nonzero partition function.  From now on we will
therefore assume overall charge neutrality, $\sum_\alpha q_\alpha = 0$,
so that $N_+ = N_- = N/2$, where $N$ must be an even integer.  To
regularize the remaining divergence, we replace $\ln(0)$ with an
undetermined, negative constant $-c$.  The exponent of (\ref{kt:Zx})
thus becomes
\begin{equation} \label{kt:reg}
\frac{\beta}{2} \sum_{\alpha, \beta} q_\alpha q_\beta \ln\left(|{\bf
X}^\alpha - {\bf X}^\beta|/a \right) = \frac{\beta}{2} \sum_{\alpha \neq
\beta} q_\alpha q_\beta \ln\left(|{\bf X}^\alpha - {\bf X}^\beta|/a
\right) - \beta \epsilon_{\rm c} N,
\end{equation} 
where $\epsilon_{\rm c}= c q^2/2$ physically represents the core energy,
i.e., the energy required to create a single vortex.  In deriving this we
used the identity $\sum_{\alpha \neq \beta} q_\alpha q_\beta = -
\sum_\alpha q_\alpha^2 = -N q^2$ which follows from charge neutrality.
Having dealt with the self-interaction, we limit the integrations
$\prod_\alpha \int_{{\bf X}^\alpha}$ in (\ref{kt:Zx}) over the location of
the vortices to those regions where they are more than a distance $a$ apart,
$|{\bf X}^\alpha - {\bf X}^\beta| >a$.  The grand-canonical partition
function of the system can now be cast in the form
\begin{equation} \label{kt:coul}
Z = \sum_{N=0}^\infty \frac{z^{N}}{[(N/2)!]^2}
\prod_{\alpha} \int_{{\bf X}^\alpha} \exp\left[ \frac{\beta}{2}\sum_{\alpha
\neq \beta} q_\alpha q_\beta \ln\left(|{\bf X}^\alpha - {\bf X}^\beta|/a\right)
\right], 
\end{equation}
where $z = \exp(-\beta \epsilon_{\rm c})$ is the fugacity.  The system is
known to undergo a phase transition at the Kosterlitz-Thouless
temperature
\cite{Berezinskii,KT73}
\begin{equation} \label{jump}
T_{\rm KT} = \frac{1}{4} q^2 =  \frac{\pi}{2} \frac{\rho_{\rm
s}}{m^2}, 
\end{equation} 
triggered by the unbinding of vortex-antivortex pairs.  It follows from this
equation that the two-dimensional superfluid mass density $\rho_{\rm
s}(T)$, which varies from sample to sample, terminates on a line with
universal slope as the temperature approaches the Kosterlitz-Thouless
temperature from below \cite{NeKo}. 
\section{Dual Theory}
Let us proceed to represent the partition function (\ref{kt:coul}) by a
field theory---a so-called dual theory.  The idea behind such a dual
transformation is to obtain a formulation in which the vortices are not
described as singular objects as is the case in the original
formulation, but by ordinary fields.  To derive it we note that
$\ln(|{\bf x}|)$ is the inverse of the Laplace operator $\nabla^2$,
\begin{equation} 
\frac{1}{2 \pi} \nabla^2 \ln(|{\bf x}|) = \delta({\bf x}).
\end{equation} 
This allows us to represent the exponential function in (\ref{kt:coul})
as a functional integral over an auxiliary field $\phi$:
\begin{equation}  \label{kt:aux}
\exp\left[ \frac{\beta}{2} \sum_{\alpha \neq \beta} q_\alpha q_\beta
\ln\left(|{\bf X}^\alpha - {\bf X}^\beta|/a\right) \right] =
\int \DD \phi \exp\left\{ - \int_{\bf x} \left[ \frac{1}{4 \pi \beta} (\nabla
\phi)^2 + i \rho_q \phi \right] \right\}, 
\end{equation} 
where $\rho_q({\bf x}) = \sum_\alpha q_\alpha \delta({\bf x} - {\bf
X}^\alpha)$ is the charge density.  In this way, the partition function
becomes
\begin{equation} \label{kt:phi}
Z = \sum_{N=0}^\infty \frac{z^{N}}{[(N/2)!]^2}
\prod_{\alpha=1}^N \int_{{\bf X}^\alpha} 
\int \DD \phi \exp\left\{ - \int_{\bf x} \left[ \frac{1}{4 \pi \beta} (\nabla
\phi)^2 + i \rho_q \phi \right] \right\}.
\end{equation}
In a mean-field treatment, the functional integral over the auxiliary field
introduced in (\ref{kt:aux}) is approximated by the saddle point determined
by the field equation
\begin{equation} \label{kt:feq}
i T \nabla^2 \phi = - 2 \pi \rho_q.
\end{equation} 
When we introduce the scalar variable $\Phi := i T \phi$, this equation
becomes formally Gauss' law, with $\Phi$ the electrostatic scalar
potential.  The auxiliary field introduced in (\ref{kt:aux}) may
therefore be thought of as representing the scalar potential of the
equivalent two-dimensional Coulomb gas \cite{GFCM}.

On account of charge neutrality, we have the identity
\begin{equation} 
\left[ \int_{\bf x} \left( {\rm e}^{iq \phi({\bf x})} + {\rm e}^{-iq \phi({\bf
x})} \right) \right]^N = \frac{N!}{[(N/2)!]^2} \prod_{\alpha=1}^{N}
\int_{{\bf X}^\alpha} {\rm e}^{-i \sum_{\alpha} q_\alpha \phi({\bf
X}^\alpha)},
\end{equation} 
where we recall that $N$ is an even number.  The factor $N!/[(N/2)!]^2$
is the number of charge-neutral terms contained in the binomial
expansion of the left-hand side.  The partition function (\ref{kt:phi})
may thus be written as \cite{GFCM}
\begin{eqnarray} \label{kt:sG} 
Z &=& \sum_{N=0}^\infty \frac{(2z)^{N}}{N!}
\int \DD \phi \exp\left[ - \int_{\bf x} \frac{1}{4 \pi \beta} (\nabla 
\phi)^2 \right] \left[\cos\left(\int_{\bf x} q \phi \right)
\right]^N \nonumber \\ &=& 
\int \DD \phi \exp\left\{ - \int_{\bf x} \left[ \frac{1}{4 \pi \beta}
(\nabla \phi)^2 - 2z \cos(q \phi) \right] \right\},
\end{eqnarray} 
where in the final form we recognize the sine-Gordon model.  This is the
dual theory we were seeking.  Contrary to the original formulation
(\ref{kt:Zorig}), which contains the vortices as singular objects, the dual
formulation has no singularities.  To see how the vortices and the
Kosterlitz-Thouless phase transition are represented in the dual theory we
note that the field equation of the auxiliary field now reads
\begin{equation}  \label{kt:gauss}
i T \nabla^2 \phi = 2 \pi z q \left({\rm e}^{iq \phi} - {\rm e}^{-iq
\phi} \right).
\end{equation} 
On comparison with the previous field equation (\ref{kt:feq}), it follows
that the right-hand side represents the charge density of the Coulomb gas.
In terms of the scalar potential $\Phi$, Eq.~(\ref{kt:gauss}) becomes the
Poisson-Boltzmann equation
\begin{equation} \label{kt:PB} 
\nabla^2 \Phi = - 2 \pi q \left(z \, {\rm e}^{- \beta q \Phi} - z
\, {\rm e}^{\beta q \Phi} \right),
\end{equation}  
describing, at least for temperatures above the Kosterlitz-Thouless
temperature, a plasma of positive and negative charges with
density $n_\pm$,
\begin{equation} \label{kt:spatiald} 
n_\pm = z \, {\rm e}^{\mp \beta q \Phi},
\end{equation} 
respectively.  The fugacity $z$ is the density at zero scalar potential.
(It is to recalled that we suppress factors of $a$ denoting the core
size of the vortices.)  Equation (\ref{kt:PB}) is a self-consistent
equation for the scalar potential $\Phi$ giving the spatial distribution
of the charges via (\ref{kt:spatiald}).  It follows from this argument
that the interaction term $2z \cos(q \phi)$ of the sine-Gordon model
represents a plasma of vortices.

The renormalization group applied to the sine-Gordon model reveals that at
the Kosterlitz-Thouless temperature $T_{\rm KT} = \tfrac{1}{4}q^2$
there is a phase transition between a low-temperature phase of tightly bound
neutral pairs and a high-temperature plasma phase of unbound vortices
\cite{Schenker}.  In the low-temperature phase, the (renormalized) fugacity
scales to zero in the large-scale limit so that the interaction term,
representing the plasma of unbound vortices, is suppressed.  The
long-distance behavior of the low-temperature phase is therefore well
described by the free theory $(\nabla \phi)^2/4 \pi \beta$, representing
a gapless mode---the so-called Kosterlitz-Thouless mode.  This is the
superfluid state.  The expectation value of a single vortex vanishes
because in this gapless state its energy diverges in the infrared.

An important characteristic of a charged plasma is that it has no gapless
excitations, the photon being transmuted into a massive plasmon.  To see
this we assume that $q \Phi << T$, so that $\sinh(\beta q \Phi)
\approx \beta q \Phi$.  In this approximation, the Poisson-Boltzmann equation
(\ref{kt:PB}) can be linearized to give
\begin{equation} \label{kt:mpoi}
(\nabla^2 - m_{\rm D}^2) \Phi = 0, \;\;\; m_{\rm D}^2 = 4 \pi
\beta z q^2.
\end{equation} 
This shows us that, in contradistinction to the low-temperature phase,
in the high-temperature phase, the scalar potential describes a massive
mode---the plasmon.  In other words, the Kosterlitz-Thouless mode
acquires an energy gap $m_{\rm D}$.  Since it provides the
high-temperature phase with an infrared cutoff, isolated vortices have a
finite energy now and accordingly a finite probability to be created.
This Debeye mechanism of mass generation for the photon should be
distinguished from the Higgs mechanism which operates in superconductors
(see below) and which also generates a photon mass.

Another property of a charged plasma is that it screens charges.  This
so-called Debeye screening may be illustrated by adding an external
charge to the system.  The linearized Poisson-Boltzmann equation
(\ref{kt:mpoi}) then becomes
\begin{equation} \label{kt:pois}
(\nabla^2 - m_{\rm D}^2) \Phi({\bf x}) = - 2 \pi q_0 \delta ({\bf x}), 
\end{equation} 
with $q_0$ the external charge which we have placed at the origin.  The
solution of this equation is given by $\Phi ({\bf x}) = q_0
K_0(m_{\rm D}|{\bf x}|)$ with $K_0$ a modified Bessel function.  The mass
term in (\ref{kt:pois}) is ($2 \pi$ times) the charge density induced by the
external charge, i.e.,
\begin{equation} 
\rho_{\rm ind}({\bf x}) = - \frac{1}{2 \pi} q_0 m_{\rm D}^2
K_0(m_{\rm D}|{\bf x}|).  
\end{equation} 
By integrating this density over the entire system, we see that the total
induced charge $\int_{\bf x} \rho_{\rm ind} = -q_0$ completely screens the
external charge---at least in the linear approximation we are using here.
The inverse of the plasmon mass is the so-called Debeye screening length.

To see that the sine-Gordon model gives a dual description of a
superfluid film we cast the field equation (\ref{kt:feq}) in the form
\begin{equation} 
i T \nabla^2 \phi = - m q \nabla \times {\bf v}_{\rm s},
\end{equation} 
where we employed Eq.\ (\ref{qm:vort}).  On integrating this
equation, we obtain up to an irrelevant integration constant
\begin{equation} 
i T \partial_i \phi = - q \epsilon_{i j} (\partial_j \varphi -
\varphi_j^{\rm P}).
\end{equation} 
This relation, involving the antisymmetric Levi-Civita symbol, is a typical
one between dual variables.  It also nicely illustrates that although the
dual variable $\phi$ is a regular field, it nevertheless contains the
information about the vortices which in the original formulation are
described via the singular vortex gauge field $\bbox{\varphi}^{\rm P}$.

Given this observation it is straightforward to calculate the
current-current correlation function $\langle g_i ({\bf k}) g_j(-{\bf
k}) \rangle$, with
\begin{equation} 
{\bf g} = \rho_{\rm s} {\bf v}_{\rm s}
\end{equation} 
the mass current.  We find
\begin{equation} 
\langle g_i ({\bf k}) g_j(-{\bf k}) \rangle = - \frac{\rho_{\rm s}}{2
\pi \beta^2} \epsilon_{ik} \epsilon_{jl} k_k k_l \langle \phi({\bf k})
\phi(-{\bf k}) \rangle,
\end{equation} 
where the average is to be taken with respect to the partition function
\begin{equation} 
Z_0 = \int \DD \phi \exp\left[ - \frac{1}{4 \pi \beta} \int_{\bf x} (\nabla
\phi)^2 \right],
\end{equation} 
which is obtained from (\ref{kt:sG}) by setting the interaction term to
zero.  We obtain in this way the standard expression for a superfluid
\begin{equation} \label{kt:jj} 
\langle g_i ({\bf k}) g_j(-{\bf k}) \rangle = - \frac{\rho_{\rm s}}{\beta} 
 \frac{1}{{\bf k}^2} \left( \delta_{ij} {\bf k}^2 - k_i k_j \right).
\end{equation}  
The $1/{\bf k}^2$ reflects the gaplessness of the $\phi$-field in the
low-temperature phase, while the combination $\delta_{ij} {\bf k}^2 - k_i
k_j$ arises because the current is divergent free, $\nabla \cdot {\bf
g}({\bf x}) = 0$, or ${\bf k} \cdot {\bf g}({\bf k}) = 0$.

%% file: Superconductor.tex
\chapter{Superconductivity \label{chap:sc}}
In this chapter we shall demonstrate a close connection between the
Bogoliubov theory of superfluidity discussed in the previous chapter and
the strong-coupling limit of the BCS theory of superconductivity.  The
phase-only effective theory governing the superconducting state is
derived.  It is also pointed out that a superconducting film at finite
temperature undergoes a Kosterlitz-Thouless phase transition.
\section{BCS Theory}
Our starting point is the famous microscopic model of Bardeen, Cooper,
and Schrieffer (BCS) defined by the Lagrangian \cite{BCS}
\begin{eqnarray}  \label{bcs:BCS}
     {\cal L} &=& \psi^{\ast}_{\uparrow} [i\partial_0 - \xi(-i \nabla)]
\psi_{\uparrow}  
     + \psi_{\downarrow}^{\ast} [i \partial_0 - \xi(-i
     \nabla)]\psi_{\downarrow} - \lambda_0
     \psi_{\uparrow}^{\ast}\,\psi_{\downarrow}
     ^{\ast}\,\psi_{\downarrow}\,\psi_{\uparrow} \nonumber \\ &:=& {\cal
     L}_{0} + {\cal L}_{\rm i},
\end{eqnarray} 
where ${\cal L}_{\rm i} = - \lambda_0 \psi_{\uparrow}^{\ast} \,
\psi_{\downarrow}^{\ast}\,\psi_{\downarrow}\,\psi_{\uparrow}$ is a
contact interaction term, representing the effective, phonon mediated,
attraction between electrons with coupling constant $\lambda_0 < 0$, and
${\cal L}_{0}$ is the remainder.  In (\ref{bcs:BCS}), the field
$\psi_{\uparrow (\downarrow )}$ is an anticommuting field describing the
electrons with mass $m$ and spin up (down); $\xi(-i \nabla) = \epsilon(-i
\nabla) - \mu_0$, with $\epsilon(-i \nabla) = - \nabla^2/2m$, is the kinetic
energy operator with the chemical potential $\mu_0$ subtracted.

The Lagrangian (\ref{bcs:BCS}) is invariant under global U(1)
transformations.  Under such a transformation, the electron fields pick up
an additional phase factor
\begin{equation} \label{bcs:3g}
\psi_{\sigma} \rightarrow \mbox{e}^{i \alpha }
\psi_{\sigma}                                
\end{equation}
with $\sigma = \uparrow, \downarrow$ and $\alpha$ a constant.
Notwithstanding its simple form, the microscopic model (\ref{bcs:BCS}) is a
good starting point to describe BCS superconductors.  The reason is that the
interaction term allows for the formation of Cooper pairs which below a
critical temperature condense.  This results in a nonzero expectation value
of the field $\Delta$ describing the Cooper pairs, and a spontaneous
breakdown of the global U(1) symmetry.  This in turn gives rise to the
gapless Anderson-Bogoliubov mode which---after incorporating the
electromagnetic field---lies at the root of most startling properties of
superconductors \cite{Weinberg}.

To obtain the effective theory governing the Anderson-Bogoliubov mode,
let us integrate out the fermionic degrees of freedom.  To this end we
introduce Nambu's notation and rewrite the Lagrangian (\ref{bcs:BCS}) in
terms of a two-component field
\begin{equation} \label{bcs:32}
\psi = \left( \begin{array}{c} \psi_{\uparrow} \\ 
           \psi_{\downarrow}^{\ast} \end{array} \right) \:\:\:\:\:\:
    \psi^{\dagger} = (\psi_{\uparrow}^{\ast},\psi_{\downarrow}).
\end{equation} 
In this notation, ${\cal L}_{0}$ becomes 
\begin{equation}    \label{bcs:33}
{\cal L}_{0} = \psi^{\dagger}\,
\left(\begin{array}{cc}
i \partial_0 - \xi(-i \nabla) & 0              \\
0           &  i \partial_0 + \xi(-i \nabla) 
    \end{array}\right) \, \psi,                        
\end{equation}
where we explicitly employed the anticommuting character of the electron
fields and neglected terms which are a total derivative.  The partition
function,
\begin{equation}     \label{bcs:34}
Z = \int \DD \psi^{\dagger} \DD \psi \exp \left( i \int_x
\,{\cal L} \right),                                            
\end{equation} 
must for our purpose be manipulated in a form bilinear in the electron
fields.  This is achieved by rewriting the quartic interaction term as a
functional integral over auxiliary fields $\Delta$ and $\Delta^*$ (for
details see Ref.\ \cite{KleinertFS}):
\begin{eqnarray}   \label{bcs:35} 
\lefteqn{
\exp \left( -i \lambda_0 \int_x \psi_{\uparrow}^{\ast}
\, \psi_{\downarrow}^{\ast} \, \psi_{\downarrow}\, \psi_{\uparrow} 
\right)  = }                                           \\
& & \!\!\!\!\!\! \int \DD \Delta^* \DD \Delta \exp \left[ -i
\int_x \left( \Delta^* \, \psi_{\downarrow}\,\psi_{\uparrow} +
\psi_{\uparrow}^{\ast} \, \psi_{\downarrow}^{\ast} \, \Delta -
\frac{1}{\lambda_0 } \Delta^* \Delta \right) \right], \nonumber 
\end{eqnarray} 
where, as always, an overall normalization factor is omitted.  Classically,
$\Delta$ merely abbreviates the product of two electron fields
\begin{equation}  \label{bcs:del}
\Delta = \lambda_0 \psi_{\downarrow} \psi_{\uparrow}.       
\end{equation} 
It would therefore be more appropriate to give $\Delta$ two spin labels
$\Delta_{\downarrow \uparrow}$. Since $\psi_{\uparrow}$ and
$\psi_{\downarrow}$ are anticommuting fields, $\Delta$ is antisymmetric in
these indices.  Physically, it describes the Cooper pairs of the
superconducting state.

By employing (\ref{bcs:35}), we can cast the partition function in the desired
bilinear form:
\begin{eqnarray}  \label{bcs:36}
Z = \int \DD \psi^{\dagger} \DD \psi \int \DD
\Delta^* \DD \Delta  \!\!\!\!\! && \!\!\!\!\!
\exp\left(\frac{i}{\lambda_0} \int_x \Delta^*  \Delta \right)  \\  &&
\!\!\!\!\!\!\!\!\!\!\! \times \exp  \left[  
    i \int_x \, \psi^{\dagger} \left( \begin{array}{cc} i \partial_{0} -
\xi(-i \nabla) & -\Delta \\ -\Delta^* & i \partial_{0} + \xi(-i \nabla)
\end{array} \right)   \psi \right] \nonumber .  
\end{eqnarray} 
Changing the order of integration and performing the Gaussian integral over
the Grassmann fields, we obtain
\begin{equation}   \label{bcs:37}
Z = \int \DD \Delta^* \DD \Delta \, \exp \left(i S_{\rm eff} [
\Delta^*, \Delta] + \frac{i}{\lambda_0}
\int_x \Delta^* \Delta \right),  
\end{equation}
where $S_{\rm eff}$ is the one-loop effective action which, using the
identity Det($A$) = exp[Tr ln($A$)], can be cast in the form
\begin{equation}  \label{bcs:312}
S_{\rm eff}[\Delta^*, \Delta] = -i \, {\rm Tr} \ln \left(
\begin{array}{cc} p_{0} - \xi ({\bf p}) & -\Delta \\ -\Delta^* &
p_{0} + \xi ({\bf p})
\end{array}\right),
\end{equation} 
where $p_0 = i \partial_0$ and $\xi({\bf p}) = \epsilon({\bf p}) - \mu_0$,
with $\epsilon({\bf p}) = {\bf p}^2/2m$.

In the mean-field approximation, the functional integral (\ref{bcs:37}) is
approximated by the saddle point:
\begin{equation}   \label{bcs:38}
Z =  \exp \left(i S_{\rm eff}
[ \Delta^*_{\rm mf}, \Delta_{\rm mf} ]  + \frac{i}{\lambda_0} \int_x
\Delta^*_{\rm mf} \Delta_{\rm mf}  \right),                   
\end{equation} 
where $\Delta_{\rm mf}$ is the solution of mean-field equation
\begin{equation}     \label{bcs:gap}
\frac{\delta S_{\rm eff} }{\delta \Delta^* (x) } = - \frac{1}{\lambda_0} \Delta.
\end{equation}
If we assume the system to be spacetime independent so that $\Delta_{\rm
mf}(x) = \bar{\Delta}$, Eq.\ (\ref{bcs:gap}) yields the celebrated BCS gap
\cite{BCS} equation:
\begin{eqnarray}   \label{bcs:gape} 
\frac{1}{\lambda_0} &=& - i  \int_k \frac{1}{k_{0}^{2} - E^{2}(k) + i \eta}
\nonumber \\ &=& - \frac{1}{2} \int_{\bf k} \frac{1}{E({\bf k})},
\end{eqnarray} 
where $\eta$ is an infinitesimal positive constant that is to be set to
zero at the end of the calculation, and 
\begin{equation}  \label{bcs:spec}
E({\bf k}) = \sqrt{\xi^2({\bf k}) + |\bar{\Delta}|^2}
\end{equation}  
is the spectrum of the elementary fermionic excitations.  If this equation
yields a solution with $\bar{\Delta} \neq 0$, the global U(1) symmetry
(\ref{bcs:3g}) is spontaneously broken since
\begin{equation}
\bar{\Delta} \rightarrow \mbox{e}^{2i \alpha } \bar{\Delta} \neq  
\bar{\Delta}                             
\end{equation}
under this transformation.  The factor $2$ in the exponential function arises
because $\Delta$, describing the Cooper pairs, is built from two electron
fields.  It satisfies Landau's definition of an order parameter as its value
is zero in the symmetric, disordered state and nonzero in the state with
broken symmetry.  It directly measures whether the U(1) symmetry is
spontaneously broken.

In the case of a spacetime-independent system, the effective action
(\ref{bcs:312}) is readily evaluated.  Writing 
\begin{equation} 
\left(
\begin{array}{cc} p_{0} - \xi ({\bf p}) & -\bar{\Delta} \\ -\bar{\Delta}^* &
p_{0} + \xi ({\bf p}) \end{array}\right) = 
\left(
\begin{array}{cc} p_{0} - \xi ({\bf p}) & 0 \\ 0 &
p_{0} + \xi ({\bf p}) \end{array}\right) - \left(
\begin{array}{cc} 0 & \bar{\Delta} \\ \bar{\Delta}^* & 0 \end{array}\right),
\end{equation} 
and expanding the second logarithm in a Taylor series, we recognize the
form 
\begin{equation}   
S_{\rm eff}[\bar{\Delta}^*, \bar{\Delta}] = -i \, {\rm Tr} \ln \left(
\begin{array}{cc} p_{0} - \xi ({\bf p}) & 0 \\ 0 &
p_{0} + \xi ({\bf p}) \end{array}\right) - i \, {\rm Tr}
\ln \left(1 - \frac{|\bar{\Delta}|^2}{p_0^2 - \xi^2({\bf p})} \right),
\end{equation} 
up to an irrelevant constant.  The integral over the loop energy $k_0$ gives
for the corresponding effective Lagrangian
\begin{equation} 
{\cal L}_{\rm eff} = \int_{\bf k} \left[ E({\bf k}) - \xi({\bf k})
\right].
\end{equation} 
To this one-loop result we have to add the tree term
$|\bar{\Delta}|^2/\lambda_0$.  Expanding $E({\bf k})$ in $\bar{\Delta}$, we see
that the effective Lagrangian also contains a term quadratic in
$\bar{\Delta}$.  This term amounts to a renormalization of the coupling
constant; we find to this order for the renormalized coupling constant
$\lambda$:
\begin{equation} \label{bcs:ren} 
\frac{1}{\lambda} = \frac{1}{\lambda_0} + \frac{1}{2} \int_{\bf k}
\frac{1}{|\xi({\bf k})|},
\end{equation} 
where it should be remembered that the bare coupling constant $\lambda_0$ is
negative, so that there is an attractive interaction between the fermions.
We shall analyze this equation later on, for the moment it suffice to note
that we can distinguish two limits.  One, the limit where the bare coupling
constant is taken to zero, $\lambda_0 \rightarrow 0^-$, which is the famous
weak-coupling BCS limit.  Second, the limit where the bare coupling is taken
to minus infinity $\lambda_0 \rightarrow - \infty$.  This is the
strong-coupling limit, where the attractive interaction is such that the
fermions form tightly bound pairs \cite{Leggett}.  These composite bosons
have a weak repulsive interaction and can undergo Bose-Einstein
condensation (see succeeding section).

Since there are two unknowns contained in the theory, viz., $\bar{\Delta}$
and $\mu_0$, we need a second equation to determine these variables in the
mean-field approximation \cite{Leggett}.  To find the second equation we
note that the average fermion number $N$, which is obtained by
differentiating the effective action (\ref{bcs:312}) with respect to $\mu$
\begin{equation} 
N = \frac{\partial S_{\rm eff}}{\partial \mu},
\end{equation} 
is fixed.  If the system is spacetime independent, this reduces in the
one-loop approximation to
\begin{equation} \label{bcs:n}
\bar{n} = - i\, {\rm tr} \int_k \, G_0(k) \tau_3,
\end{equation} 
where $\bar{n}=N/V$, with $V$ the volume of the system, is the constant
fermion number density, $\tau_3$ is the diagonal Pauli matrix in Nambu
space,
\begin{equation} 
\tau_3 = \left(
\begin{array}{cr} 1 & 0 \\ 0 & -1
\end{array} \right),
\end{equation} 
and $G_0(k)$ is the Feynman propagator,
\begin{eqnarray}    \label{bcs:prop}
G_0(k) &=&
\left( \begin{array}{cc} k_0 - \xi  ({\bf k}) 
& -\bar{\Delta} \\ -\Delta^*_0  & k_0 + \xi ({\bf k}) 
\end{array} \right)^{-1}  \\ &=& 
\frac{1}{k_0^2 - E^2({\bf k}) + i  \eta } 
\left( \begin{array}{cc} k_{0} \, {\rm e}^{i k_0 \eta } + \xi
({\bf k})  & 
\bar{\Delta} \\ \bar{\Delta}^* & k_{0} \, {\rm e}^{-i k_0 \eta}- \xi
({\bf k}) \end{array} \right). \nonumber 
\end{eqnarray}
Here, $\eta$ is an infinitesimal positive constant that is to be set to zero
at the end of the calculation.  The exponential functions in the diagonal
elements of the propagator are an additional convergence factor needed in
nonrelativistic theories \cite{Mattuck}.  If the integral over the loop
energy $k_0$ in the particle number equation (\ref{bcs:n}) is carried out,
it takes the familiar form
\begin{equation} \label{bcs:ne} 
\bar{n} = \int_{\bf k} \left(1 - \frac{\xi({\bf k})}{E({\bf k})} \right)
\end{equation} 
The two equations (\ref{bcs:gape}) and (\ref{bcs:ne}) determine $\bar{\Delta}$
and $\mu_0$.  They are usually evaluated in the weak-coupling BCS limit.
However, as was first pointed out by Leggett \cite{Leggett}, they can also
be easily solved in the strong-coupling limit (see succeeding section),
where the fermions are tightly bound in pairs.  More recently, also the
crossover between the weak-coupling BCS limit and the strong-coupling
composite boson limit has been studied in detail
\cite{Haussmann,DrZw,MRE,MPS}.

We are now in a position to derive the effective theory governing the
Anderson-Bogoliubov mode.  To this end we write the order parameter
$\Delta_{\rm mf}$ as
\begin{equation} \label{bcs:London} 
\Delta_{\rm mf}(x) = \bar{\Delta} \, {\rm e}^{2i \varphi (x)}, 
\end{equation}   
where $\bar{\Delta}$ is a spacetime-independent solution of the mean-field
equation (\ref{bcs:gap}) and $\varphi(x)$ represents the
Anderson-Bogoliubov mode, i.e., the Goldstone mode of the spontaneously
broken U(1) symmetry.  This approximation, where the phase of the order
parameter is allowed to vary in spacetime while the modulus is kept
fixed, is called the London limit.  This limit is relevant for our
discussion of the zero-temperature superconductor-to-insulator phase
transition in Ch.\ \ref{chap:qpt} because this transition is driven by
phase fluctuations; the modulus of the order parameter remains finite
and constant at the transition.  The critical behavior can thus be
studied with this effective theory formulated solely in terms of the
phase field.  We proceed by decomposing the Grassmann field as, cf.\
\cite{ESA}
\begin{equation}   \label{bcs:decompose}
\psi_\sigma(x) = {\rm e}^{i \varphi(x)} \chi_\sigma(x)
\end{equation}
and substituting the specific form (\ref{bcs:London}) of the order
parameter in the partition function (\ref{bcs:36}).  Instead of the
effective action (\ref{bcs:312}) we now obtain
\begin{equation} 
S_{\rm eff} = -i {\rm Tr} \ln \left( 
\begin{array}{cc} p_{0} - \partial_0 \varphi - \xi ({\bf p} +
\nabla \varphi) & -\bar{\Delta} \\
-\Delta^*_0 &  p_{0} + \partial_0 \varphi + \xi  ({\bf p} -
\nabla \varphi) 
\end{array} \right),
\end{equation}
where the derivative $\tilde{\partial}_\mu \varphi$ of the Goldstone field
plays the role of an Abelian gauge field.  This expression can be handled
with the help of the derivative expansion outlined in Sec.\ \ref{sec:der},
to yield the phase-only effective theory.  We shall not give any details
here and merely state the result \cite{effBCS}, that the effective theory is
again of the form (\ref{eff:Leff}).
\section{Composite Boson Limit}
\label{sec:comp}
In this section we shall investigate the strong-coupling limit of the
pairing theory.  In this limit, the attractive interaction between the
fermions is such that they form tightly bound pairs of mass $2m$.  To
explicate this limit in arbitrary dimension $d$, we swap the bare coupling
constant for a more convenient parameter---the binding energy $\epsilon_a$
of a fermion pair in vacuum \cite{RDS}.  Both parameters characterize the
strength of the contact interaction.  To see the connection between the two,
let us consider the Schr\"odinger equation for the problem at hand.  In
reduced coordinates it reads
\begin{equation} 
\left[- \frac{\nabla^2}{m} + \lambda_0 \, \delta({\bf x}) \right] \psi({\bf
x}) = - \epsilon_a,
\end{equation} 
where the reduced mass is $m/2$ and the delta-function potential, with
$\lambda_0 < 0$, represents the attractive contact interaction ${\cal
L}_{\rm i}$ in (\ref{bcs:BCS}).  We stress that this is a two-particle
problem in vacuum; it is not the famous Cooper problem of two interacting
fermions on top of a filled Fermi sea.  The equation is most easily solved
by Fourier transforming it.  This yields the bound-state equation
\begin{equation} 
\psi({\bf k}) = - \frac{\lambda_0}{{\bf k^2}/m + \epsilon_a} \psi(0),
\end{equation} 
or
\begin{equation} 
- \frac{1}{\lambda_0} = \int_{\bf k} \frac{1}{{\bf k^2}/m + \epsilon_a} .
\end{equation} 
This equation allows us to replace the coupling constant with the binding energy
$\epsilon_a$.  When substituted in the gap equation (\ref{bcs:gape}),  
the latter becomes
\begin{equation} \label{bcs:reggap}
\int_{\bf k} \frac{1}{{\bf k^2}/m + \epsilon_a} = \frac{1}{2}
\int_{\bf k} \frac{1}{E({\bf k})}.
\end{equation} 
By inspection, it is easily seen that this equation has a solution given
by \cite{Leggett}
\begin{equation} \label{comp:self}
\bar{\Delta} \rightarrow 0, \;\;\;\;\; \mu_0 \rightarrow - \tfrac{1}{2}
\epsilon_a,
\end{equation}   
where it should be noted that the chemical potential is negative here.  This
is the strong-coupling limit.  To appreciate the physical significance of
the specific value found for the chemical potential in this limit, we note
that the spectrum $E_{\rm b}({\bf q})$ of the two-fermion bound state
measured relative to the pair chemical potential $2\mu_0$ reads
\begin{equation} 
E_{\rm b}({\bf q}) = - \epsilon_a + \frac{{\bf q}^2}{4m} -2 \mu_0.
\end{equation} 
The negative value for $\mu_0$ found in (\ref{comp:self}) is precisely the
condition for a Bose-Einstein condensation of the composite bosons in the
${\bf q} = 0$ state.

To investigate this limit further, we consider the effective action
(\ref{bcs:312}) and expand $\Delta(x)$ around a constant value $\bar{\Delta}$
satisfying the gap equation (\ref{bcs:gape}),
\begin{equation} 
\Delta(x) = \bar{\Delta} + \tilde{\Delta}(x).
\end{equation} 
We obtain in this way,
\begin{equation} 
S_{\rm eff} = i \, {\rm Tr} \sum_{l =1}^\infty \frac{1}{l} \left[ G_0(p) 
\left( \begin{array}{cc} 0 & \tilde{\Delta} \\
\tilde{\Delta}^* & 0 \end{array} \right) \right]^l,
\end{equation} 
where $G_0$ is given in (\ref{bcs:prop}).  We are interested in terms
quadratic in $\tilde{\Delta}$.  Employing the derivative expansion
outlined in Sec.\ \ref{sec:der}, we find
\begin{eqnarray}  \label{comp:Seff}
S_{\rm eff}^{(2)}(q) \!\!\!\! &=& \!\!\!\! \tfrac{1}{2}i \, {\rm Tr} \,
\frac{1}{p_0^2 - E^2({\bf p})} 
\frac{1}{(p_0 + q_0)^2 - E^2({\bf p} - {\bf q})}  \\ &&
\;\;\;\; \times 
\Bigr\{ \bar{\Delta}^2 \, \tilde{\Delta}^* \tilde{\Delta}^*  
+ [p_0 + \xi({\bf p})] [p_0 + q_0 - \xi({\bf p} - {\bf q})] \tilde{\Delta}
\tilde{\Delta}^* \nonumber \\ && \;\;\;\;\;\;\;\;\; +  \bar{\Delta}^*\mbox{}^2
\tilde{\Delta} \tilde{\Delta}  
+ [p_0 - \xi({\bf p})] [p_0 + q_0 + \xi({\bf p} - {\bf q})] \tilde{\Delta}^*
\tilde{\Delta} \Bigl\}, \nonumber 
\end{eqnarray} 
where $q_\mu = i\tilde{\partial}_\mu$.  It is to be recalled here that the
derivative $p_\mu$ operates on everything to its right, while
$\tilde{\partial}_\mu$ operates only on the first object to its right.  Let
us for a moment ignore the derivatives in this expression.  After carrying
out the integral over the loop energy $k_0$ and using the gap equation
(\ref{bcs:gape}), we then obtain
\begin{equation} \label{comp:Lag1}
{\cal L}^{(2)}(0) = -\frac{1}{8} \int_{\bf k} \frac{1}{E^3({\bf k})}
\left(\bar{\Delta}^2 \, \tilde{\Delta}^*\mbox{}^2 + \bar{\Delta}^*\mbox{}^2
\tilde{\Delta}^2 + 2 |\bar{\Delta}|^2 |\tilde{\Delta}|^2 \right).
\end{equation} 
In the composite boson limit $\bar{\Delta} \rightarrow 0$, so that the
spectrum (\ref{bcs:spec}) of the elementary fermionic excitations can be
approximated by
\begin{equation} 
E({\bf k}) \approx  \epsilon({\bf k}) + \tfrac{1}{2} \epsilon_a.
\end{equation} 
The remaining integrals in (\ref{comp:Lag1}) then become elementary,
\begin{equation}
\int_{\bf k} \frac{1}{E^3({\bf k})} = \frac{4 \Gamma(3-d/2)}{(4 \pi)^{d/2}}
m^{d/2} \epsilon_a^{d/2-3}.
\end{equation} 

We next consider the terms involving derivatives in (\ref{comp:Seff}).
Following Ref.\ \cite{Haussmann} we set $\bar{\Delta}$ to zero here.  The
integral over the loop energy is easily carried out, with the result
\begin{eqnarray} 
{\cal L}^{(2)}(q) &=& - \frac{1}{2} \int_{\bf k}
\frac{1}{q_0 - {\bf k}^2/m + 
2 \mu_0 - {\bf q}^2/4m} \tilde{\Delta} \tilde{\Delta}^* \nonumber \\ && 
 - \frac{1}{2} \int_{\bf k} \frac{1}{-q_0 - {\bf k}^2/m +
2 \mu_0 - {\bf q}^2/4m} \tilde{\Delta}^* \tilde{\Delta}.
\end{eqnarray} 
The integral over the loop momentum ${\bf k}$ gives in the strong-coupling
limit using dimensional regularization
\begin{equation}   \label{comp:Lag2}
\int_{\bf k} \frac{1}{q_0 - {\bf k}^2/m -\epsilon_a - {\bf q}^2/4m} =
- \frac{\Gamma(1-d/2)}{(4 \pi)^{d/2}} m^{d/2} (-q_0 + \epsilon_a + {\bf
q}^2/4m)^{d/2-1},
\end{equation} 
or expanded in derivatives
\begin{eqnarray} 
\lefteqn{\int_{\bf k} \frac{1}{q_0 - {\bf k}^2/m - \epsilon_a - {\bf
q}^2/4m} =} \\ && 
- \frac{ \Gamma(1-d/2)}{(4 \pi)^{d/2}} m^{d/2} \epsilon_a^{d/2-1} -
\frac{ \Gamma(2-d/2)}{(4 \pi)^{d/2}} m^{d/2}
\epsilon_a^{d/2-2} \left(q_0 - \frac{{\bf q}^2}{4m} \right) + \cdots. 
\nonumber 
\end{eqnarray}  
The first term at the right-hand side yields as contribution to the
effective theory
\begin{equation} \label{bcs:con}
{\cal L}^{(2)}_\lambda = \frac{\Gamma(1-d/2)}{(4 \pi)^{d/2}} m^{d/2}
\epsilon_a^{d/2-1} |\tilde{\Delta}|^2.
\end{equation} 
To this we have to add the contribution $|\tilde{\Delta}|^2/\lambda_0$
coming from the tree potential, i.e., the last term in the partition
function (\ref{bcs:37}).  But this combination is no other than the one
needed to defined the renormalized coupling constant via (\ref{bcs:ren}),
which in the strong-coupling limit reads explicitly
\begin{equation} 
\frac{1}{\lambda} = \frac{1}{\lambda_0} + 
\frac{\Gamma(1-d/2)}{(4 \pi)^{d/2}} m^{d/2} \epsilon_a^{d/2-1}.
\end{equation} 
In other words, the contribution (\ref{bcs:con}) can be combined with the
tree contribution to yield the term $|\tilde{\Delta}|^2/\lambda$.  Expanding
the square root in (\ref{comp:Lag2}) in powers of the derivative $q_\mu$
using the value (\ref{comp:self}) for the chemical potential, and pasting
the pieces together, we obtain for the terms quadratic in $\tilde{\Delta}$
\cite{Haussmann}, 
\begin{equation} 
{\cal L}^{(2)} = \frac{1}{2} \frac{\Gamma(2-d/2)}{(4 \pi)^{d/2}} m^{d/2}
\epsilon_a^{d/2-2}\, \tilde{\Psi}^\dagger \, 
M_0(q) \, \tilde{\Psi}, \;\;\;\;\;  \tilde{\Psi} = \left(\begin{array}{l}
\tilde{\Delta} \\ \tilde{\Delta}^* \end{array} \right),
\end{equation} 
where $M_0(q)$ is the $2 \times 2$ matrix,
\begin{eqnarray}    \label{comp:M} 
\lefteqn{M_0(q) =} \\ && \!\!\!\!\!\!\!\!
\left( \begin{array}{cc}
q_0 - {\bf q}^2/4m - (2-d/2) |\bar{\Delta}|^2/ \epsilon_a & 
\!\!\!\! - (2-d/2) \bar{\Delta}^2/ \epsilon_a   \\
- (2-d/2) \bar{\Delta}^*\mbox{}^2/ \epsilon_a
& \!\!\!\! -q_0 - {\bf q}^2/4m - (2-d/2) |\bar{\Delta}|^2/ \epsilon_a
\end{array} \right). \nonumber 
\end{eqnarray} 
This Lagrangian is precisely of the form found in (\ref{eff:L0})
describing an interacting Bose gas.  On comparing with Eq.\ (\ref{eff:M}
), we conclude that the composite bosons have---as expected---a mass
$m_{\rm b}=2m$ twice the fermion mass $m$, and a small chemical
potential
\begin{equation} 
\mu_{0,{\rm b}} = (2-d/2) \frac{|\bar{\Delta}|^2}{\epsilon_a}.
\end{equation} 
From (\ref{comp:M}) one easily extracts the Bogoliubov spectrum and the
velocity $c_0$ of the sound mode it describes,
\begin{equation} 
c_0^2 = \frac{\mu_{0,{\rm b}}}{m_{\rm b}} = (1-d/4)
\frac{|\bar{\Delta}|^2}{m \epsilon_a}. 
\end{equation} 
Also the number density $\bar{n}_{0,{\rm b}}$ of condensed composite bosons, 
\begin{equation} 
\bar{n}_{0,{\rm b}} = \frac{\Gamma(2-d/2)}{(4 \pi)^{d/2}} m^{d/2}
\epsilon_a^{d/2-2} |\bar{\Delta}|^2
\end{equation} 
as well as the weak repulsive interaction $\lambda_{0,{\rm b}}$ between the
composite bosons,
\begin{equation} \label{comp:lambda}
\lambda_{0,{\rm b}} = (4 \pi)^{d/2} \frac{1-d/4}{\Gamma(2-d/2)}
\frac{\epsilon_a^{1-d/2}}{m^{d/2}}
\end{equation}   
follow immediately.  We in this way have explicitly demonstrated that
the BCS theory in the composite boson limit maps onto the Bogoliubov
theory.

In concluding this section, we remark that in $d=2$ various integrals we
encountered become elementary for arbitrary values of $\bar{\Delta}$.  For
example, the gap equation (\ref{bcs:reggap}) reads explicitly in $d=2$
\begin{equation} 
\epsilon_a = \sqrt{\mu_0^2 + |\bar{\Delta}|^2} - \mu_0,
\end{equation} 
while the particle number equation (\ref{bcs:ne}) becomes
\begin{equation} 
\bar{n} = \frac{m}{2 \pi} \left(\sqrt{\mu_0^2 + |\bar{\Delta}|^2} + \mu_0 \right).
\end{equation} 
Since in two dimensions, 
\begin{equation} 
\bar{n} = \frac{k_{\rm F}^2}{2 \pi} = \frac{m}{\pi} \epsilon_{\rm F},
\end{equation} 
with $k_{\rm F}$ and $\epsilon_{\rm F} = k_{\rm F}^2/2m$ the Fermi momentum
and energy, the two equations can be combined to yield \cite{RDS}
\begin{equation}
\frac{\epsilon_a}{\epsilon_{\rm F}} = 2 \frac{\sqrt{\mu_0^2 +
|\bar{\Delta}|^2} - \mu_0}{\sqrt{\mu_0^2 + |\bar{\Delta}|^2} + \mu_0}.
\end{equation}  
The composite boson limit we have been discussing in this section is
easily retrieved from these more general equations.  Also note that in
this limit, $\bar{n} = 2 \bar{n}_{0,{\rm b}}$, while the renormalization
of the coupling constant takes the same form as for an interacting Bose
gas
\begin{equation} 
\frac{1}{\lambda_0} = \frac{1}{\kappa^\epsilon} \left( \frac{1}{\hat{\lambda}} -
\frac{m}{4 \pi \epsilon} \right),
\end{equation} 
cf.\ (\ref{eff:lambdar}).
\section{Dual Theory}
\label{sec:2sc}
We now turn to the dual description of a superconducting film at finite
temperature.  We thereto minimally couple the model of Sec.\
\ref{sec:KT} to a magnetic field described by the magnetic vector
potential ${\bf A}$.  For the time being we ignore vortices by setting
the vortex gauge field $\bbox{\varphi}^{\rm P}$ to zero.  The partition
function of the system then reads
\begin{equation}  \label{2sc:znovor}  
Z = \int \DD\varphi \int \DD {\bf A} \, \Xi ({\bf A}) 
\, \exp\left( -\beta \int_{\bf x}{\cal H} \right),
\end{equation}
where $\Xi({\bf A})$ is a gauge-fixing factor for the gauge field ${\bf A}$,
and ${\cal H}$ is the Hamiltonian
\begin{equation} \label{2sc:H}
{\cal H} = \tfrac{1}{2} \rho_{\rm s} {\bf v}_{\rm s}^2 + \tfrac{1}{2}
(\nabla \times {\bf A})^2 
\end{equation} 
with
\begin{equation} \label{2sc:vs}
{\bf v}_{\rm s} = \frac{1}{m} (\nabla \varphi - 2e {\bf A}).   
\end{equation}
The double charge $2e$ stands for the charge of the Cooper pairs which are
formed at the bulk transition temperature.  The functional integral over
$\varphi$ in (\ref{2sc:znovor}) is easily carried out with the result
\begin{equation}    \label{2sc:schwlike} 
Z = \int \DD {\bf A} \, \Xi({\bf A}) \, {\rm exp}\left\{-\frac{\beta}{2}
\int_{\bf x} \left[(\nabla \times {\bf A})^2 + m_A^2 A_i \left( \delta_{i j}
- \frac{\partial _i
\partial_j}{\nabla^2} \right) A_j \right] \right\}, 
\end{equation} 
where the last term, with $m^2_A = 4 e^2 \rho_{\rm s}/m^2$, is a
gauge-invariant, albeit nonlocal mass term for the gauge field
generated by the Higgs mechanism.  The number of degrees of freedom does
not change in the process.  This can be seen by noting that a gapless
gauge field in two dimensions represents no physical degrees of freedom.
(In Minkowski spacetime, this is easily understood by recognizing that
in $1+1$ dimensions there is no transverse direction.)  Before the Higgs
mechanism took place, the system therefore contains only a single
physical degree of freedom described by $\varphi$.  This equals the
number of degrees of freedom contained in (\ref{2sc:schwlike}).

We next introduce an auxiliary field $\tilde{h}$ to linearize the first term
in (\ref{2sc:schwlike}),
\begin{equation}     \label{2sc:efield}
\exp \left[-\frac{\beta}{2} \int_{\bf x}  (\nabla \times {\bf A})^2
\right] = \int \DD \tilde{h} \, {\rm exp}\left[-\frac{1}{2 \beta} \int_{\bf x}
\tilde{h}^2 + i \int_{\bf x} \tilde{h} (\nabla
\times {\bf A}) \right],
\end{equation} 
and integrate out the gauge-field fluctuations [with a gauge-fixing term
$(1/2\alpha)(\nabla \cdot {\bf A})^2$].  The result is a manifestly
gauge-invariant expression for the partition function in terms of a massive
scalar field $\tilde{h}$, representing the single degree of freedom
contained in the theory:
\begin{equation}   \label{2sc:massivescalar}
Z = \int \DD \tilde{h} \, {\rm exp}\left\{-\frac{1}{2 \beta} \int_{\bf x}
\left[ \frac{1}{m_A^2} (\nabla \tilde{h})^2 + \tilde{h}^2 \right] \right\}.
\end{equation} 
To understand the physical significance of this field, we note from
(\ref{2sc:efield}) that it satisfies the field equation
\begin{equation}  \label{2sc:id} 
\tilde{h} = i \beta \nabla \times {\bf A}.
\end{equation} 
That is, the fluctuating field $\tilde{h}$ represents the local magnetic
induction, which is a scalar in two space dimensions.  Equation
(\ref{2sc:massivescalar}) shows that the magnetic field has a finite
penetration depth $\lambda_{\rm L} = 1/m_A$.  In contrast to the original
description where the functional integral runs over the gauge potential, the
integration variable in (\ref{2sc:massivescalar}) is the physical field.

We next include vortices.  The penetration depth $\lambda_{\rm L}$ provides
the system with an infrared cutoff so that a single magnetic vortex in the
charged theory has a finite energy.  Vortices can therefore be thermally
activated.  This is different from the superfluid phase of the neutral
model, where the absence of an infrared cutoff permits only tightly bound
vortex-antivortex pairs to exist.  We expect, accordingly, the
superconducting phase to describe a plasma of vortices, each carrying one
magnetic flux quantum $\pm \pi/e$.  The partition function now reads
\begin{equation}     \label{2sc:vincluded}
Z = \sum_{N_{+},N_{-}=0}^{\infty} \frac{z^{N_{+}+N_{-}}}{N_{+}!\, N_{-}!}
\prod_{\alpha} \int_{{\bf x}^\alpha} \, \int \DD \varphi
\int \DD {\bf A} \,  \Xi({\bf A}) \, \exp \left(- \beta \int_{\bf
x} {\cal H} \right)
\end{equation} 
where $z$ is the fugacity, i.e., the Boltzmann factor associated with the
vortex core energy.  The velocity appearing in the Hamiltonian (\ref{2sc:H})
now includes the vortex gauge field
\begin{equation} 
{\bf v}_{\rm s} = \frac{1}{m} (\nabla \varphi - 2e {\bf A} -
\bbox{\varphi}^{\rm P}).    
\end{equation}
This field can be shifted from the first to the second term in the
Hamiltonian (\ref{2sc:H}) by applying the transformation ${\bf A}
\rightarrow {\bf A} - \bbox{\varphi}^{\rm P}/2e$.  This results in the shift
\begin{equation} 
\nabla \times {\bf A} \rightarrow \nabla \times {\bf A} - B^{\rm P},
\end{equation} 
with the plastic field 
\begin{equation}  \label{2sc:BP}
B^{\rm P} = -\Phi_0 \sum_{\alpha} w_{\alpha} \, \delta({\bf x} - {\bf
x}^{\alpha })
\end{equation} 
representing the magnetic flux density.  Here, $\Phi_0 = \pi/e$ is the
elementary flux quantum.  Repeating the steps of the previous paragraph
we now obtain instead of (\ref{2sc:massivescalar})
\begin{equation}   \label{2sc:vortexsum}
Z = \sum_{N_\pm=0}^\infty \frac{z^{N_{+}+N_{-}}}{N_{+}!\, N_{-}! }
\prod_{\alpha} \int_{{\bf x}^\alpha} \int \DD \tilde{h} \, {\rm
exp}\left\{-\frac{1}{2\beta} \int_{\bf x} \left[\frac{1}{m_A^2} (\nabla
\tilde{h})^2 + \tilde{h}^2 \right] + i \int_{\bf x} B^{\rm P}
\tilde{h}\right\}, 
\end{equation} 
where $\tilde{h}$ represents the physical local magnetic induction $h$
\begin{equation} 
\tilde{h} = i \beta (\nabla \times {\bf A} - B^{\rm P}) = i \beta h.
\end{equation} 
The field equation for $\tilde{h}$ obtained from (\ref{2sc:vortexsum})
yields for the magnetic induction:
\begin{equation} \label{2sc:fam}
- \nabla^2 h + m_A^2 h = m_A^2 B^{\rm P},
\end{equation} 
which is the familiar equation in the presence of magnetic vortices.

The last term in (\ref{2sc:vortexsum}) shows that the charge $g$ with which
a magnetic vortex couples to the fluctuating $\tilde{h}$-field is the
product of an elementary flux quantum (contained in the definition of
$B^{\rm P}$) and the inverse penetration depth $m_A = 1/\lambda_{\rm L}$,
\begin{equation} \label{2sc:g}
g = \Phi_0 m_A.
\end{equation} 
For small fugacities the summation indices $N_{+}$ and $N_{-}$ can be
restricted to the values $0,1$ and we arrive at the partition function
of the massive sine-Gordon model \cite{Schaposnik}
\begin{equation}   \label{2sc:sineGordon}
Z = \int \DD \tilde{h} \, {\rm exp} \left( - \int_{\bf x}
\left\{\frac{1}{2 \beta} \left[\frac{1}{m_A^2} (\nabla \tilde{h})^2 + 
\tilde{h}^2\right]- 2z \cos \left( \Phi_0 \tilde{h} \right) \right\}
\right).
\end{equation} 
This is the dual formulation of a two-dimensional superconductor.  The
magnetic vortices of unit winding number $w_\alpha = \pm 1$ turned the
otherwise free theory (\ref{2sc:massivescalar}) into an interacting one.

The final form (\ref{2sc:sineGordon}) demonstrates the rationales for going
over to a dual theory.  First, it is a formulation directly in terms of a
physical field representing the local magnetic induction.  There is no
redundancy in this description and therefore no gauge invariance.  Second,
the magnetic vortices are accounted for in a nonsingular fashion.  This is
different from the original formulation of the two-dimensional
superconductor where the local magnetic induction is the curl of an
unphysical gauge potential ${\bf A}$, and where the magnetic vortices appear
as singular objects.

Up to this point we have discussed a genuine two-dimensional superconductor.
As a model to describe superconducting films this is, however, not adequate.
The reason is that the magnetic interaction between the vortices takes place
mostly not through the film but through free space surrounding the film
where the photon is gapless.  This situation is markedly different from a
superfluid film.  The interaction between the vortices there is mediated by
the Kosterlitz-Thouless mode which is confined to the film.  A genuine
two-dimensional theory therefore gives a satisfactory description of a
superfluid film.

To account for the fact that the magnetic induction is not confined to the
film and can roam in outer space, the field equation (\ref{2sc:fam}) is
modified in the following way \cite{Pearl,deGennes}
\begin{equation}  \label{2sc:mod}
- \nabla^2 h({\bf x}_\perp,x_3) + \frac{1}{\lambda_\perp} \delta_d(x_3) h({\bf
x}_\perp,x_3) = \frac{1}{\lambda_\perp}  \delta_d(x_3) B^{\rm P}({\bf x}).
\end{equation}  
Here, $1/\lambda_\perp = d m_A^2 $, with $d$ denoting the thickness of the
superconducting film, is an inverse length scale, ${\bf x}_\perp$ denotes
the coordinates in the plane, $h$ the component of the induction field
perpendicular to the film, and $\delta_d(x_3)$ is a smeared delta function
of thickness $d$ along the $x_3$-axis
\begin{equation}  
\delta_d(x_3) \left\{ \begin{array}{cc} = 0 & {\rm for} \;\;\;\; |x_3| > d/2
\\ \neq 0 &  {\rm for} \;\;\;\; |x_3| \leq d/2 \end{array} \right. .
\end{equation} 
The reason for including the smeared delta function at the right-hand side
of (\ref{2sc:mod}) is that the vortices are confined to the film.  The delta
function in the second term at the left-hand side is included because this
term is generated by screening currents which are also confined to the film.

To be definite, we consider a single magnetic vortex located at the origin.
The induction field found from (\ref{2sc:mod}) reads
\begin{equation} 
h({\bf x}_\perp,0) = \frac{\Phi_0}{2 \pi} \int_0^\infty \dd q \frac{q}{1+ 2
\lambda_\perp q} J_0(q |{\bf x}_\perp|),
\end{equation} 
with $J_0$ the 0th Bessel function of the first kind.  At small distances
from the vortex core ($\lambda_\perp q >> 1$)
\begin{equation} \label{2sc:vincin}
h({\bf x}_\perp,0) \sim \frac{\Phi_0}{4 \pi \lambda_\perp |{\bf x}_\perp|},
\end{equation} 
while far away  ($\lambda_\perp q << 1$)
\begin{equation} 
h({\bf x}_\perp,0) \sim \frac{\Phi_0 \lambda_\perp}{\pi |{\bf x}_\perp|^3}.
\end{equation} 
This last equation shows that the field does not exponentially decay as
would be the case in a genuine two-dimensional system.  The reason for
the long range is that most of the magnetic interaction takes place in
free space outside the film where the photon is gapless.  If, as is
often the case, the length $\lambda_\perp =1/d m_A^2$ is much larger
than the sample size, it can be effectively set to infinity.  In this
limit, the effect of the magnetic interaction diminishes, as can be seen
from (\ref{2sc:vincin}), and the vortices behave as in a superfluid
film.  One therefore expects a superconducting film to also undergo a
Kosterlitz-Thouless transition at a temperature $T_{\rm KT}$
characterized by an unbinding of vortex-antivortex pairs.  The first
experiment to study this possibility was carried out in Ref.\
\cite{BMO}.  Because the transition temperature $T_{\rm KT}$ is well
below the bulk temperature $T_{\rm c}$ where the Cooper pairs form, the
energy gap of the fermions remains finite at the critical point
\cite{CFGWY}.  This prediction has been corroborated by experiments
performed by Hebard and Paalanen on superconducting films \cite{HPsu1}.
For temperatures $T_{\rm KT} \leq T \leq T_{\rm c}$, there is a plasma
of magnetic vortices which disorder the superconducting state.  At
$T_{\rm KT}$ vortices and antivortices bind into pairs and
algebraic long-range order sets in.

%% file: FQHE.tex
\chapter{Fractional Quantized Hall Effect \label{chap:fqhe}}
The nonrelativistic $|\phi|^4$-theory describing an interacting Bose gas
is also of importance for the description of the fractional quantized
Hall effect (FQHE).  As a function of the applied magnetic field, this
two-dimensional system undergoes a zero-temperature transition between a
so-called quantum Hall liquid, where the Hall conductance is quantized
in odd fractions of $e^2/2 \pi$, or, reinstalling Planck's constant,
$e^2/h$, and an insulating phase.  Here, the nonrelativistic
$|\phi|^4$-theory describes---after coupling to a Chern-Simons
term---the original electrons bound to an odd number of flux quanta.
The Hall liquid is in this picture characterized by a condensate of
composite particles.
\section{Chern-Simons-Ginzburg-Landau Theory}
\label{sec:CSGL}
The fractional quantized Hall effect (FQHE) is the hallmark of a new,
intrinsically two-dimensional condensed-matter state---the quantum Hall
liquid.  Many aspects of this state are well understood in the framework
of the quantum-mechanical picture developed by Laughlin
\cite{Laughlin}.  Considerable effort has nevertheless been invested in
formulating an effective field theory which captures the essential
low-energy, small-momentum features of the liquid.  A similar approach
in the context of superconductors has proven most successful.
Initially, only the phenomenological model proposed by Ginzburg and
Landau \cite{GL} in 1950 was known here.  Most of the fundamental
properties of the superconducting state such as superconductivity---the
property that gave this condensed-matter state its name, Meissner
effect, magnetic flux quantization, Abrikosov flux lattice, and
Josephson effect, can be explained by the model.  The microscopic theory
was given almost a decade later by Bardeen, Cooper, and Schrieffer
\cite{BCS}.  Shortly here after, Gorkov \cite{Gorkov} made the
connection between the two approaches by deriving the Ginzburg-Landau
model from the microscopic BCS theory, thus giving the phenomenological
model the status of an effective field theory.

A first step towards an effective field theory of the quantum Hall liquid
was taken by Girvin and MacDonald \cite{GMac} and has been developed further
by Zhang, Hansson and Kivelson \cite{ZHK}, who also gave an explicit
construction starting from a microscopic Hamiltonian.  Their formulation
(for a review see Ref.\ \cite{Zhang}) incorporates time dependence which is
important for the study of quantum phase transitions.  

An essential ingredient for obtaining an effective theory of the FQHE
was the identification by Girvin and MacDonald \cite{GMac} of a bosonic
operator $\phi$ exhibiting (algebraic) off-diagonal long-range order of a
type known to exist in two-dimensional bosonic superfluids.  They argued
that this field should be viewed as an order parameter in terms of which
the effective field theory should be formulated.  To account for the
incompressibility of the quantum Hall liquid they suggested to minimally
couple $\phi$ to a so-called statistical gauge field $(a_0, {\bf a})$
governed solely by a Chern-Simons term
\begin{equation}
\label{CSGL:CS}
{\cal L}_{\rm CS} = \tfrac{1}{2} e^2 \theta \partial_0 {\bf a} \times {\bf
a} - e^2 \theta a_0 \nabla \times {\bf a}, 
\end{equation}
with $\nabla \times {\bf a}$ the statistical magnetic field and $\theta$
a constant.  As we will see below, the gapless Bogoliubov spectrum of
the neutral system changes as a result of this coupling into one with an
energy gap \cite{ZHK}, thus rendering the charged system incompressible.

Because of the absence of a kinetic term (the usual Maxwell term), the
statistical gauge field does not represent a physical degree of freedom.  In
a relativistic setting, a Maxwell term is usually generated by quantum
corrections so that the statistical gauge field becomes dynamical at the
quantum level.  The quantum theory then differs qualitatively from the
classical theory.  On the other hand, as we shall see below, this need not
be the case in a nonrelativistic setting.  That is to say, the {\it Ansatz}
of the absence of a Maxwell term is here not necessarily obstructed by
quantum corrections.

The effective theory of the quantum Hall liquid is given by the
so-called Chern-Simons-Ginzburg-Landau (CSGL) Lagrangian \cite{ZHK}
\begin{equation}
\label{CSGL:L}
{\cal L} = i \phi^* D_0 \phi -
\frac{1}{2m} |{\bf D} \phi|^2 + \mu_0 |\phi|^2 - \lambda_0 |\phi|^4 +
{\cal L}_{\rm CS}.
\end{equation}
The covariant derivatives $D_0 = \partial_0 + i e A_0 + i e a_0$ and
${\bf D} = \nabla - i e {\bf A} - i e {\bf a}$ give a minimal coupling
to the applied magnetic and electric field described by the gauge field
$(A_0,{\bf A})$ and also to the statistical gauge field.  For
definiteness we will assume that our two-dimensional sample is
perpendicular to the applied magnetic field, defining the $z$-direction,
and we choose the electric field to point in the $x$-direction.  The
charged field $\phi$ represents the Girvin-MacDonald order parameter
describing the original electrons bound to an odd number $2l+1$ of flux
quanta.  To see that it indeed does, let us consider the field equation
for $a_0$:
\begin{equation} \label{CSGL:a0}
|\phi|^2 = - e \theta  \nabla \times {\bf a}.
\end{equation} 
The simplest solution of the CSGL Lagrangian is the uniform mean-field
solution
\begin{equation} 
|\phi|^2 = \bar{n}, \;\;\;\; {\bf a} = - {\bf A}, \;\;\;\; a_0 = - A_0 =
0,
\end{equation} 
where $\bar{n}$ indicates the constant fermion number density.  The
statistical gauge field is seen to precisely cancel the applied field.
The constraint equation (\ref{CSGL:a0}) then becomes
\begin{equation}  \label{CSGL:n}
\bar{n} = e \theta H,
\end{equation} 
with $H$ the applied magnetic field.  Now, if we choose $\theta^{-1} = 2
\pi (2l+1)$, it follows on integrating this equation that, as required,
with every electron there is associated $2l+1$ flux quanta:
\begin{equation} 
N = \frac{1}{2l+1} N_\otimes,
\end{equation} 
where $N_\otimes = \Phi/\Phi_0$, with $\Phi = \int_{\bf x} H$ the
magnetic flux, indicates the number of flux quanta.  Equation
(\ref{CSGL:n}) implies an odd-denominator filling factor $\nu_H$ which
is defined by
\begin{equation} 
\nu_H = \frac{\bar{n}}{H/\Phi_0}= \frac{1}{2l+1}.
\end{equation} 

The coupling constant $\lambda_0 \, (>0)$ in (\ref{CSGL:L}) is the
strength of the repulsive contact interaction between the composite
particles, and $\mu_0$ is a chemical potential introduced to account for
a finite number density of composite particles.  

It is well known from anyon physics that the inclusion of the
Chern-Simons term changes the statistics of the field $\phi$ to which
the statistical gauge field is coupled \cite{Wilczek}.  If one composite
particle circles another, it picks up an additional Aharonov-Bohm
factor, representing the change in statistics.  The binding of an odd
number of flux quanta changes the fermionic character of the electrons
into a bosonic one for the composite particles, allowing them to Bose
condense.  The algebraic off-diagonal long-range order of a quantum Hall
liquid can in this picture be understood as resulting from this
condensation.  Conversely, a flux quantum carries $1/(2l+1)$th of
an electron's charge \cite{Laughlin}, and also $1/(2l+1)$th of an
electron's statistics \cite{ASW}.

The defining phenomenological properties of a quantum Hall liquid are
easily shown to be described by the CSGL theory \cite{ZHK,Zhang}.  From
the lowest-order expression for the induced electromagnetic current one
finds
\begin{equation}
\label{CSGL:inducedji}
e j_i = \frac{\delta {\cal L}}{\delta A_i} = - \frac{\delta {\cal
L}_\phi}{\delta a_i} = \frac{\delta {\cal L}_{\rm CS}}{\delta a_i} = - e^2
\theta \epsilon_{ij} (\partial_0 a_j - \partial_j a_0) = e^2 \theta 
\epsilon_{ij} E_j,
\end{equation}
with ${\bf E}$ the applied electric field and where we have written the
Lagrangian (\ref{CSGL:L}) as a sum ${\cal L} = {\cal L}_\phi + {\cal L}_{\rm
CS}$.  It follows that the Hall conductance $\sigma_{xy}$ is quantized in
odd fractions of $e^2/2 \pi$, or, reinstalling Planck's constant, $e^2/h$.
This result can also be understand in an intuitive way as follows.  Since
the composite particles carry a charge $e$, the applied electric field gives
rise to an electric current
\begin{equation} 
I = e \frac{\dd N}{\dd t}
\end{equation} 
in the direction of ${\bf E}$, i.e., the $x$-direction.  This is not the
end of the story because the composite objects carry in addition to
electric charge also $2l+1$ flux quanta.  When the Goldstone field
$\varphi$ encircles $2l+1$ flux quanta, it picks up a factor $2 \pi$ for
each of them
\begin{equation} 
\oint_\Gamma \nabla \cdot \varphi = 2 \pi (2l+1).
\end{equation} 
Now, consider two points across the sample from each other.  Let the
phase of these points initially be equal.  As a composite particle moves
downstream, and crosses the line connecting the two points, the relative
phase $\Delta \varphi$ between them changes by $2 \pi (2l+1)$.
This phase slippage \cite{PWA} leads to a voltage drop across the
sample given by 
\begin{equation} 
V_{\rm H} =  \frac{1}{e} \partial_0 \Delta \varphi = (2l+1) \Phi_0
\frac{\dd N}{\dd t}, 
\end{equation} 
where the first equation can be understood by recalling that due to minimal
coupling $\partial_0 \varphi \rightarrow \partial_0 \varphi + e A_0$.  For
the Hall resistance we thus obtain the expected value
\begin{equation} 
\rho_{xy} = \frac{V_{\rm H}}{I} = (2l+1) \frac{2 \pi}{e^2}.
\end{equation} 

If the CSGL theory is to describe an incompressible liquid, the spectrum
of the single-particle excitations must have a gap.  Without the
coupling to the statistical gauge field, the spectrum is given by the
gapless Bogoliubov spectrum (\ref{eff:bogo}).  To obtain the
single-particle spectrum of the coupled theory, we integrate out the
statistical gauge field.  The integration over $a_0$ was shown to yield
the constraint (\ref{CSGL:a0}) which in the Coulomb gauge $\nabla \cdot
{\bf a} = 0$ is solved by
\begin{equation}
\label{CSGL:solution}
a_i = \frac{1}{e \theta} \epsilon_{ij} \frac{\partial_j}{\nabla^2} |\phi|^2.
\end{equation}
The integration over the remaining components of the statistical gauge field
is now simply performed by substituting (\ref{CSGL:solution}) back into the
Lagrangian.  The only nonzero contribution arises from the term $- e^2
|\phi|^2 {\bf a}^2/2m$.  The spectrum of the charged system acquires as a
result an energy gap $\omega_{\rm c}$
\begin{equation}
E({\bf k}) = \sqrt{\omega_{\rm c}^2 + \epsilon^2({\bf k}) + 2 \mu_0
\epsilon( {\bf k}) },
\end{equation}
with $\omega_{\rm c} = \mu_0 /2\theta m\lambda_0$.  To lowest order, the gap
equals the cyclotron frequency of a free charge $e$ in a magnetic field $H$
\begin{equation}
\omega_{\rm c} = \frac{\bar{n}}{\theta m} = \frac{e H}{m}.
\end{equation}
The presence of this energy gap results in dissipationless flow with
$\sigma_{xx} =0$.

These facts show that the CSGL theory captures the essentials of a
quantum Hall liquid.  Given this success, it is tempting to investigate
if the theory can also be employed to describe the field-induced
Hall-liquid-to-insulator transitions.  This will be done in
Sec. \ref{sec:QHL}.  It should however be borne in mind that both the
$1/|{\bf x}|$-Coulomb potential as well as impurities should be
incorporated into the theory in order to obtain a realistic description
of the FQHE.  The repulsive Coulomb potential is believed to play a
decisive role in the formation of the the composite particles, while the
impurities are responsible for the width of the Hall plateaus.  As the
magnetic field moves away from the magic filling factor, magnetic
vortices will materialize in the system to make up the difference
between the applied field and the magic field value.  In the presence of
impurities, these defects get pinned and do not contribute to the
resistivities, so that both $\sigma_{xx}$ and $\sigma_{xy}$ are
unchanged.  Only if the difference becomes too large, the system reverts
to an other quantum Hall state with a different filling factor.

%% file: QPT.tex
\chapter{Quantum Phase Transitions \label{chap:qpt}}
This chapter is devoted to continuous phase transitions at the absolute zero
of temperature; so-called quantum phase transitions.  Unlike in classical
phase transitions taking place at finite temperature and in equilibrium,
time plays an important role in quantum phase transitions.  Put differently,
whereas the critical behavior of classical 2nd-order phase transitions is
governed by thermal fluctuations, that of 2nd-order quantum transitions is
controlled by quantum fluctuations.  These transitions, which have attracted
much attention in recent years (for an introductory review, see Ref.\
\cite{SGCS}), are triggered by varying not the temperature, but some
other parameter in the system, like the applied magnetic field, the
charge carrier density, or the disorder strength.  The quantum phase
transitions we will be discussing here are all dominated by phase
fluctuations.
\section{Scaling}
The natural language to describe quantum phase transitions
is quantum field theory.  In addition to a diverging correlation length
$\xi$, quantum phase transitions also have a diverging correlation time
$\xi_t$.  They indicate, respectively, the distance and time period over
which the order parameter characterizing the transition fluctuates
coherently.  The way the diverging correlation time relates to the
diverging correlation length, 
\begin{equation} \label{zcrit}
\xi_t \sim \xi^z, 
\end{equation} 
defines the so-called dynamic exponent $z$.  It is a measure for the
asymmetry between the time and space directions and tells us how long it
takes for information to propagate across a distance $\xi$.  The traditional
scaling theory of classical 2nd-order phase transitions, first put forward
by Widom \cite{Widom}, is easily extended to include the time dimension
\cite{Ma} because relation (\ref{zcrit}) implies the presence of only one
independent diverging scale.  Let $\delta = K - K_{\rm c}$, with $K$ the
parameter that drives the phase transition, measure the distance from
the critical coupling $K_{\rm c}$.  A physical observable at the
absolute zero of temperature $O(k_0,|{\bf k}|,K)$ can in the critical
region close to the transition be written as
\begin{equation} \label{scaling0}
O(k_0,|{\bf k}|,K) = \xi^{d_O} {\cal O}(k_0 \xi_t, |{\bf k}| \xi),
\;\;\;\;\;\;\;\; (T=0),
\end{equation} 
where $d_O$ is the dimension of the observable $O$.  The right-hand side
does not depend explicitly on $K$; only implicitly through $\xi$ and
$\xi_t$.  The closer one approaches the critical coupling $K_{\rm c}$, the
larger the correlation length and time become.

Since a physical system is always at some finite temperature, we have to
investigate how the scaling law (\ref{scaling0}) changes when the
temperature becomes nonzero.  The easiest way to include temperature in
a quantum field theory is to go over to imaginary time $\tau = it$, with
$\tau$ restricted to the interval $0 \leq \tau \leq \beta$.  The
temporal dimension becomes thus of finite extend.  The critical behavior
of a phase transition at finite temperature is still controlled by the
quantum critical point provided $\xi_t < \beta$.  If this condition is
fulfilled, the system does not see the finite extend of the time
dimension.  This is what makes quantum phase transitions experimentally
accessible.  Instead of the zero-temperature scaling (\ref{scaling0}),
we now have the finite-size scaling
\begin{equation} \label{scalingT}
O(k_0,|{\bf k}|,K,T) = \beta^{d_O/z} {\cal O}(k_0 \beta, |{\bf k}|
\beta^{1/z},\beta/\xi_t), \;\;\;\;\;\;\;\; (T \neq 0).
\end{equation} 
The distance to the quantum critical point is measured by the ratio
$\beta/\xi_t \sim |\delta|^{z\nu}/T$.
\section{Repulsively Interacting Bosons}
\label{sec:BT}
The first quantum phase transition we wish to investigate is the
superfluid-to-Mott-insulating transition of interacting bosons in the
absence of impurities \cite{FWGF}.  The transition is described by the
nonrelativistic $|\phi|^4$-theory (\ref{eff:Lagr}), which becomes critical
at the absolute zero of temperature at some (positive) value $\mu_{\rm c}$
of the renormalized chemical potential.  The Mott insulating phase is
destroyed and makes place for the superfluid phase as $\mu$ increases.
Whereas in the superfluid phase the single-particle (Bogoliubov) spectrum is
gapless and the system compressible, the single-particle spectrum of the
insulating phase has an energy gap and the compressibility $\kappa$
vanishes here.

\begin{figure}
\begin{center}
\epsfxsize=8.cm
\mbox{\epsfbox{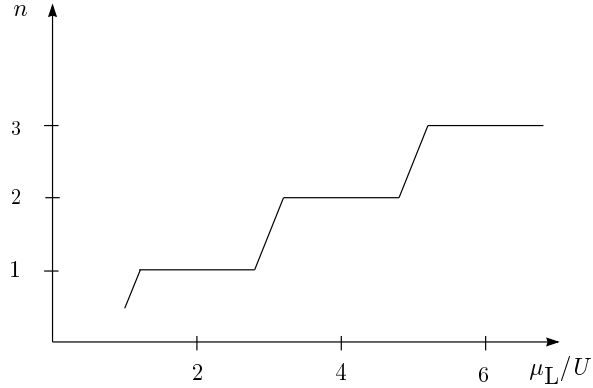}}
\end{center}
\caption{Schematic representation of the average number $n$ of particles per
site as function of the chemical potential $\mu_{\rm L}$ at some finite
value of the hopping parameter $t < t_{\rm c}$.  \label{fig:occu}}
\end{figure}

The nature of the insulating phase can best be understood by putting the
theory on a lattice.  The lattice model is defined by the Hamiltonian
\begin{equation}    \label{BT:hu}
H_{\rm H} = - t  \sum_j (a^{\dagger}_j a_{j+1} + {\rm 
h.c.}) + \sum_j (- \mu_{\rm L} \hat{n}_j + U  \hat{n}_j^2),
\end{equation} 
where the sum $\sum_j$ is over all lattice sites.  The operator
$a^{\dagger}_j$ creates a boson at site $j$ and $\hat{n}_j = a^{\dagger}_j
a_j$ is the particle number operator at that site; $t$ is the hopping
parameter, $U$ the interparticle repulsion, and $\mu_{\rm L}$ is the
chemical potential on the lattice.  The zero-temperature phase
diagram is as follows \cite{FWGF}.  In the limit $t/U \rightarrow 0$, each
site is occupied by an integer number $n$ of bosons which minimizes the
on-site energy (see Fig.\ \ref{fig:occu})
\begin{equation} 
\epsilon(n) = -\mu_{\rm L} n + U n^2.
\end{equation} 
It follows that within the interval $2n-1 < \mu_{\rm L}/U < 2n+1$, each
site is occupied by exactly $n$ bosons.  When the chemical potential is
negative, $n=0$.  The intervals become smaller when $t/U$ increases.
Within such an interval, where the particles are pinned to the lattice
sites, the single-particle spectrum has an energy gap, and the system is
in the insulating phase with zero compressibility, $\kappa =
n^{-2}\partial n/\partial \mu_{\rm L} =0$.  Outside these intervals, the
particles delocalize and can hop through the lattice.  Being at zero
temperature, the delocalized bosons condense in a superfluid state.  The
single-particle spectrum is gapless here and the system compressible
($\kappa \neq 0$).

\begin{figure}
\epsfxsize=10.cm
\mbox{\epsfbox{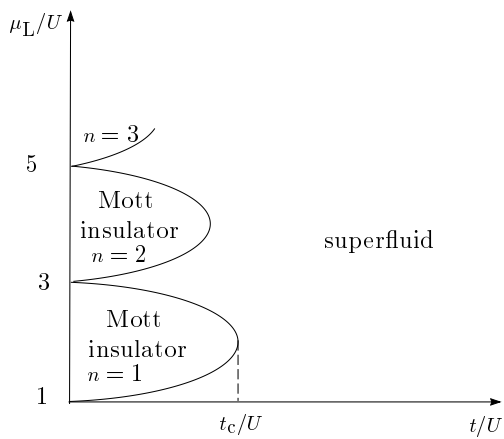}}
\caption{Schematic representation of the  phase diagram  of the lattice
model (\protect\ref{BT:hu}) at the absolute zero of temperature
\protect\cite{FWGF}.
\label{fig:qphase}}
\end{figure}

As $t/U$ increases, the gap in the single-particle spectrum as well as
the width of the intervals decrease and eventually vanish at some
critical value $t_{\rm c}$.  For values $t>t_{\rm c}$ of the hopping
parameter, the superfluid phase is the only phase present (see Fig.\
\ref{fig:qphase}).
The continuum model (\ref{eff:Lagr}), with $\mu > \mu_{\rm c}$ describes the
condensed delocalized lattice bosons which are present when the density
deviates from integer values (see Fig.\ \ref{fig:occu}).  In the limit $\mu
\rightarrow \mu_{\rm c}$ from above, the number of delocalized bosons
decreases and eventually becomes zero at the phase boundary $\mu=\mu_{\rm
c}$ between the superfluid and insulating phases.

Various quantum phase transitions belong to the universality class
defined by the zero-density transition of repulsively interacting
bosons.  For example, itinerant quantum antiferromagnets
\cite{Hertz,Ian,KB} as well as lower-dimensional (clean) superconductors
belong to this universality class.  As we have seen in Sec.\
\ref{sec:comp}, Cooper pairs become tightly bound composite particles in
the strong-coupling limit, which are described by the nonrelativistic
$|\phi|^4$-theory with a weak repulsive interaction.  For $\mu >
\mu_{\rm c}$, the field $\phi$ now describes the condensed delocalized
Cooper pairs.  When the chemical potential decreases, the condensate
diminishes, and the system again becomes insulating for $\mu < \mu_{\rm
c}$ \cite{CFGWY}.  By continuity, we expect also the
superconductor-to-insulator transition of a (clean) weakly interacting
BCS superconductor to be in this universality class.  The restriction to
lower dimensions is necessary for two different reasons.  First, only
for $d \leq 2$ the penetration depth is sufficiently large [see, for
example, below Eq.\ (\ref{2sc:mod})], so that it is appropriate to work
in the limit $\lambda_{\rm L} \rightarrow \infty$ with no fluctuating
gauge field
\cite{FGG}.  Second, in lower dimensions, the energy gap which the fermionic
excitations face remains finite at the critical point, so that it is
appropriate to ignore these degrees of freedom.  Moreover, since also
the coherence length remains finite at the critical point, the Cooper
pairs look like point particles on the scale of the diverging
correlation length associated with the phase fluctuations,
even in the weak-coupling limit \cite{CFGWY}.

In the preceding chapter, we argued that the nonrelativistic
$|\phi|^4$-theory is also of importance for the description of the
fractional quantized Hall effect (FQHE), where it describes---after
coupling to the Chern-Simons term---the original electrons bound to an
odd number of flux quanta.  As function of the applied magnetic field,
this two-dimensional system undergoes a zero-temperature transition
between a quantum Hall liquid, where the Hall conductance is quantized
in odd fractions of $e^2/2 \pi$, and an insulating phase.  The Hall
liquid corresponds to the phase with $\mu > \mu_{\rm c}$, while the
other phase again describes the insulating phase.

It should be noted however that in most of the applications of the
nonrelativistic $|\phi|^4$-theory mentioned here, impurities play an
important role; this will be the main subject of Sec.\ \ref{sec:Dirt}.

\begin{figure}
\begin{center}
\epsfxsize=2.cm
\mbox{\epsfbox{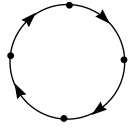}}
\end{center}
\caption{A closed oriented loop. \label{fig:oriented1}}
\end{figure}

\begin{figure}
\begin{center}
\epsfxsize=8.cm
\mbox{\epsfbox{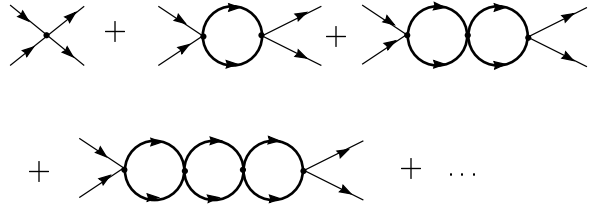}}
\end{center}
\caption{Ring diagrams renormalizing the vertex function of the neutral
$|\phi|^4$-theory. \label{fig:ring}}
\end{figure}

The critical properties of the zero-density transition of the
nonrelativistic $|\phi|^4$-theory were first studied by Uzunov
\cite{Uzunov}.  To facilitate the discussion let us make use of the fact
that in nonrelativistic theories the mass is---as far as critical phenomena
concerned---an irrelevant parameter which can be transformed away.  This
transformation changes, however, the scaling dimensions of the $\phi$-field
and the coupling constant which is of relevance to the renormalization-group
theory.  The engineering dimensions become
\begin{equation}  \label{BT:scale} 
[{\bf x}] = -1, \;\;\;\; [t] = -2, \;\;\;\; [\mu_0] = 2, \;\;\;\;
[\lambda_0] = 2-d,  \;\;\;\; [\phi] = \tfrac{1}{2}d,
\end{equation} 
with $d$ denoting the number of space dimensions.  In two space dimensions
the coupling constant $\lambda_0$ is dimensionless, showing that the
$|\phi|^4$-term is a marginal operator, and $d_{\rm c}=2$ the upper critical
space dimension.  Uzunov showed that below the upper critical dimension
there appears a non-Gaussian infrared-stable (IR) fixed point.  He computed
the corresponding critical exponents to all orders in perturbation theory
and showed them to have Gaussian values, $\nu=\tfrac{1}{2}, \; z=2, \;
\eta=0$.  Here, $\nu$ characterizes the divergence of the
correlation length, $z$ is the dynamic exponent, and $\eta$ is the
correlation-function exponent which determines the anomalous dimension of
the field $\phi$.  The unexpected conclusion that a non-Gaussian fixed point
has nevertheless Gaussian exponents is rooted in the analytic structure of
the nonrelativistic propagator at zero bare chemical potential ($\mu_0=0$):
\begin{equation} \label{BT:Green}
\raisebox{-0.3cm}{\epsfxsize=2.5cm
\epsfbox{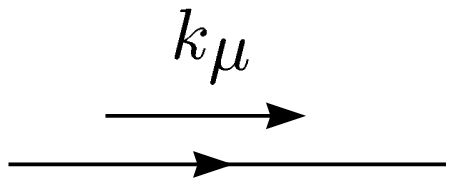}  }
= G(k) = \frac{i {\rm e}^{i k_0 \eta}}{k_0 - \tfrac{1}{2}{\bf k}^2 + i \eta },
\end{equation}
where, as before, $\eta$ is a small positive constant that has to be
taken to zero after the loop integrations over the energies have been
carried out.  The factor $\exp(i k_0 \eta)$ is an additional convergence
factor typical for nonrelativistic theories, which is needed for Feynman
diagrams involving only one $\phi$-propagator.  The rule $k_0
\rightarrow k_0 + i \eta$ in (\ref{BT:Green}) expresses the fact that in
this nonrelativistic theory particles propagate only forward in time.
In diagrams involving loops with more than one propagator, the integrals
over the loop energy are convergent and can be evaluated by contour
integration with the contour closed in either the upper or the lower
half plane.  If a diagram contains a loop which has all its poles in the
same half plane, it consequently vanishes.  Pictorially, such a loop has
all its arrows, representing the Green functions contained in the loop,
oriented in a clockwise or anticlockwise direction \cite{OB} (see Fig.\
\ref{fig:oriented1}).
We will refer to them as closed oriented loops.  Owing to this property most
diagrams are zero.  In particular, all self-energy diagrams vanish.  The
only surviving ones are the so-called ring diagrams which renormalize the
vertex (see Fig.\ \ref{fig:ring}).  
Because this class of diagrams constitute a geometric series, the one-loop
result is already exact.  The vertex renormalization leads to a non-Gaussian
fixed point in $d < 2$, while the vanishing of all the self-energy diagrams
asserts that the exponents characterizing the transition are not affected by
quantum fluctuations and retain their Gaussian values \cite{Uzunov}.  These
results have been confirmed by numerical simulations in $d=1$ \cite{BSZ} and
also by general scaling arguments \cite{FF,FWGF}.  

To understand the scaling arguments, let us consider the two terms in
the effective theory (\ref{eff:Leff}) quadratic in the Goldstone field
$\varphi$ with $m$ effectively set to 1 \cite{FF}:
\begin{equation} \label{general} 
{\cal L}_{\rm eff}^{(2)} = - \tfrac{1}{2} \rho_{\rm s} (\nabla
\varphi)^2 +  \tfrac{1}{2} \bar{n}^2 \kappa (\partial_0 \varphi)^2.
\end{equation}  
We have written this in the most general form.  The coefficient $\rho_{\rm
s}$ is the superfluid mass density which in the presence of, for example,
impurities does not equal $m \bar{n}$---even at the absolute zero of
temperature.  The other coefficient, 
\begin{equation} 
\bar{n}^2 \kappa = \frac{\partial \bar{n}}{\partial \mu} = \lim_{k
\rightarrow 0} \Pi_{0 0} (0,{\bf k}) , 
\end{equation} 
with $\Pi_{0 0}$ the (0 0)-component of the full polarization tensor
(\ref{bcs:cruc}), involves the full compressibility and particle number
density of the system at rest.  This is because the chemical potential
is according to (\ref{jo-pwa}) represented in the effective theory by
$\mu = -\partial_0 \varphi$ and
\begin{equation} 
\frac{\partial^2 {\cal L}_{\rm eff}}{\partial \mu^2} = \bar{n}^2 \kappa.
\end{equation} 
Equation (\ref{general}) leads to the general expression of the sound
velocity
\begin{equation} 
c^2 = \frac{\rho_{\rm s}}{\bar{n}^2 \kappa}
\end{equation} 
at the absolute zero of temperature.

Let $\delta \propto \mu - \mu_{\rm c}$ denote the distance from the phase
transition, so that $\xi \sim |\delta|^{-\nu}$.  Now, on the one hand,
the singular part of the free energy density $f_{\rm sing}$ arises from
the low-energy, long-wavelength fluctuations of the Goldstone field.
(Here, we adopted the common practice of using the symbol $f$ for the
density $\Omega/V$ and of referring to it as the free energy density.)
The ensemble averages give
\begin{equation} 
\langle (\nabla \varphi)^2 \rangle \sim \xi^{-2}, \;\;\;\;
\langle (\partial_0 \varphi)^2 \rangle \sim \xi_t^{-2} \sim \xi^{-2z} .
\end{equation} 
On the other hand, dimensional analysis shows that the singular part of
the free energy density scales near the transition as
\begin{equation} 
f_{\rm sing} \sim \xi^{-(d+z)}.
\end{equation} 
Combining these hyperscaling arguments, we arrive at the following
conclusions:
\begin{equation} \label{hyperrho} 
\rho_{\rm s} \sim \xi^{-(d+z-2)}, \;\;\;\; \bar{n}^2 \kappa \sim
\xi^{-(d-z)} \sim |\delta|^{(d-z)\nu}.
\end{equation} 
The first conclusion is consistent with the universal jump (\ref{jump})
predicted by Nelson and Kosterlitz \cite{NeKo} which corresponds to
taking $z=0$ and $d=2$.  Since $\xi \sim |\delta|^{-\nu}$, $f_{\rm
sing}$ can also be directly differentiated with respect to the chemical
potential to yield for the the singular part of the compressibility
\begin{equation} 
\bar{n}^2 \kappa_{\rm sing} \sim |\delta|^{(d+z)\nu -2}. 
\end{equation} 
Fisher and Fisher \cite{FF} continued to argue that there are two
alternatives.  Either $\kappa \sim \kappa_{\rm sing}$, implying $z \nu
=1$; or the full compressibility $\kappa$ is constant, implying $z=d$.
The former is consistent with the Gaussian values $\nu=\tfrac{1}{2}, \; z=2$ 
found by Uzunov \cite{Uzunov} for the pure case in $d < 2$.  The latter is
believed to apply to repulsively interacting bosons in a random media.  These
remarkable simple arguments thus predict the exact value $z=d$ for the
dynamic exponent in this case.

For later reference, let us consider the charged case and calculate the
conductivity $\sigma$.  The only relevant term for this purpose is the
first one in (\ref{general}) with $\nabla \varphi$ replaced by $\nabla
\varphi - e {\bf A}$.  We allow the superfluid mass density to vary
in space and time.  The term in the action quadratic in ${\bf A}$ then
becomes in the Fourier representation
\begin{equation} 
S_\sigma = - \tfrac{1}{2} e^2 \int_{k_0,{\bf k}} {\bf A}(-k_0,-{\bf k})
\rho_{\rm s} (k_0,{\bf k}) {\bf A}(k_0,{\bf k}).
\end{equation} 
The electromagnetic current,
\begin{equation} 
{\bf j}(k_0,{\bf k}) = \frac{\delta S_\sigma}{\delta {\bf A}(-k_0,-{\bf k})}
\end{equation} 
obtained from this action can be written as
\begin{equation} 
{\bf j}(k_0,{\bf k}) = \sigma(k_0,{\bf k}) {\bf E}(k_0,{\bf k})
\end{equation} 
with the conductivity
\begin{equation} \label{conductivity}
\sigma(k) = i e^2 \frac{\rho_{\rm s}(k)}{k_0}
\end{equation} 
essentially given by the superfluid mass density divided by $k_0$,  where
it should be remembered that the mass $m$ is effectively set to 1 here.

The above hyperscaling arguments have been extended by Fisher, Grinstein, and
Girvin \cite{FGG} to include the $1/|{\bf x}|$-Coulomb potential.  The
quadratic terms in the effective theory (\ref{effCoul}) may be cast in the
general form
\begin{equation} 
{\cal L}_{\rm eff}^{(2)} = \frac{1}{2} \left(\rho_{\rm s} {\bf k}^2 -
\frac{|{\bf k}|^{d-1}}{\hat{e}^2} k_0^2\right) |\varphi(k)|^2,
\end{equation}  
where $\hat{e}$ is the renormalized charge.  From (\ref{effCoul}) we find
that to lowest order:
\begin{equation} 
\hat{e}^2 = 2^{d-1} \pi^{(d-1)/2} \Gamma\left[\tfrac{1}{2}(d-1)\right] e_0^2.
\end{equation} 
The renormalized charge is connected to the (0 0)-component of the full
polarization tensor (\ref{bcs:cruc}) via
\begin{equation} 
\hat{e}^2 = \lim_{|{\bf k}| \rightarrow 0} \frac{|{\bf k}|^{d-1}}{\Pi_{0
0} (0,{\bf k})} . 
\end{equation} 
A simple hyperscaling argument like the ones given above shows that near
the transition, the renormalized charge scales as
\begin{equation} 
\hat{e}^2 \sim \xi^{1-z}.
\end{equation} 
They then argued that in the presence of random impurities this charge
is expected to be finite at the transition so that $z=1$.  This again is
an exact results which  replaces the value $z=d$ of the neutral system.

We have seen that $d_{\rm c}=2$ is the upper critical dimension of the
nonrelativistic $|\phi|^4$-theory.  Dimensional analysis shows that for an
interaction term of the form
\begin{equation}
{\cal L}_{\rm i} = - g_0 |\phi|^{2k}
\end{equation}
the upper critical dimension is
\begin{equation}
\label{BT:dcnr}
d_{\rm c} = \frac{2}{k-1}.
\end{equation}
The two important physical cases are $d_{\rm c}=2$, $k=2$ and $d_{\rm c}=1$,
$k=3$, while $d_{\rm c} \rightarrow 0$ when $k \rightarrow \infty$.  For space
dimensions $d > 2$ only the quadratic term, $|\phi|^2$, is relevant so that
here the critical behavior is well described by the Gaussian theory. 

In the corresponding relativistic theory, the scaling dimensions of $t$ and
${\bf x}$ are, of course, equal $[t] = [{\bf x}] = -1$ and $[\phi] =
\tfrac{1}{2} (d-1)$.  This leads to different upper critical (space)
dimensions, viz., 
\begin{equation}
d_{\rm c} = \frac{k+1}{k-1} = \frac{2}{k-1} + 1,
\end{equation}
instead of (\ref{BT:dcnr}).  The two important physical cases are here $d_{\rm
c}=3$, $k=2$ and $d_{\rm c}=2$, $k=3$, while $d_{\rm c} \rightarrow 1$ when $k
\rightarrow \infty$.  On comparison with the nonrelativistic results, we see
that the nonrelativistic theory has an upper critical space dimension which
is one lower than that of the corresponding relativistic theory (see Table
\ref{table:1}).  Heuristically, this can be understood by noting that in a
nonrelativistic context the time dimension counts double in that it
has a scaling dimension twice that of a space dimension [see Eq.\
(\ref{BT:scale})], thereby increasing the {\it effective} spacetime
dimensionality by one.
\begin{table}
\caption{The upper critical space dimension $d_{\rm c}$ of a
nonrelativistic (NR) and a relativistic (R) quantum theory with a
$|\phi|^{2k}$ interaction term.}
\label{table:1}
\begin{center}
\vspace{.5cm}
\begin{tabular}{ccc|cccccccc} \hline \hline 
&  & & & &  & & &  & & \\[-.2cm] 
& $k$ & & & & $d_{\rm c}$(NR)&  & & $d_{\rm c}$(R) & & \\[.1cm]
\hline  
&  & & & &  & & &  & & \\[-.2cm] 
& 2 & & & & 2 & & & 3  & & \\ 
& 3 & & & & 1 & & & 2 & & \\ 
& $\infty$ & & & & 0 & & & 1 & &   \\[.1cm]
\hline \hline
\end{tabular}
\end{center}
\end{table}

From this analysis it follows that for a given number of space
dimensions the critical properties of a nonrelativistic theory are unrelated
to those of the corresponding relativistic extension.

In closing this section we recall that in a one-dimensional relativistic
theory---corresponding to the lowest upper critical dimension ($d_{\rm
c}=1)$---a continuous symmetry cannot be spontaneously broken.  However, the
theory can nevertheless have a phase transition of the 
Kosterlitz-Thouless type.  Given the connection between the relativistic and
nonrelativistic theories discussed above, it seems interesting to study the
nonrelativistic theory at zero space dimension ($d = 0$) to see if a similar
rich phenomenon as in the lower critical dimension of the relativistic
theory occurs here.  This may be of relevance to so-called quantum dots.
\section{Quantum Hall Liquid}
\label{sec:QHL}
In this section we shall argue that the effective theory of a quantum
Hall liquid can be used to describe its liquid-to-insulator transition
as the applied magnetic field changes, and study its critical properties.

Experimentally, if the external field is changed so that the filling
factor $\nu_H$ moves away from an odd-denominator value, the system
eventually becomes critical and undergoes a transition to an insulating
phase.  Elsewhere \cite{NP}, we have argued that this feature is encoded
in the CSGL theory.  In the spirit of Landau, we took a phenomenological
approach towards this field-induced phase transition.  And assumed that
when the applied magnetic field $H$ is close to the upper critical field
$H^+_{\nu_H}$ at which the quantum Hall liquid with filling factor
$\nu_H$ is destroyed, the chemical potential of the composite particles
depends linearly on $H$, i.e., $\mu_0
\propto eH^+_{\nu_H}- eH$.  This state can of course also be destroyed by
lowering the applied field.  If the system is near the lower critical
field $H^-_{\nu_H}$, we assumed that the chemical potential is instead
given by $\mu_0 \propto eH - eH^-_{\nu_H}$.  This is the basic postulate of
our approach.

We modify the CSGL Lagrangian (\ref{CSGL:L}) so that it only includes
the fluctuating part of the statistical gauge field.  That is, we ignore
the classical part of $a$ which yields a magnetic field that precisely
cancels the externally applied field.  We can again transform the mass
$m$ of the nonrelativistic $|\phi|^4$-theory away.  In addition to the
engineering dimensions (\ref{BT:scale}), we have for the Chern-Simons
field
\begin{equation} \label{CSGL:dim}
[e a_i] = 1, \;\;\;\; [e a_0] = 2, \;\;\;\; [\theta] = 0.
\end{equation} 
In two space dimensions, the coupling constant $\lambda_0$ was seen to be
dimensionless, implying that the $|\phi|^4$-term is a marginal operator.
From (\ref{CSGL:dim}) it follows that also the Chern-Simons term is a
marginal operator.  Hence, the CSGL theory contains---apart from a
random and a Coulomb term---precisely those terms relevant to the
description of the liquid-to-insulator transition in a quantized Hall
system.  

It is well-known \cite{CH,LSW}, that the coefficient of the Chern-Simons
term is not renormalized by quantum corrections.

To first order in a loop expansion, the theory in two space dimensions
was known to have an IR fixed point determined by the zero of the beta
function \cite{LBL},
\begin{equation}
\label{RG:beta1loop}
\beta(\lambda) = \frac{1}{\pi} \left(4\lambda^2 - \frac{1}{\theta^2} \right).
\end{equation}
The calculation of $\beta(\lambda)$ has been extended to fourth order in the
loop expansion \cite{NP}.  This study revealed that the one-loop result
(\ref{RG:beta1loop}) is unaffected by these higher-order loops.  Presumably,
this remains true to all orders in perturbation theory, implying that---just
as in the neutral system which corresponds to taking the limit $\theta
\rightarrow \infty$---the one-loop beta function (\ref{RG:beta1loop}) is
exact.

\begin{figure}
\vspace{-1.cm}
\begin{center}
\epsfxsize=7.cm
\mbox{\epsfbox{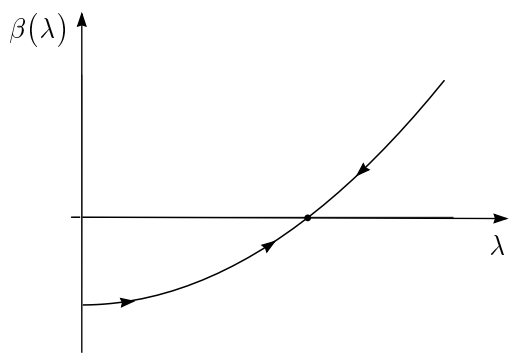}}
\end{center}
\vspace{-1.5cm}
\caption{Schematic representation of the beta function
(\ref{RG:beta1loop}). \label{fig:beta}} 
\end{figure}

It is schematically represented in Fig.\ \ref{fig:beta},
and is seen to yield a nontrivial IR fixed point
$\lambda^*\mbox{}^2= 1/4\theta^2$ determined by the filling
factor.  More precisely, the strength of the repulsive coupling at the
fixed point $\lambda^{*} = \pi (2l+1)$ increases with the number $2l+1$
of flux quanta bound to the electron.  The presence of the fixed point
shows that the CSGL theory undergoes a 2nd-order phase transition when
the chemical potential of the composite particles tends to a critical
value.  As in the neutral case, it can be shown that the boson
self-energy $\Sigma$ also vanishes at every loop order in the charged
theory, and that the self-coupling parameter $\lambda$ is the only
object that renormalizes.  The 2nd-order phase transition described by
the nontrivial IR fixed point has consequently again Gaussian exponents
$\nu=\tfrac{1}{2}, \; z=2$ \cite{NP}.  It should be noted that only the
location of the fixed point depends on $\theta$, the critical
exponents---which in contrast to the strength of the coupling at the
fixed point are independent of the regularization and renormalization
scheme---are universal and independent of the filling factor.  This
``triviality'' is in accord with the experimentally observed
universality of the FQHE.  A dependence of the critical exponents on
$\theta$ could from the theoretical point of view hardly be made
compatible with the hierarchy construction
\cite{HH} which implies a cascade of phase transitions.  From this
viewpoint the present results are most satisfying: the CSGL theory is
shown to encode a new type of field-induced 2nd-order quantum phase
transition that is simple enough not to obscure the observed
universality of the FQHE.  

We stress again that in order to arrive at a realistic description of
the FQHE, the CSGL theory has to be extended to include 
a $1/|{\bf x}|$-Coulomb potential and impurities.  Both will change
the critical behavior we found in this section. In particular, the
Coulomb potential will change the Gaussian value $z=2$ into $z=1$.
\section{Random Theory}
\label{sec:Dirt}
In Sec.\ \ref{sec:BT}, we saw that in the absence of impurities
repulsively interacting bosons undergo a 2nd-order quantum phase
transition as function of the chemical potential.  As was pointed out
there, this universality class is of relevance to various
condensed-matter systems.  However, in most of the systems mentioned
there, as well in $^4$He in porous media, impurities play an essential
if not decisive role.  For example, the two-dimensional
superconductor-to-insulator transition investigated by Hebard and
Paalanen
\cite{HPsu1} is driven by impurities.  This means that, e.g., the correlation
length $\xi$ diverges as $|\Delta_{\rm c} - \Delta|^{-\nu}$ when
the disorder strength $\Delta$ characterizing the randomness approaches
the critical value $\Delta_{\rm c}$.  Hence, a realistic description
of the critical behavior of these systems should include impurities.

To include these, we proceed as before and add the random term
(\ref{Dirt:dis}) to the nonrelativistic $|\phi|^4$-theory (\ref{eff:Lagr}).
The random field $\psi({\bf x})$ has the Gaussian distribution
(\ref{random}).  We shall study the theory in the symmetrical state where
the bare chemical potential is negative and the global U(1) symmetry
unbroken.  We therefore set $\mu_0 = - r_0$ again, with $r_0>0$.  We leave
the number of space dimensions $d$ unspecified for the moment.  As we
remarked before, since $\psi({\bf x})$ depends only on the $d$ spatial
dimensions, the impurities it describes should be considered as grains
randomly distributed in space.  When---as is required for the study of
quantum critical phenomena---time is included, the static grains trace out
straight worldlines.  That is to say, these impurities are linelike in the
quantum theory.  It has been shown by Dorogovtsev \cite{Dorogovtsev} that
the critical properties of systems with extended defects must be studied in
a double $\epsilon$-expansion, otherwise no IR fixed point is found.  The
method differs from the usual $\epsilon$-expansion, in that it also includes
an expansion in the defect dimensionality $\epsilon_{\rm d}$.  To carry out
this program in the present context, where the defect dimensionality is
determined by the dimensionality of time, the theory has to be formulated in
$\epsilon_{\rm d}$ time dimensions.  The case of interest is $\epsilon_{\rm
d}=1$, while in the opposite limit, $\epsilon_{\rm d}\rightarrow 0$, the
random nonrelativistic $|\phi|^4$-theory reduces to the classical spin model
with random (pointlike) impurities.  Hence, $\epsilon_{\rm d}$ is a
parameter with which quantum fluctuations can be suppressed.  An expansion
in $\epsilon_{\rm d}$ is a way to perturbatively include the effect of
quantum fluctuations on the critical behavior.  Ultimately, we will be
interested in the case $\epsilon_{\rm d}=1$.

To calculate the quantum critical properties of the random theory, which
have first been studied in \cite{KU}, we will not employ the replica
method \cite{GrLu}, but instead follow Lubensky
\cite{Lubensky}.  In this approach, the averaging over impurities is carried
out for each Feynman diagram separately.  The upshot is that only those
diagrams are to be included which remain connected when $\Delta_0$, the
parameter characterizing the Gaussian distribution of the impurities, is set
to zero \cite{Hertzrev}.  To obtain the relevant Feynman rules of the random
theory we average the interaction term (\ref{Dirt:dis}) over the
distribution (\ref{random}):
\begin{eqnarray}  \label{Dirt:int}
\lefteqn{\int \DD \psi \, P(\psi)
\exp\left[i^{\epsilon_{\rm d}} \int
\dd^{\epsilon_{\rm d}} t \, \dd^d x \, \psi({\bf x}) \, |\phi(x)|^2  \right]
= } \nonumber \\ && \exp \left[\tfrac{1}{4} i^{2 \epsilon_{\rm d}} \Delta_0
\int \dd^{\epsilon_{\rm d}} t \, \dd^{\epsilon_{\rm d}} t' \, \dd^d x \,
|\phi(t,{\bf x})|^2 |\phi(t',{\bf x})|^2 \right].
\end{eqnarray} 
The randomness is seen to result in a quartic interaction term which is
nonlocal in time.  The factor $i^{\epsilon_{\rm d}}$ appearing in
(\ref{Dirt:int}) arises from the presence of $\epsilon_{\rm d}$ time
dimensions, each of which is accompanied by a factor of $i$.  The Feynman
rules of the random theory are now easily obtained
\begin{eqnarray} \label{Dirt:Feynart}
\raisebox{-0.3cm}{\epsfxsize=2.5cm
\epsfbox{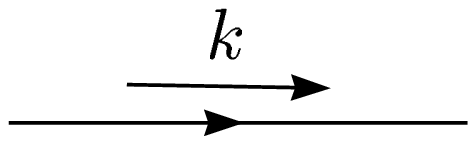}  }
 &=& \frac{-i^{- \epsilon_{\rm d}} {\rm e}^{i(\omega_1 + \omega_2 + \cdots +
\omega_{\epsilon_{\rm d}}) \eta}}{\omega_1 + \omega_2 + \cdots +
\omega_{\epsilon_{\rm d}} -{\bf k}^2 - r_0 + i \eta} \nonumber \\
\raisebox{-0.5cm}{\epsfxsize=2.5cm
\epsfbox{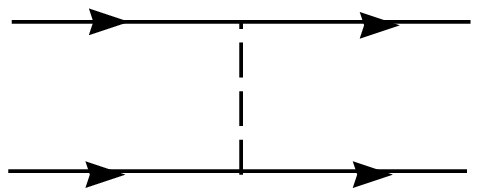}  }
&=& -4 i^{\epsilon_{\rm d}} \lambda_0 \nonumber \\
\raisebox{-0.5cm}{\epsfxsize=2.5cm
\epsfbox{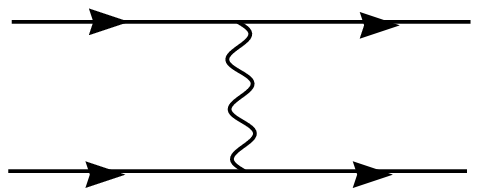}  }
&=& i^{\epsilon_{\rm d}} (2 \pi)^{\epsilon_{\rm d}} \delta^{\epsilon_{\rm
d}}(\omega_1 + \omega_2 + \cdots + \omega_{\epsilon_{\rm d}})
\Delta_0,
\end{eqnarray} 
where we note that the Lagrangian in $\epsilon_{\rm d}$ time dimensions
involves instead of just one time derivative, a sum of $\epsilon_{\rm d}$
derivatives: $\partial_t \rightarrow \partial_{t_1} +
\partial_{t_2} + \cdots + \partial_{t_{\epsilon_{\rm d}}}$.  

Following Weichman and Kim \cite{WK}, we evaluate the integrals over loop
energies assuming that all energies are either positive or negative.  This
allows us to employ Schwinger's propertime representation of propagators
\cite{proptime}, which is based on the integral representation (\ref{gamma})
of the gamma function.  The energy integrals we encounter to the one-loop
order can be carried out with the help of the equations
\begin{eqnarray}  \label{Dirt:inta} 
\lefteqn{\int' \frac{\dd^{\epsilon_{\rm d}} \omega}{(2\pi)^{\epsilon_{\rm d}}}
\frac{1}{\omega_1 + \omega_2 + \cdots + \omega_{\epsilon_{\rm d}} -x
\pm i \eta} =} \nonumber \\ && -\frac{\Gamma(1-\epsilon_{\rm
d})}{(2\pi)^{\epsilon_{\rm d}}} {\rm sgn}(x) |x|^{\epsilon_{\rm d}-1}
\left({\rm e}^{\pm i \, {\rm sgn}(x) \pi
\epsilon_{\rm d}} + 1 \right),  \\ 
\lefteqn{\int' \frac{\dd^{\epsilon_{\rm d}} \omega}{(2\pi)^{\epsilon_{\rm d}}}
\frac{{\rm e}^{i(\omega_1 + \omega_2 + \cdots + \omega_{\epsilon_{\rm
d}})\eta}}{\omega_1 + \omega_2 + \cdots + \omega_{\epsilon_{\rm d}} -x + i x
\eta} =} \nonumber \\ &&  \frac{i \pi}{(2\pi)^{\epsilon_{\rm
d}}\Gamma(\epsilon_{\rm d})} (i|x|)^{\epsilon_{\rm d}-1} \left[
\sin(\tfrac{1}{2} \pi \epsilon_{\rm d}) - \frac{{\rm
sgn}(x)}{\sin(\tfrac{1}{2} \pi \epsilon_{\rm d} )} \right].
\label{Dirt:intb} 
\end{eqnarray}  
The prime on the integrals is to remind the reader that the energy integrals
are taken over only two domains with either all energies positive or
negative.  The energy integrals have been carried out by using again the
integral representation (\ref{gamma}) of the gamma function.  In doing
so, the integrals are regularized and---as is always the case with analytic
regularizations---irrelevant divergences suppressed.

By differentiation with respect to $x$, Eq.\ (\ref{Dirt:inta}) can,
for example, be employed to calculate integrals involving integrands of the
form $1/(\omega_1 + \omega_2 + \cdots + \omega_{\epsilon_{\rm d}} -x + i
\eta)^2$.  Is is easily checked that in the limit $\epsilon_{\rm d}
\rightarrow 1$, where the energy integral can be performed with help of
contour integration, Eqs.\ (\ref{Dirt:inta}) and (\ref{Dirt:intb}) reproduce
the right results.  When considering the limit of zero time dimensions
($\epsilon_{\rm d} \rightarrow 0$), it should be borne in mind that the
energy integrals were taken over two separate domains with all energies
either positive or negative.  Each of these domains is contracted to a
single point in the limit $\epsilon_{\rm d} \rightarrow 0$, so that one
obtains a result which is twice that obtained by simply purging any
reference to the time dimensions.  The integral (\ref{Dirt:intb}) contains
an additional convergence factor $\exp(i\omega \eta)$ for each of the
$\epsilon_{\rm d}$ energy integrals.  This factor, which---as we remarked
before---is typical for nonrelativistic quantum theories \cite{Mattuck}, is
to be included in self-energy diagrams containing only one
$\phi$-propagator.

Before studying the random theory, let us briefly return to the repulsively
interacting bosons in the absence of impurities.  In this case, there is no
need for an $\epsilon_{\rm d}$-expansion and the formalism outlined above
should yield results for arbitrary time dimensions $0 \leq
\epsilon_{\rm d} \leq 1$, interpolating between the classical and quantum
limit.  After the energy integrals have been performed with the help of
Eqs.\ (\ref{Dirt:inta}) and (\ref{Dirt:intb}), the standard technique of
integrating out a momentum shell can be applied to obtain the
renormalization-group equations.  For the correlation-length exponent
$\nu$ we obtain in this way \cite{pla}
\begin{equation} \label{Dirt:nupure}
\nu = \frac{1}{2} \left[1 + \frac{\epsilon}{2} \frac{m+1}{(m+4) -
(m+3) \epsilon_{\rm d}} \cos^2( \tfrac{1}{2} \pi \epsilon_{\rm d}) \right].
\end{equation}   
Here, $\epsilon = 4-2\epsilon_{\rm d}-d$ is the deviation of the {\it
effective} spacetime dimensionality from 4, where it should be noted that in
(canonical) nonrelativistic theories, time dimensions have an engineering
dimension twice that of space dimensions.  (This property is brought out by
the Gaussian value $z=2$ for the dynamic exponent $z$.)  For comparison we
have extended the theory (\ref{eff:Lagr}) to include $m$ complex
$\phi$-fields instead of just one field.  In the classical limit, Eq.\
(\ref{Dirt:nupure}) gives the well-known one-loop result for a classical
spin model with $2m$ real components \cite{Ma},
\begin{equation} 
\nu \rightarrow \frac{1}{2} \left(1 + \frac{\epsilon}{2}
\frac{m+1}{m+4} \right), 
\end{equation} 
while in the quantum limit it gives the result $\nu \rightarrow
\frac{1}{2}$, as required.  

The exponent (\ref{Dirt:nupure}), and also the location of the fixed point,
diverges when the number of time dimensions becomes $\epsilon_{\rm
d}\rightarrow (m+4)/(m+3)$.  Since this value is always larger than one, the
singularity is outside the physical domain $0 \leq \epsilon_{\rm d} \leq 1$.
This simple example illustrates the viability of the formalism developed
here to generate results interpolating between the classical and quantum
limit.

We continue with the random theory.  After the energy integrals have been
carried out, it is again straightforward to derive the renormalization-group
equations by integrating out a momentum shell $\Lambda/b<k<\Lambda$, where
$\Lambda$ is a high-momentum cutoff and $b=\exp(l)$, with $l$ infinitesimal.
Defining the dimensionless variables
\begin{equation} 
\hat{\lambda} = \frac{K_d}{(2 \pi)^{\epsilon_{\rm d}}} \lambda 
\Lambda^{-\epsilon}; \;\;\;
\hat{\Delta} = K_d \Delta \Lambda^{d-4}; \;\;\;
\hat{r} = r \Lambda^{-2},
\end{equation} 
where 
\begin{equation} 
K_d = \frac{2}{(4\pi)^{d/2}\Gamma(\tfrac{d}{2})} 
\end{equation}  
is the area of a unit sphere in $d$ spatial dimensions divided by
$(2\pi)^d$, we find \cite{pla}
\begin{eqnarray} \label{Dirt:reneq}
\frac{\dd \hat{\lambda}}{\dd l} &=&  \epsilon \hat{\lambda}  -8
\left[\Gamma(1-\epsilon_{\rm d}) + (m+3) \Gamma(2-\epsilon_{\rm d}) \right]
\cos(\tfrac{1}{2}\pi \epsilon_{\rm d}) 
\hat{\lambda}^2 + 6 \hat{\Delta} \hat{\lambda} \nonumber  \\
\frac{\dd \hat{\Delta}}{\dd l} &=&  (\epsilon + 2\epsilon_{\rm
d})\hat{\Delta }  + 4 \hat{\Delta}^2 - 16 (m+1)
\Gamma(2-\epsilon_{\rm d}) \cos(\tfrac{1}{2}\pi \epsilon_{\rm d} ) 
\hat{\lambda} \hat{\Delta} \nonumber  \\
\frac{\dd \hat{r}}{\dd l} &=& 2 \hat{r} + 4 \pi \frac{m+1}{
\Gamma(\epsilon_{\rm d})} \frac{\cos^2(\tfrac{1}{2} \pi \epsilon_{\rm d})}
{\sin(\tfrac{1}{2} \pi \epsilon_{\rm d} )}  \hat{\lambda} -
 \hat{\Delta}. 
\end{eqnarray} 
These results are to be trusted only for small values of $\epsilon_{\rm d}$.
For illustrative purposes we have, however, kept the full $\epsilon_{\rm d}$
dependence.  The set of equations yields the fixed point
\begin{eqnarray} \label{Dirt:fp}
\hat{\lambda}^* &=& \frac{1}{16 \cos(\tfrac{1}{2}\pi \epsilon_{\rm d} )
\Gamma(1-\epsilon_{\rm d})} \, \frac{\epsilon + 6 \epsilon_{\rm d}} 
{2m(1-\epsilon_{\rm d}) -1}  \\
\hat{\Delta}^* &=& \frac{1}{4} \frac{
m(1-\epsilon_{\rm d}) (2 \epsilon_{\rm d} -\epsilon) + 2 \epsilon_{\rm d}
(4-3\epsilon_{\rm d}) + \epsilon (2 -\epsilon_{\rm d})}{2m(1-\epsilon_{\rm d})
-1}, \nonumber 
\end{eqnarray}   
and the critical exponent 
\begin{equation} \label{Dirt:nufull} 
\nu = \frac{1}{2} + \frac{\epsilon +2 \epsilon_{\rm d}}{16} + 
\frac{m+1}{16} \frac{
(6\epsilon_{\rm d} + \epsilon )  [\epsilon_{\rm d}+\cos( \pi
\epsilon_{\rm d})]}{2m(1-\epsilon_{\rm d})-1}.
\end{equation} 
The dynamic exponent is given by $z = 2 + \hat{\Delta}^*$.  When the
equations are expanded to first order in $\epsilon_{\rm d}$, we recover
the IR fixed point found by Weichman and Kim \cite{WK} using an
high-energy cutoff:
\begin{equation} \label{dirt:WK}
\hat{\lambda}^*= \frac{1}{16} \frac{\epsilon + 6 \epsilon_{\rm d}}{2m-1};
\;\;\;  \hat{\Delta}^*= \frac{1}{4} \frac{(2-m)\epsilon + 2(m+4)
\epsilon_{\rm d}}{2m-1}, 
\end{equation}
with the critical exponent
\begin{equation} \label{Dirt:nuqm}
\nu = \frac{1}{2} \left[1 + \frac{1}{8} \frac{3m \epsilon + (5m +2)
2\epsilon_{\rm d}}{2m-1} \right].
\end{equation} 

The value of the critical exponent (\ref{Dirt:nuqm}) should be compared with
that of the classical spin model with $2m$ components in the presence of
random impurities of dimension $\epsilon_{\rm d}$ \cite{Dorogovtsev}:
\begin{equation} 
\nu = \frac{1}{2} \left[1 + \frac{1}{8} \frac{3m \epsilon + (5m +2)
\epsilon_{\rm d}}{2m-1} \right].
\end{equation}  
Taking into account that in a nonrelativistic quantum theory, time
dimensions count double as compared to space dimensions, we see that both
results are equivalent.  As to the dynamic exponent, notice that the
perturbative result $z = 2 + \hat{\Delta}^*$, with $\hat{\Delta}^*$ given by
(\ref{dirt:WK}), is far away from the exact value $z=d$ for $\epsilon_{\rm
d}=1$ \cite{FF}.
\section{Experiments}
\subsection{Superconductor-To-Insulator Transition}
The first experiments we wish to discuss are those performed by Hebard
and Paalanen on superconducting films in the presence of random
impurities \cite{HPsu1,HPsu2}.  It has been predicted by Fisher
\cite{MPAFisher} that with increasing applied magnetic field such
systems undergo a zero-temperature transition into an insulating state.
(For a critical review of the experimental data available in 1993, see
Ref.\ \cite{LG}.) 

Let us restrict ourselves for the moment to the $T\Delta$-plane of the
phase diagram by setting the applied magnetic field $H$ to zero.  For
given disorder strength $\Delta$, the system then undergoes a
Kosterlitz-Thouless transition induced by the unbinding of magnetic
vortex pairs at a temperature $T_{\rm KT}$ well below the bulk
transition temperature (see Sec.\ \ref{sec:2sc}).  The
Kosterlitz-Thouless temperature is gradually suppressed to zero when the
disorder strength approaches criticality $\Delta \rightarrow
\Delta_{\rm c}$.  The transition temperature scales with the correlation
length $\xi \sim |\Delta_{\rm c} - \Delta|^{-\nu}$ as
$T_{\rm KT} \sim \xi^{-z}$.

In the $H\Delta$-plane, i.e., at $T=0$, the situation is as follows.
For given disorder strength, there is now at some critical value $H_{\rm
c}$ of the applied magnetic field a phase transition from a
superconducting state of pinned vortices and condensed Cooper pairs to
an insulating state of pinned Cooper pairs and condensed vortices.  The
condensation of vortices disorder the ordered state as happens in
classical, finite temperature superfluid- and superconductor-to-normal
phase transitions \cite{GFCM}.  When the disorder strength approaches
criticality again, $H_{\rm c}$ is gradually suppressed to zero.  The
critical field scales with $\xi$ as $H_{\rm c} \sim \Phi_0/\xi^2$.  In
fact, this expresses a more fundamental result, namely that the scaling
dimension $d_{\bf A}$ of ${\bf A}$ is one,
\begin{equation} 
d_{\bf A} = 1,
\end{equation} 
so that $|{\bf A}| \sim \xi^{-1}$.  From this it in turn
follows that $E \sim \xi_t^{-1} \xi^{-1} \sim \xi^{-(z+1)}$, and that
the scaling dimension $d_{A_0}$ of $A_0$ is $z$, 
\begin{equation} 
d_{A_0} = z,
\end{equation} 
so that $A_0 \sim \xi_t^{-1} \sim \xi^{-z}$.  Together, the scaling
results for $T_{\rm KT}$ and $H_{\rm c}$ imply that
\cite{MPAFisher}
\begin{equation}  \label{H-T}
H_{\rm c} \sim T_{\rm KT}^{2/z}.
\end{equation} 
This relation, linking the critical field of the zero-temperature
transition to the Kosterlitz-Thouless temperature, provides a direct way
to measure the dynamic exponent $z$ at the $H=0$, $T=0$ transition.
This has been first done by Hebard and Paalanen \cite{HPsu1,HPsu2}.
Their experimental determination of $T_{\rm KT}$ and $H_{\rm c}$ for
five different films with varying amounts of impurities confirmed the
relation (\ref{H-T}) with $2/z = 2.04 \pm 0.09$.  The zero-temperature
critical fields were obtained by plotting $\dd \rho_{xx}/\dd T|_H$
versus $H$ at the lowest accessible temperature and interpolating to the
the field where the slope is zero. The resulting value $z= 0.98 \pm .04$
is in accordance with Fisher's prediction \cite{MPAFisher}, $z=1$, for a
random system with a $1/|{\bf x}|$-Coulomb potential.

Hebard and Paalanen \cite{HPsu1} also investigated the field-induced
zero-temperature transition.  The control parameter is here $\delta \propto
H -H_{\rm c}$.  When plotted as function of $|H -H_{\rm c}|/T^{1/\nu_H
z_H}$ they saw their resistivity data collapsing onto two branches; an upper
branch tending to infinity for the insulating state, and a lower branch
bending down for the superconducting state.  The unknown product $\nu_H z_H$
is experimentally determined by looking for which value the best scaling
behavior is obtained.  Further experiments carried out by Yazdani and
Kapitulnik \cite{YaKa} also determined the product $\nu_H (z_H+1)$ (see
below).  The two independent measurements together fix the critical
exponents $\nu_H$ and $z_H$ separately.  From their best data, Yazdani and
Kapitulnik extracted the values \cite{YaKa}
\begin{equation} \label{zHnuH}
z_H = 1.0 \pm 0.1, \;\;\;\; \nu_H = 1.36 \pm 0.05.
\end{equation} 
\subsection{Quantum-Hall Systems}
We continue to discuss the field-induced quantum phase transitions in
quantum Hall systems.  Since an excellent discussion recently appeared in
the literature \cite{SGCS}, we shall be brief, referring the reader to
that review for a more thorough discussion and additional references.

One can image transitions from one Hall liquid to another Hall liquid with a
different (integer or fractional) filling factor, or to the insulating
state.  Experiments seem to suggest that all the quantum-Hall transitions
are in the same universality class.  The transitions are probed by measuring
the resistivities $\rho_{xx}$ and $\rho_{xy}$.  From the dependence of the
conductivity $\sigma$ on the superfluid mass density, Eq.\
(\ref{conductivity}), and the scaling relation (\ref{hyperrho}), it follows
that it scales as \cite{AALR}
\begin{equation} 
\sigma \sim \xi^{-(d-2)}. 
\end{equation} 
In other words, the scaling dimension of the conductivity and therefore
that of the resistivity is zero in two space dimensions.  On account of
the general finite-size scaling form (\ref{scalingT}), we then have in
the limit $|{\bf k}| \rightarrow 0$:
\begin{equation} 
\rho_{xx/y}(k_0,H,T) = \varrho _{xx/y}(k_0/T, |\delta|^{\nu z}/T),
\end{equation} 
where the distance to the zero-temperature critical point is measured by
$\delta \propto H - H_{\nu_H}^\pm \sim T^{1/\nu z}$.  This scaling of
the width of the transition regime with temperature has been
corroborated by DC or $k_0=0$ experiments on various transitions between
integer quantum-Hall states which were all found to yield the value
$1/\nu z = 0.42 \pm 0.04$ \cite{WTPP}.

A second measurement of the critical exponents involves the applied
electric field.  As we have seen above, it scales as $E 
\sim \xi^{-(z+1)}$, so that for the DC resistivities we now obtain the
scaling form: 
\begin{equation}  \label{scalingE}
\rho_{xx/y}(H,T,E) = \varrho
_{xx/y}(|\delta|^{\nu z}/T,|\delta|^{\nu (z+1)}/E).
\end{equation}
The scaling $|\delta| \sim E^{1/\nu (z+1)}$ has again been corroborated by
experiment which yielded the value $\nu (z+1)  \approx 4.6$ \cite{WET}.
Together with the previous result obtained from the temperature scaling
this gives
\begin{equation} 
z \approx 1, \;\;\;\; \nu \approx 2.3.
\end{equation} 
The value of the dynamic exponent strongly suggests that it is a result
of the presence of the $1/|{\bf x}|$-Coulomb potential.  The correlation
length exponent $\nu$ is large.
\subsection{$2d$ Electron Systems}
Recently, silicon MOSFET's at extremely low electron number densities
has been studied \cite{MIT,KSSMF,SKS,PFW}.  Earlier experiments at
higher densities seemed to confirm the general believe, based on the
work by Abrahams {\it et al.} \cite{AALR}, that such two-dimensional
electron systems do not undergo a quantum phase transition.  In that
influential paper, it was demonstrated that even weak disorder is
sufficient to localize the electrons at the absolute zero of temperature
thus excluding conducting behavior.  Electron-electron interactions were
however not included.  As we saw in Sec.\
\ref{sec:ET}, at low densities, the $1/|{\bf x}|$-Coulomb interaction
becomes important and the analysis of Abrahams {\it et al.}
\cite{AALR} no longer applies.

The recent experiments have revealed a zero-temperature
conductor-to-insulator transition triggered by a change in the charge
carrier density $\bar{n}$.  That is, the distance to the critical point
is in these systems measured by $\delta \propto \bar{n} - \bar{n}_{\rm c}$.
Like in the quantum-Hall systems, these transitions are probed by
measuring the resistivity.  It scales with temperature near the
transition according to the scaling form (\ref{scalingE}) with $H$ set
to zero.  For $\bar{n} < \bar{n}_{\rm c}$, where the Coulomb interaction
is dominant and fluctuations in the charge carrier density are
suppressed, the electron system is insulating.  On increasing the
density, these fluctuations intensify and at the critical value
$\bar{n}_{\rm c}$, the system reverts to a conducting phase.  By plotting
their conductivity data as function of $T/|\delta|^{\nu z}$ with $\nu z =
1.6 \pm 0.1$, Popov\'{\i}c, Fowler, and Washburn \cite{PFW} saw it
collapse onto two branches; the upper branch for the conducting side of the
transition, and the lower one for the insulating side.  A similar
collapse with a slightly different value $1/\nu z = 0.83 \pm 0.08$ was found
in Ref.\ \cite{KSSMF}, where also the collapse of the data when plotted
as function of $|\delta|/E^{1/\nu (z+1)}$ was obtained.  The best collapse
resulted for $1/(z+1) \nu = 0.37 \pm 0.01$, leading to
\begin{equation} z = 0.8 \pm 0.1, \;\;\;\; \nu = 1.5 \pm 0.1.
\end{equation} 
The value for the dynamic exponent is close to the expected value $z=1$ for
a charged system with a $1/|{\bf x}|$-Coulomb interaction, while that of
$\nu$ is surprisingly close to the value (\ref{zHnuH}) found for the
superconductor-to-insulator transition.

A further experimental result for these two-dimensional electron systems
worth mentioning is the suppression of the conducting phase by an
applied magnetic field found by Simonian, Kravchenko, and Sarachik
\cite{SKS}.  They applied the field {\it parallel} to the plane of the
electrons instead of perpendicular as is done in quantum-Hall
measurements.  In this way, the field presumably couples only to the
spin of the electrons and the complications arising from orbital effects
do not arise.  At a fixed temperature, a rapid initial raise in the
resistivity was found with increasing field.  Above a value of about 20
kOe, the resistivity saturates.  It was pointed out that both the
behavior in a magnetic filed, as well as in zero field strongly
resembles that near the superconductor-to-insulator transition discussed
above, suggesting that the conducting phase might in fact be
superconducting.
\subsection{Conclusions}
We have seen that general scaling arguments can be employed to understand
the scaling behavior observed in various quantum phase transitions.  Most
of the experiments seem to confirm the expected value $z=1$ for a random
system with a $1/|{\bf x}|$-Coulomb interaction.  The number of different
universality classes present is yet not known.  Even if the
conductor-to-insulator transition observed in silicon MOSFET's at low
electron number densities turns out to be in the same universality class as
the superconductor-to-insulator transition, there are still the
field-induced transitions in quantum-Hall systems, which have a larger
correlation-length exponent.  

The paradigm provided by a repulsively interacting Bose gas, seems to be a
good starting point to describe the various systems.  However,
high-precision estimates calculated from this theory with impurities and
a $1/|{\bf x}|$-Coulomb interaction included are presently lacking.